\documentclass[]{aa}

\newcommand{\cha}{\textit{Chandra\/}}
\def\xmm{{XMM-{\it Newton\/}}}

\def\lu{{erg s$^{-1}$}}

\def\athena{{\it Athena}}
\def\cgs{{ erg cm$^{-2}$ s$^{-1}$}}

\def\lu{{ erg s$^{-1}$}}
\def\flu{{ erg s$^{-1}$} cm$^{-2}$}

\def\nustar{{\it NuSTAR}}

\def\myt{\texttt{MYTorus}}
\def\borus{\texttt{borus02}}
\def\sixte{\texttt{SIXTE}}

\def\aap{A\&A}

\def\apjl{ApJL}
\def\apjs{ApJS}

\usepackage{natbib}
\usepackage{float}
\usepackage{color}
\usepackage{graphicx}
\usepackage{textcomp,gensymb}
\usepackage{array}
\usepackage{mathtools}
\usepackage{multirow,makecell}
\usepackage{enumitem}
\usepackage{longtable}
\usepackage{hyperref}
\begin{document}

\title{Mock catalogs for the extragalactic X-ray sky: simulating AGN surveys with \athena\ and with the AXIS probe}

\titlerunning{Mock catalogs for the extragalactic X-ray sky}

\author{S. Marchesi\inst{1,2} \and R. Gilli\inst{1} \and G. Lanzuisi\inst{1} \and T. Dauser\inst{3} \and S. Ettori\inst{1,4} \and F. Vito\inst{5,6} \and N. Cappelluti\inst{7} \and A. Comastri\inst{1} \and R. Mushotzky\inst{8} \and A. Ptak\inst{9,10} \and C. Norman\inst{11,10}
}

\institute{INAF - Osservatorio di Astrofisica e Scienza dello Spazio di Bologna, Via Piero Gobetti, 93/3, 40129, Bologna, Italy \and
Department of Physics and Astronomy, Clemson University,  Kinard Lab of Physics, Clemson, SC 29634, USA \and
Dr Karl Remeis-Observatory and Erlangen Centre for Astroparticle Physics, Sternwartstr. 7, D-96049 Bamberg, Germany \and
INFN, Sezione di Bologna, viale Berti Pichat 6/2, 40127 Bologna, Italy \and
Instituto de Astrofisica and Centro de Astroingenieria, Facultad de Fisica, Pontificia Universidad Catolica de Chile,Casilla 306,Santiago 22, Chile \and 
Chinese Academy of Sciences South America Center for Astronomy, National Astronomical Observatories, CAS, Beijing 100012,China \and
Physics Department, University of Miami, Coral Gables, FL 33124
\and Department of Astronomy, University of Maryland, College Park, MD 20742 \and
NASA Goddard Space Flight Center, Code 662, Greenbelt, MD 20771, USA \and
Johns Hopkins University, 3400 N. Charles Street, Baltimore, MD 21218, USA \and
Space Telescope Science Institute, 3700 San Martin Dr., Baltimore, MD 21210, USA
}

\abstract{
We present a series of new, publicly available mock catalogs of X-ray selected active galactic nuclei (AGNs), non-active galaxies, and clusters of galaxies. These mocks are based on up-to-date observational results on the demographic of extragalactic X-ray sources and their extrapolations. They reach fluxes below 10$^{-20}$\,\cgs in the 0.5-2\,keV band, i.e., more than an order of magnitude below the predicted limits of future deep fields, and therefore represent an important tool for simulating extragalactic X-ray surveys with both current and future telescopes. We use our mocks to perform a set of end-to-end simulations of X-ray surveys with the forthcoming \athena\ mission and with the AXIS probe, a sub-arcsecond resolution X-ray mission concept proposed to the Astro 2020 Decadal Survey. 
We find that these proposed, next generation surveys may transform our knowledge of the deep X-ray Universe. As an example, in a total observing time of 15\,Ms, AXIS would detect $\sim$225,000 AGNs and $\sim$50,000 non-active galaxies, reaching a flux limit $f_{\rm 0.5-2}\sim$5\,$\times$\,10$^{-19}$\,\cgs in the 0.5--2\,keV band, with an improvement of over an order of magnitude with respect to surveys with current X-ray facilities. Consequently, 90\,\% of these sources would be detected for the first time in the X-rays.
Furthermore, we show that deep and wide X-ray surveys with instruments like AXIS and \textit{Athena} are expected to detect $\sim$20,000 $z>$3 AGNs and $\sim$250 sources at redshift $z>$6, thus opening a new window of knowledge on the evolution of AGNs over cosmic time and putting strong constraints on the predictions of theoretical models of black hole seed accretion in the early universe.
}

\keywords{X-rays: galaxies -- Surveys -- Galaxies: active -- Telescopes}

\maketitle

\section{Introduction}
Observations of the sky in the X-rays (i.e., in the 0.3--10\,keV energy range) are known to play a key role in the detection and understanding of different type of extragalactic sources. For example, X-ray data are an efficient tool to select accreting supermassive black holes (SMBHs), the so-called active galactic nuclei (AGNs). In fact, X-ray surveys are significantly less affected by contamination from non-AGN emission, in particular the one associated to star-formation processes, than optical and infrared surveys \citep[see, e.g.,][]{donley08,donley12,stern12}. As a consequence, only the deepest existing X-ray survey, i.e., the \cha\ Deep Field-South (CDF-S) 7\,Ms survey \citep{luo17}, reaches 0.5--2\,keV fluxes where the number of non-active galaxies becomes non negligible (i.e., $f_{0.5-2}\lesssim10^{-17}$\,erg\,s$^{-1}$\,cm$^{-2}$): in all the other X-ray surveys, AGNs make almost the totality of the detected sources. Characterizing the properties of the diffuse medium in clusters of galaxies, the largest virialized structures in the Universe, also requires deep X-ray surveys over large areas of the sky \citep[see, e.g.,][]{finoguenov15,kafer20}.

X-ray surveys are also effective in detecting obscured sources, i.e., objects where the circumnuclear dust and gas absorb the AGN optical emission. A significant fraction of these objects can even be heavily obscured, Compton thick AGNs (i.e., objects with optical depth to electron scattering $\sigma_\tau\geq$1, or column density N$_{\rm H}\geq1.5\times10^{24}\,cm^{-2}$), which are detected in X-ray surveys up to redshifts $z\sim$2--3 \citep[see, e.g.,][]{comastri11,georgantopoulos13,buchner15,lanzuisi15,lanzuisi18}, with only one known X-ray selected CT-AGN found at $z>$4 \citep[XID403, a $z$=4.76 detected in the \cha\ Deep field,][]{gilli11,gilli14}. These sources are especially interesting because they are expected to significantly contribute to the overall AGN population \citep[up to 50\%; see, e.g.,][]{gilli07,ananna19}, and X-rays represent one of the most efficient ways to find and characterize them. However, most of the CT population is too faint to be detected by current X-ray facilities.

A complete census of AGNs over a wide range of redshifts, column densities and luminosities requires a multi-tier approach. For example, intrinsically faint and/or heavily obscured sources need extremely deep surveys, which, as of today, could have been performed only on limited ($<$0.5\,deg$^2$) regions of the sky \citep[this is the case, for example, of the CDF-S 7\,Ms survey,][ AEGIS XD, \citealt{nandra15}, SSA22 \citealt{lehmer09}, or the J1030 field, \citealt{nanni20}]{luo17}; rare, luminous quasars can instead be found in significant numbers only using shallow large-area surveys that cover tens or even hundreds of square degrees \citep[see, e.g., the Stripe 82X survey,][or XMM-XXL, \citealt{pierre16}]{lamassa13a,lamassa13b}. Finally, intermediate-area surveys \citep[such as COSMOS,][]{civano16,marchesi16a} sample relatively deep fluxes over areas of a few deg$^2$, thus allowing one to detect statistically significant samples of high-$z$ sources and study large-scale structures.

Twenty years of observations with  \cha\ and \xmm\ provided us with large samples of X-ray selected AGNs, which allowed us to put constraints on the co-evolution of the accreting supermassive black holes and their host galaxies up to the peak of the AGN activity. However, our knowledge on the behavior of the first AGNs  which formed in the 2\,Gyr after the Big Bang (i.e., at redshift $z>$3) is still limited \citep[see, e.g.,][]{ueda14,aird15,buchner15,miyaji15,ananna19}. As of today, only $\sim$300 X-ray selected AGNs have been detected at $z>$3 \citep[see, e.g.,][]{vito14,vito18,marchesi16b}, and the farthest spectroscopically confirmed X-ray selected AGN has been found at $z$=5.31 \citep{capak11}. Furthermore, the vast majority of high-$z$ X-ray sources are detected with only a few source counts, thus limiting the scientific outcome of their analysis: in particular, understanding the evolution of the fraction of obscured AGNs \citep[which has been shown to increase with redshift in several works, see, e.g.,][]{treister13,vito18} becomes particularly complex, and significant incompleteness corrections need to be applied \citep[see, e.g.,][]{lanzuisi18}.  
Notably, while obscured AGN are expected to make the majority of the high-$z$ AGN population, at  $z>$6 optical/NIR surveys have so far detected only unobscured, luminous quasars \citep{banados16,wang17}. In the X-rays, a proper spectral characterization of heavily obscured AGNs requires the detection of hundreds of source counts \citep[see, e.g.,][]{marchesi18,marchesi19}, a task that cannot be achieved by current X-ray telescopes.

For these reasons, new X-ray facilities are required to improve our knowledge on AGNs and complement the information that will become available in the next decade in the other bands. Accurate survey simulations are key to properly assess the required technical layout of future facilities. Such simulations are based on accurate mock catalogs of extragalactic sources, that should be flexible enough to be mission-independent. 

One of the major issues when building an AGN mock catalog is which approach to choose to simulate the AGN population which is not detected by current surveys, particularly those sources at redshift $z>$6. For example, one can extrapolate the available observational evidence at redshift $z\sim$6 up to redshift $z\geq$10 \citep[this is for example the approach followed in the development of the \athena\ mission, see, e.g.,][]{nandra13}. Alternatively, one can use theoretical predictions from black hole seed early accretion models \citep[an approach followed by the \textit{Lynx} mission,][]{lynx18}.
Notably, since different missions often adopt different approaches to build their mock catalogs, high-$z$ AGN detection predictions cannot be directly compared, thus significantly limiting a cross-mission analysis of future X-ray surveys.

To address this issue, in this work we present a new set of mock catalogs, based on the most recent observational results on the demographic of extragalactic X-ray sources, and we then used them to simulate surveys with future X-ray facilities such as the AXIS probe \citep{mushotzky19} and \athena. The mocks are mission-independent and are made available to the public.

The paper is organized as follows: in Section \ref{sec:software_and_catalogs} we describe \texttt{SIXTE}, the software we used to simulate the X-ray surveys, and we discuss how we built our AGN, non-active galaxies and galaxy clusters mock catalogs.
In Section \ref{sec:axis} we then present the AXIS probe, its technical capabilities and its proposed scientific goals, and in Section \ref{sec:results} we present the results of the simulations of three different AXIS surveys. In Section \ref{sec:high-z} we focus on the improvement that \athena\ and AXIS would bring to our knowledge of high-$z$ and of heavily obscured AGNs. We then summarize our results in Section \ref{sec:summary}. In this work, we assume a flat $\Lambda$CDM cosmology with H$_0$=70\,km\,s$^{-1}$\,Mpc$^{-1}$, $\Omega_m$=0.3 and $\Omega_\Lambda$=0.7.

\section{New mock catalogs for X-ray simulations}\label{sec:software_and_catalogs}

\subsection{The \texttt{SIXTE} software}\label{sec:sixte}
The Monte Carlo code Simulation of X-ray Telescopes \citep[hereafter \sixte,][]{dauser19} allows one to simulate an observation with an X-ray telescope, following a three-step approach.
\begin{enumerate}
    \item At first, the tool creates a photon list containing the arrival time, energy and position of each photon. To generate this information, \sixte\ needs to be provided with the instrument effective area, field of view and pointing.
    \item The photon list created in the first step is then convoluted with the instrument point spread function (PSF) and vignetting to provide an impact list that contains the energy and arrival time of each photon, as well as its position on the detector.
    \item Finally, the charge cloud information reported in the impact list is read-out and re-combined into events, taking into account the properties of the simulated detector (e.g., read-out properties,  redistribution matrix file...). The output of this last step is an event file that can be analyzed.
\end{enumerate}

The \sixte\ webpage\footnote{\url{https://www.sternwarte.uni-erlangen.de/research/sixte/}} contains the configuration files (i.e., telescope setup, response matrices, vignetting, point spread function) for several current facilities, such as \xmm, \nustar\ and eROSITA. Furthermore, configuration files for new missions are also provided: for example, both \athena\ \citep{nandra13} instruments, i.e., the Wide Field Imager \citep[WFI,][]{meidinger17} and the X-ray Integral Field Unit \citep[X-IFU][]{barret16} are available. 
Finally, the \sixte\ manual contains all the information to build from scratch the configuration files for facilities that are not currently available on the website.

In order to perform the steps we mentioned above, \sixte\ needs a source list as an input: such a list is generated in the \texttt{SIMPUT} data format. \texttt{SIMPUT} source lists are detector-independent, i.e., the catalogs we originated for this work can be used to simulate any type of X-ray observation. 

Our catalogs are available online at \url{http://cxb.oas.inaf.it/mock.html} in FITS format and ready to be used, among other software, within \sixte: in the following sections, we describe how we built them. 

\subsection{AGN mock catalog}\label{sec:AGN_mock}
Three AGN mock catalogs are available on our webpage, for survey areas of 1, 10 and 100 deg$^2$, respectively. Random right ascension and declination have been associated to the \sixte\ catalogs.

Our mocks are made of sources with different intrinsic 0.5-2\,keV luminosities, redshifts and column densities, that have been extracted by resampling the X-ray luminosity function of unabsorbed AGN given by \citet{hasinger05}, scaled up by a luminosity--dependent factor to account for the whole AGN population \citep[see][]{gilli07}. 
At $z>$2.7, a decline in the AGN space density as parameterized in  \citet{schmidt95} has been assumed, while we did not make any assumption on AGN or host clustering. 

The resampling has been checked to reproduce the correct AGN densities as a function of luminosity, redshift and column density. To get the number of sources per unit solid angle, the source populations have been weighted for the volume element dV/d$z$/d$\Omega$. The obtained catalogs have then been checked to simultaneously reproduce the 0.5-2\,keV and 2-10\,keV input AGN number counts as a function of column density.

We point out that in \citet{gilli07} a distribution of photon indices was assumed for the primary power-law (with dispersion $\sigma_\Gamma$=0.2), while disk reflection was assumed only for lower luminosity AGN. For simplicity (i.e., to limit the number of spectral templates used in the analysis), in the generation of the mock catalogs we did not include any photon index dispersion and assumed disk reflection everywhere. Therefore, the 2--10\,keV k-correction was adjusted to an average value to correctly reproduce the total 2--10\,keV number counts.

AGNs have been simulated down to a 0.5--2\,keV luminosity L$_{0.5-2}$=10$^{40}$\,erg\,s$^{-1}$ and up to $z=$10: the source density of the mocks is $\sim$53,500 sources/deg$^2$. 
It is worth noting that several works have reported detections of AGNs with luminosities below the 10$^{40}$\,erg\,s$^{-1}$ threshold we decided to adopt, both in surveys \citep[see, e.g.,][]{aird15} and in the Local Universe \citep[e.g.,][]{,bi20,hodges20}. Therefore, our mock is likely missing a fraction of low-luminosity AGN. We will further discuss this topic in Section \ref{sec:cxb}.

\begin{figure*}
\begin{minipage}[b]{.33\textwidth}
  \centering
  \includegraphics[width=1\textwidth]{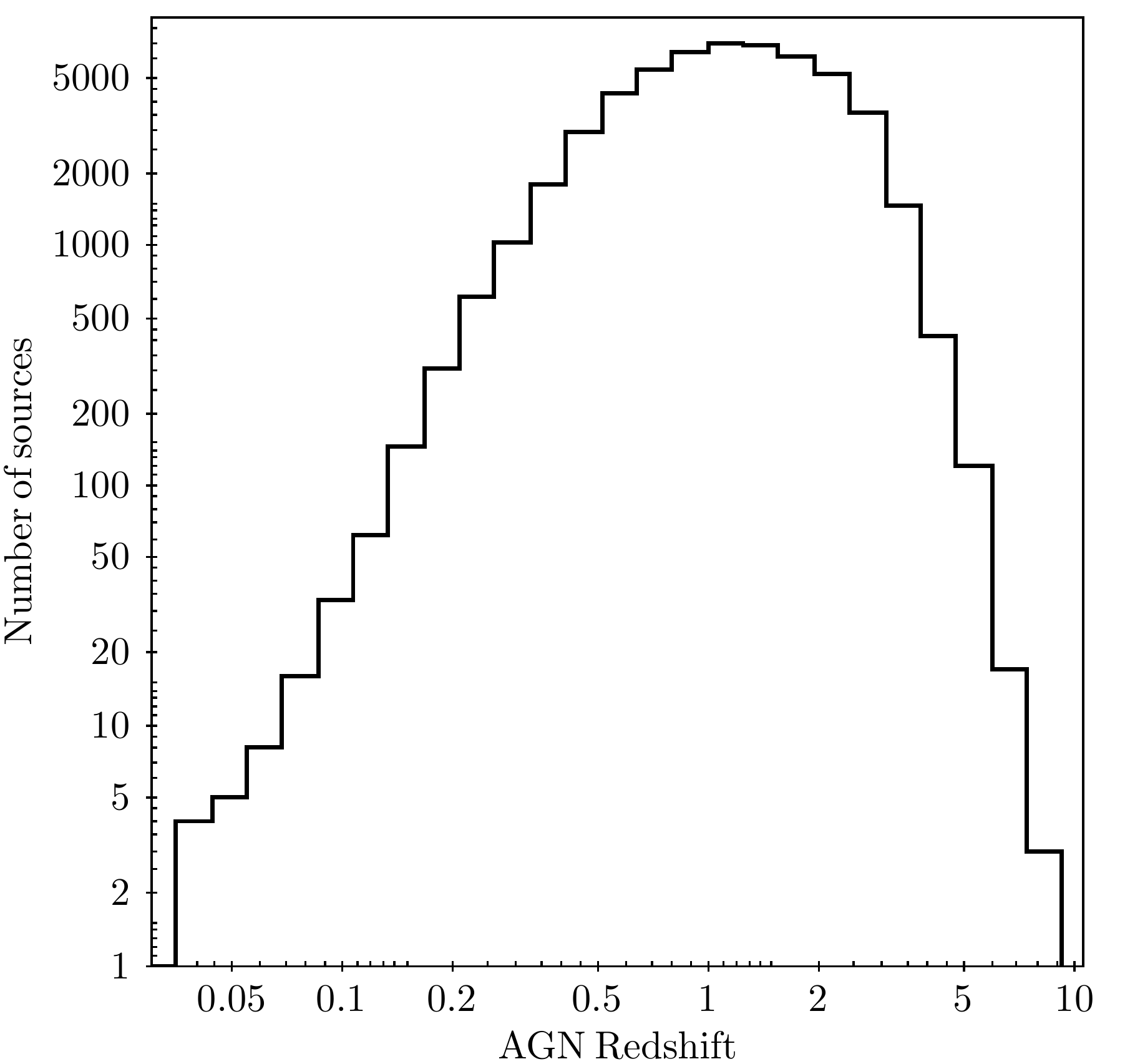}
  \end{minipage}
\begin{minipage}[b]{.33\textwidth}
  \centering
  \includegraphics[width=1\textwidth]{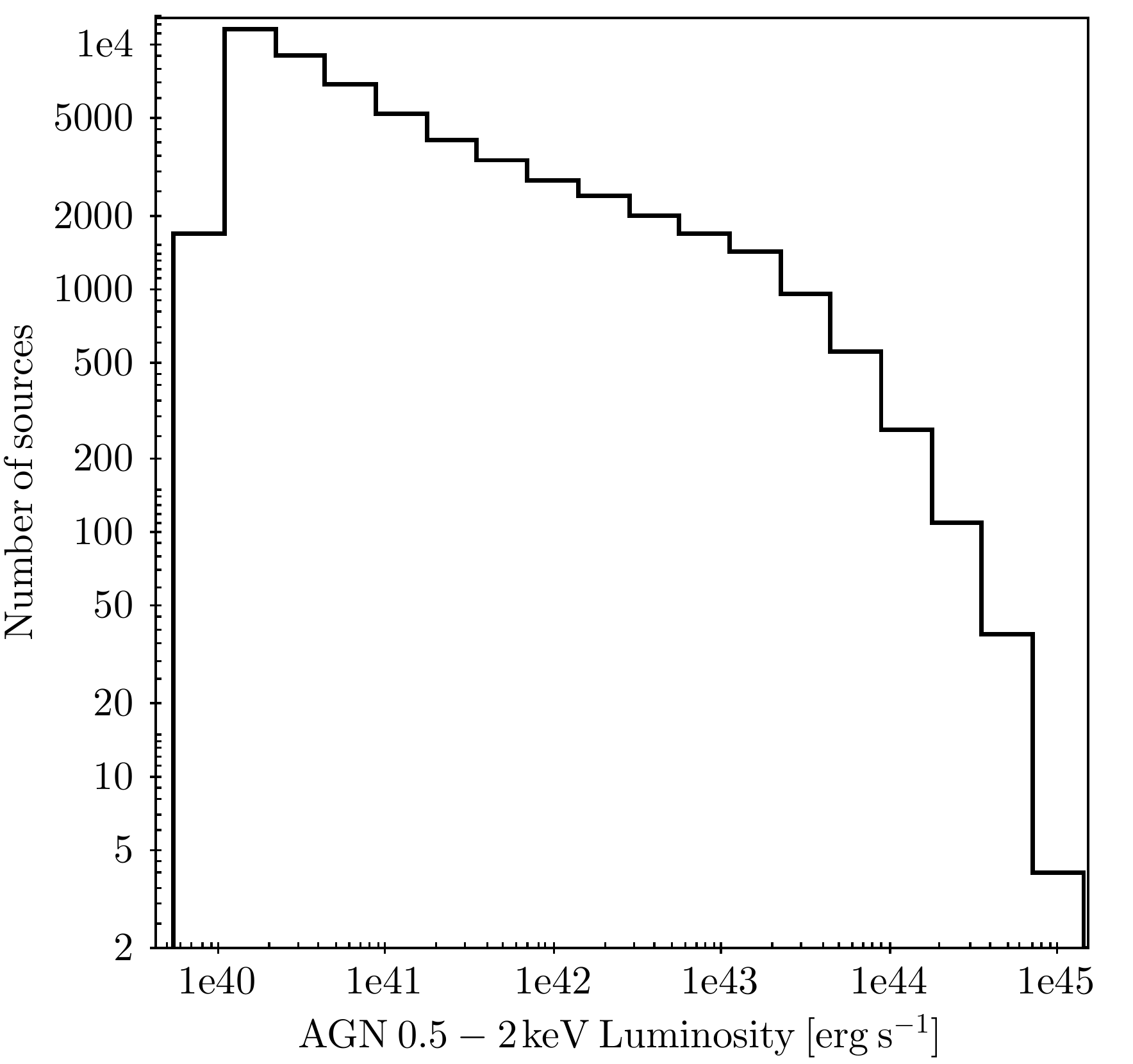}
  \end{minipage}
\begin{minipage}[b]{.31\textwidth}
  \centering
  \includegraphics[width=1\textwidth]{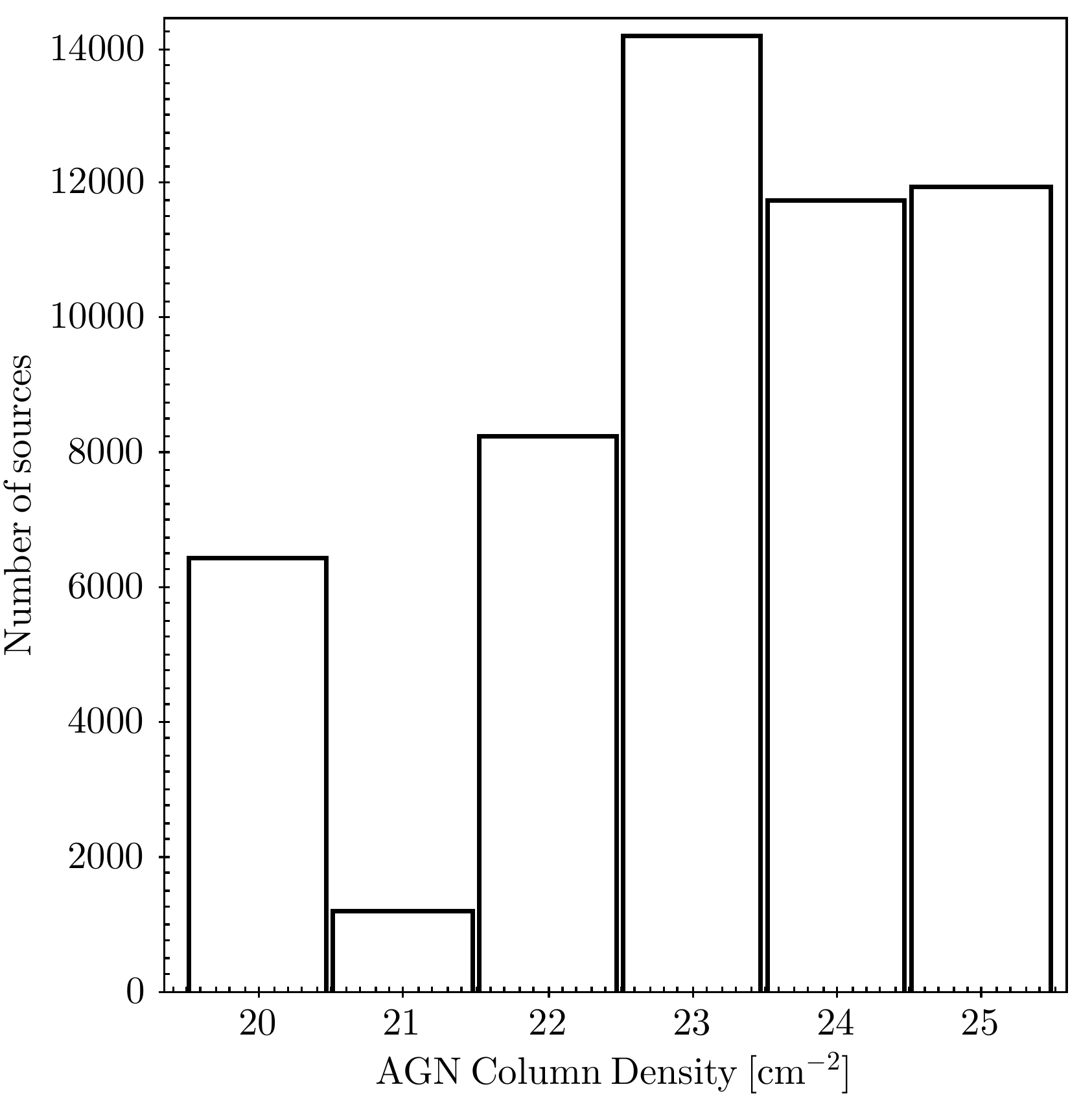}
  \end{minipage}
  \caption{\normalsize Distribution of redshift (left), 0.5--2\,keV luminosity and column density (right) for the 53,579 AGNs contained in one of the mock catalogs used in our analysis. The sources cover a 1\,deg$^2$ field and are drawn from the \citet{gilli07} model.}\label{fig:AGN_input_info}
\end{figure*}

We report in Figure \ref{fig:AGN_input_info} the redshift, 0.5--2\,keV luminosity and column density (N$_{\rm H}$) distribution of the sources in the AGN mock catalog. As it can be seen, N$_{\rm H}$ is quantized, following the same approach and column density distribution adopted in the developing of the \citet{gilli07} cosmic X-ray background model.
More in detail, we assign to each source a column density value taken from the array log(N$_{\rm H}$)=[20.5, 21.5, 22.5, 23.5, 24.5, 25.5], each representative of 1-dex wide column density bin. The average Compton thick fraction, i.e., the fraction of sources with  column density log(N$_{\rm H}$)$>$24, is $f_{\rm CT}$=39\,\%. As a reference, we report the average CT fraction values adopted in other AGN population synthesis models, such as, e.g., those by \citet[][$f_{\rm CT}$=33\,\%]{ueda14}, \citet[][$f_{\rm CT}$=38\,\%]{buchner15}, and \citet[][$f_{\rm CT}$=50\,\%]{ananna19}. We remark that these are all average values, since all models adopt a luminosity-dependent CT fraction, based on the observational evidence \citep[see, e.g.,][]{ricci15,marchesi16a}.

\begin{figure*}[htbp]
\begin{minipage}[b]{.32\textwidth}
\centering
\includegraphics[width=1.\linewidth]{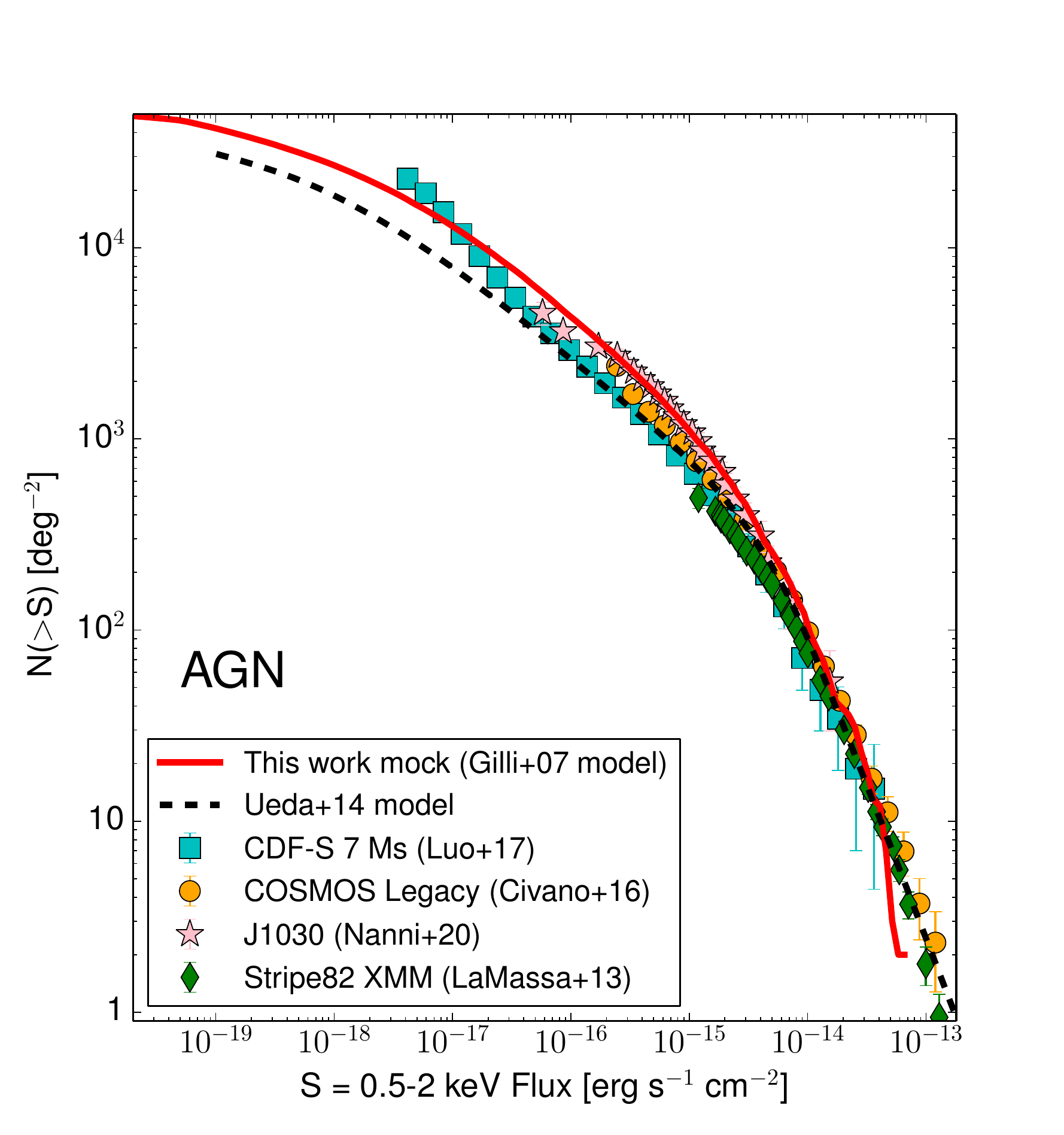}
\end{minipage}
\begin{minipage}[b]{.31\textwidth}
\centering
\includegraphics[width=1.\linewidth]{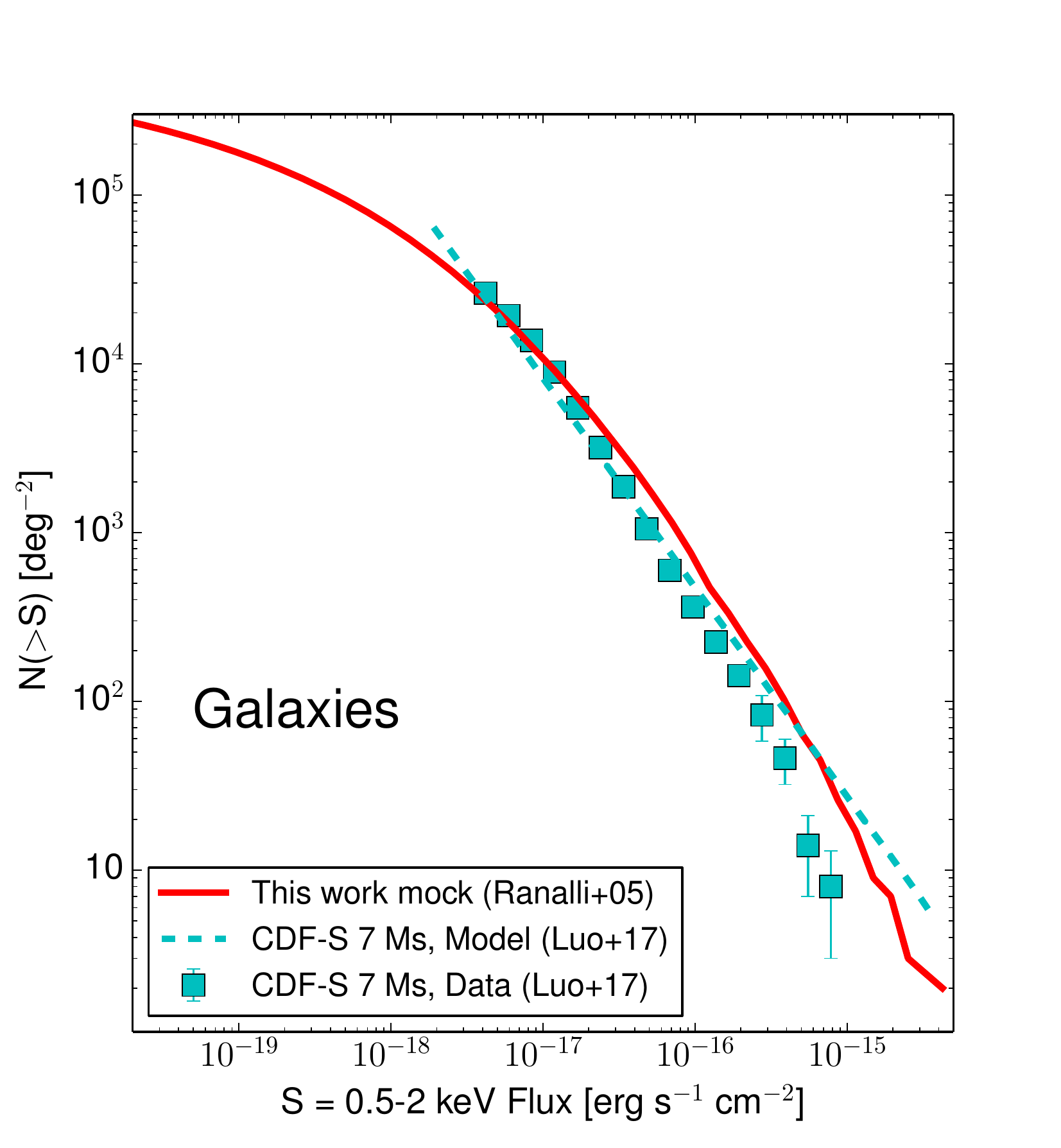}
\end{minipage}
\begin{minipage}[b]{.33\textwidth}
\centering
\includegraphics[width=1\linewidth]{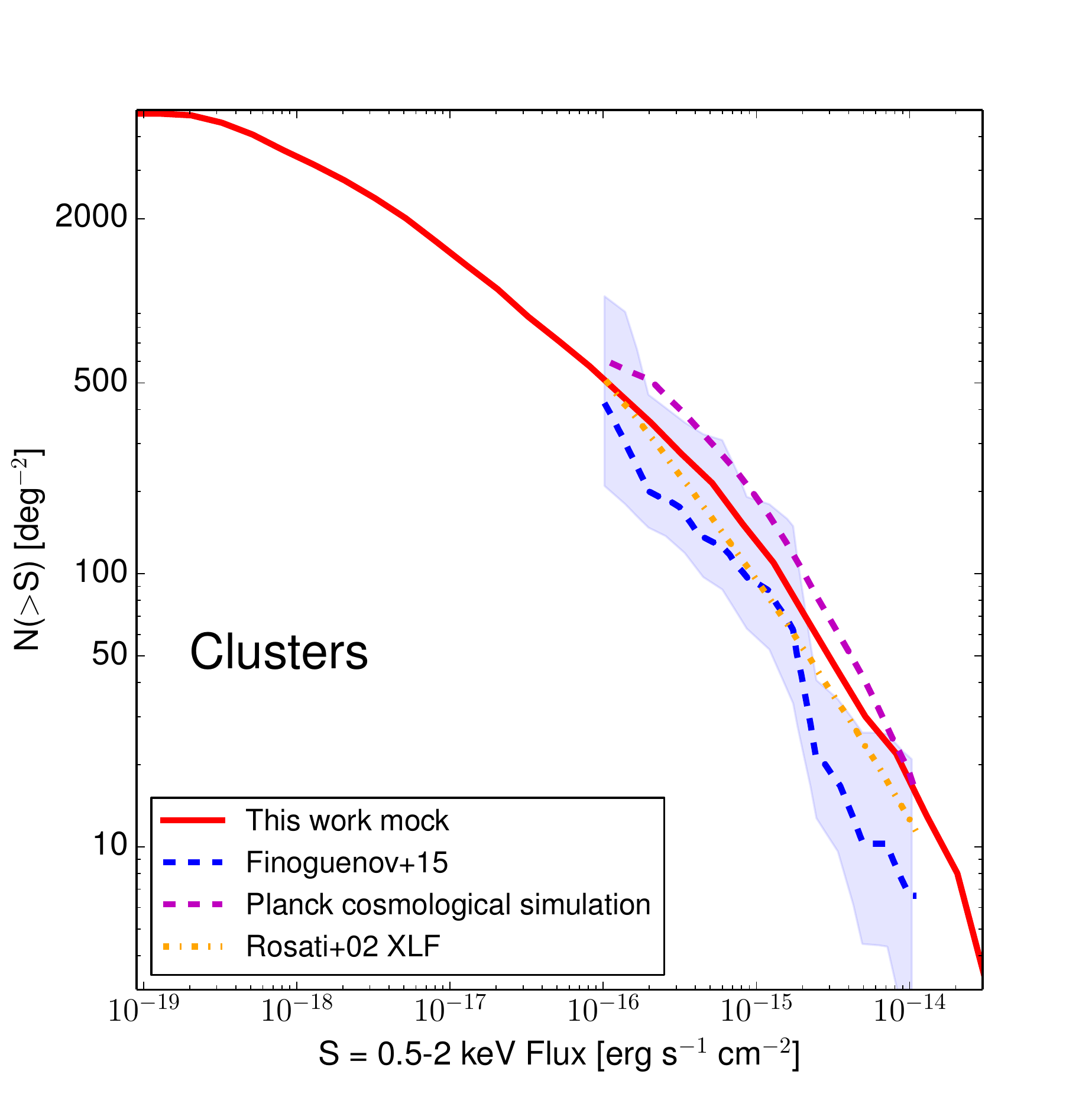}
\end{minipage}
\caption{\normalsize \textit{Left}: cumulative AGN number counts as a function of the 0.5--2\,keV flux. The number counts from our mock catalog, derived using the \citet{gilli07} AGN population synthesis model, are plotted as a solid red line. As a comparison, we also plot the number counts derived in the CDF-S 7\,Ms \citep[cyan squares;][]{luo17}, \cha\ COSMOS Legacy \citep[orange circles;][]{civano16}, J1030 \citep[pink stars;][]{nanni20}, and Stripe 82X \citep[green diamonds;][]{lamassa13a} surveys. The number counts derived from the \citet{ueda14} X-ray luminosity function are also shown with a dashed black line, for comparison. We note that the difference at faint fluxes between these number counts and those of our mock is almost completely due to the fact that \citet{ueda14} did not include AGNs with luminosities $<$10$^{41}$\,\lu\ in their analysis. 
\textit{Center}: cumulative non-active galaxies number counts as a function of the 0.5--2\,keV flux. The number counts from our mock catalog, derived using the \citet{ranalli05} luminosity function, are plotted as a solid red line. The observed (cyan squares) and model-inferred number counts (dashed cyan line) derived in the CDF-S 7\,Ms \citep[][]{luo17} are also shown for comparison.  
\textit{Right}: cumulative galaxy clusters number counts as a function of the 0.5--2\,keV flux. The number counts from our mock catalog are plotted as a solid red line. As a comparison, we also plot the number counts derived in the Extended \cha\ Deep Field--South field \citep[dashed blue line; the 1\,$\sigma$ uncertainty area is shown as a blue shaded area;][]{finoguenov15}, and those derived combining data from ROSAT, ASCA and \textit{Beppo}-SAX surveys \citep[dash-dotted orange line;][]{rosati02}. Finally, we plot as magenta dashed line the predictions of the \textit{Planck} cosmological simulation \citep[][]{planck14}. }\label{fig:number_counts_AGN_clusters}
\end{figure*}

Finally, in Figure \ref{fig:number_counts_AGN_clusters}, left panel, we plot the cumulative AGN number counts as a function of the observed 0.5--2\,keV flux for our mock catalog, and we compare them with those obtained in the deep, pencil-beam CDF-S 7\,Ms survey \citep{luo17}, in the intermediate \cha\ COSMOS Legacy survey \citep{civano16} and in the wide-area, shallow flux Stripe 82X survey \citep{lamassa13a}. Our mock nicely matches with the available observational results, over a wide range of fluxes.

In the same figure, we also report the AGN number counts derived using the \citet{ueda14} X-ray luminosity function (XLF) model. As it can be seen, our predictions are consistent with the \citet{ueda14} model. The agreement is even stronger considering that the \citet{ueda14} model accounts only for AGNs having redshift $z$=[0--5] and 0.5--2\,keV luminosity L$_{\rm 0.5-2}$=[10$^{41}$--10$^{46}$]\,erg\,s$^{-1}$\,cm$^{-2}$, while our mock contains objects up to redshift $z$=10 and, more importantly, down to L$_{\rm 0.5-2}$=10$^{40}$\,\lu. The inclusion of L$_{\rm 0.5-2}<$10$^{41}$\,\lu\ AGNs is particularly important, considering that our mock contains $\sim$30,000 of such sources over 1\,deg$^2$ at 0.5-2 keV fluxes $<$10$^{-17}$\,\flu. 
Adding these 30,000 sources to the \citet{ueda14} AGN number counts, the discrepancy between the two models is $<$30\,\% at fluxes $f_{\rm 0.5-2}\sim$10$^{-17}$\,\flu, and $<$10\,\% at $f_{\rm 0.5-2}\sim$10$^{-19}$\,\flu.  Therefore, the predictions we make in Section \ref{sec:results} on the overall number of AGNs detected by AXIS and \athena\ would not change significantly if we used the \citet{ueda14} XLF instead of the \citet{gilli07} AGN population synthesis model to generate our mock.

\subsubsection{AGN spectral models}
\texttt{SIXTE} allows one to associate a spectral model to each source in the mock catalog: this model is then used to convolve the 0.5--10\,keV flux which is given to each source when the mock is created. 

We created our spectral models using the \texttt{XSPEC} software \citep{arnaud96}: in \texttt{XSPEC}, nomenclature, the model is built as follows:

\begin{equation}
\label{eq:powerlaw}
\begin{aligned}
phabs*(zcutoffpl+pexmon
+zphabs*zcutoffpl),
\end{aligned}
\end{equation}

where $phabs$ is the Galactic absorption (fixed to $N_{\rm H,Gal}$=1.8$\times$10$^{20}$), $zcutoffpl$ is a power law with photon index $\Gamma$=1.9 \citep[a typical value for AGNs, see, e.g.,][and references therein]{marchesi16c} and high-energy cutoff $E_{\rm cut}$=200\,keV, $pexmon$ \citep{nandra07} models the emission reprocessed by cold material surrounding the accreting supermassive black hole, including  self-consistently generated fluorescence lines such as the Fe K$\alpha$, the Fe K$\beta$, the Ni K$\alpha$ and the Fe K$\alpha$ Compton shoulder. In $pexmon$, the photon index and cut-off energy are tied to the $zcutoffpl$ one, while the normalization is assumed equal to 37\,\% of the main power-law one, which corresponds to $\sim$2\,\% of the overall emission in the 2--10\,keV band \citep[see, e.g.,][]{gilli07}. While several more recent models are available to treat the reprocessed component \citep[e.g., \myt,][or \texttt{borus02}, \citealt{balokovic18}]{murphy09}, we used $pexmon$ because of its consistency  with the model originally used in \citet{gilli07}.
Finally, $zphabs$ models the photoelectric absorption of a material with column density N$_{\rm H}$. 

The first power law component is phenomenological and based on observational evidence, and it models the fraction of emission which is scattered, rather than absorbed by the obscuring material, as well as the unresolved emission lines from photoionized gas in the narrow line region, and even soft X-ray emission from star forming processes in the host galaxy.
The two power laws have same $\Gamma$ and $E_{\rm cut}$, and the normalization of the unabsorbed power law is set to be equal to 3\,\% of the main one \citep[following what is typically observed in X-ray surveys, see, e.g.,][]{marchesi16c}. For the most heavily obscured sources, i.e., those with log(N$_{\rm H}$)=25.5, we set the normalization of the power law component to 0, assuming that all emission comes from the $pexmon$ component: we do so because the $zphabs$ component does not accurately describe the absorption caused by material having column density log(N$_{\rm H}$)$>$25. 

\begin{figure}[htbp]
\centering
\includegraphics[width=1.\linewidth]{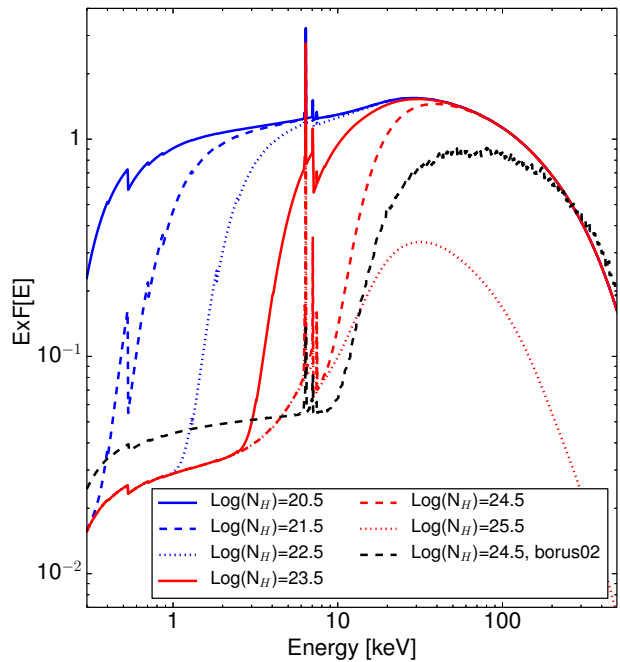}
\caption{\normalsize AGN spectra used in our simulations, at $z$=0 and for different log(N$_{\rm H}$) values.We also plot the spectrum of a CT-AGN with log(N$_{\rm H}$)=24.5 as modeled using the \borus\ model \citep[][dashed black line]{balokovic18}, as a reference.}\label{fig:AGN_spectra}
\end{figure}

We generated a set of spectra covering the redshift range $z$=[0--10], with a redshift bin $\Delta z$=0.1, adopting the same column density array used in the \citet{gilli07} AGN population synthesis model, i.e., log(N$_{\rm H}$)=[20.5, 21.5, 22.5, 23.5, 24.5, 25.5]. We then associated to each source the spectrum with the same N$_{\rm H}$ and the closest $z$. We show an example of spectrum for each Log(N$_{\rm H}$) value in Figure \ref{fig:AGN_spectra}.

While the \texttt{pexmon} model generally offers a reliable characterization of the reprocessed emission in heavily obscured AGN, in the past 10 years several models have been developed to treat this complex spectral component in a more self-consistent way \citep[e.g.,][\borus\ \citealt{balokovic18}]{murphy09}. These models have as free parameters physical quantities such as the torus average column density, $N_{\rm H}$, which is different from the line-of-sight column density if the torus is inhomogeneous, and the torus covering factor. Thus, in Figure \ref{fig:AGN_spectra} we also report the spectrum of a CT-AGN with log(N$_{\rm H}$)=24.5 as modeled by the self-consistent \borus\ model \citep{balokovic18}, assuming a torus covering factor $f_c$=0.5 and and an almost edge-on viewing angle, $\theta_{\rm obs}$=87\degree.

As shown in Figure \ref{fig:AGN_spectra}, in heavily obscured sources the $>$10\,keV flux predicted by the physically motivated \borus\ model is 30--60\,\% fainter than the one predicted by the \texttt{pexmon} one.
This discrepancy, while not extreme, can affect the number of heavily obscured sources in simulations with our mocks. We will further discuss this effect in Section \ref{sec:results-CT}.

\subsubsection{Mock catalog of high-redshift AGNs}
One of the fundamental scientific topics that can be addressed by X-ray surveys, particularly with next-generation facilities, is the study and characterization of the high-redshift AGN population. As of today, our knowledge on the $z>$3 AGN population is extremely limited, since only a few hundreds of these sources have been detected in X-ray surveys \citep[see, e.g.,][]{vito14,vito18,marchesi16b,nanni20}, and there are no X-ray selected AGN at $z>$6.

While the \citet{gilli07} model is in overall excellent agreement with the observational results obtained by the most up-to-date X-ray surveys (as shown in Figure \ref{fig:number_counts_AGN_clusters}, left panel), in the high-redshift regime (i.e., at $z>$3, where the AGN space density starts its decline) there is a more significant discrepancy between the predictions of the AGN population synthesis model and the observational evidence. 

\begin{figure*}[htbp]
\begin{minipage}[b]{.50\textwidth}
  \centering
\includegraphics[width=1.\linewidth]{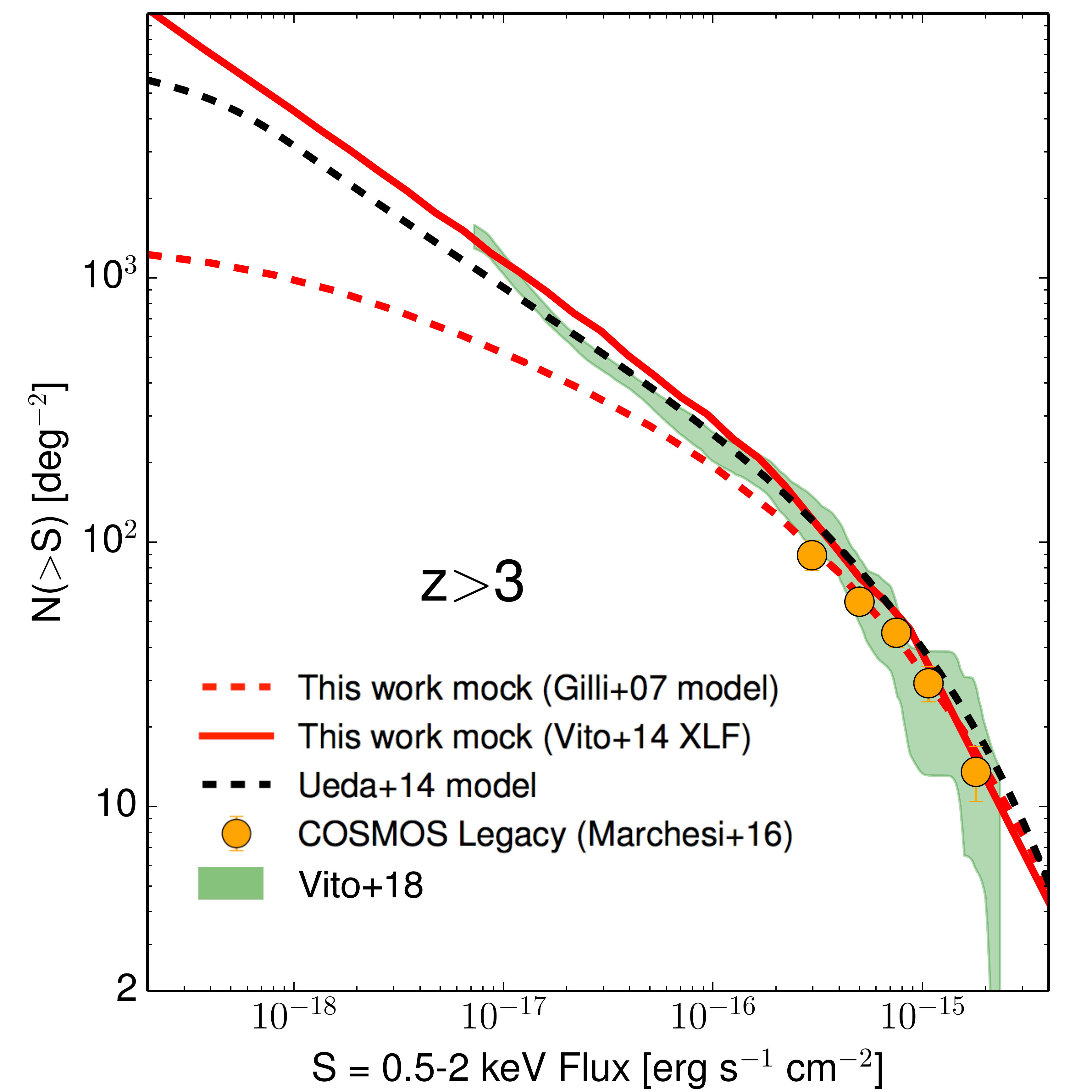}
  \end{minipage}
\begin{minipage}[b]{.47\textwidth}
  \centering
  \includegraphics[width=1\textwidth]{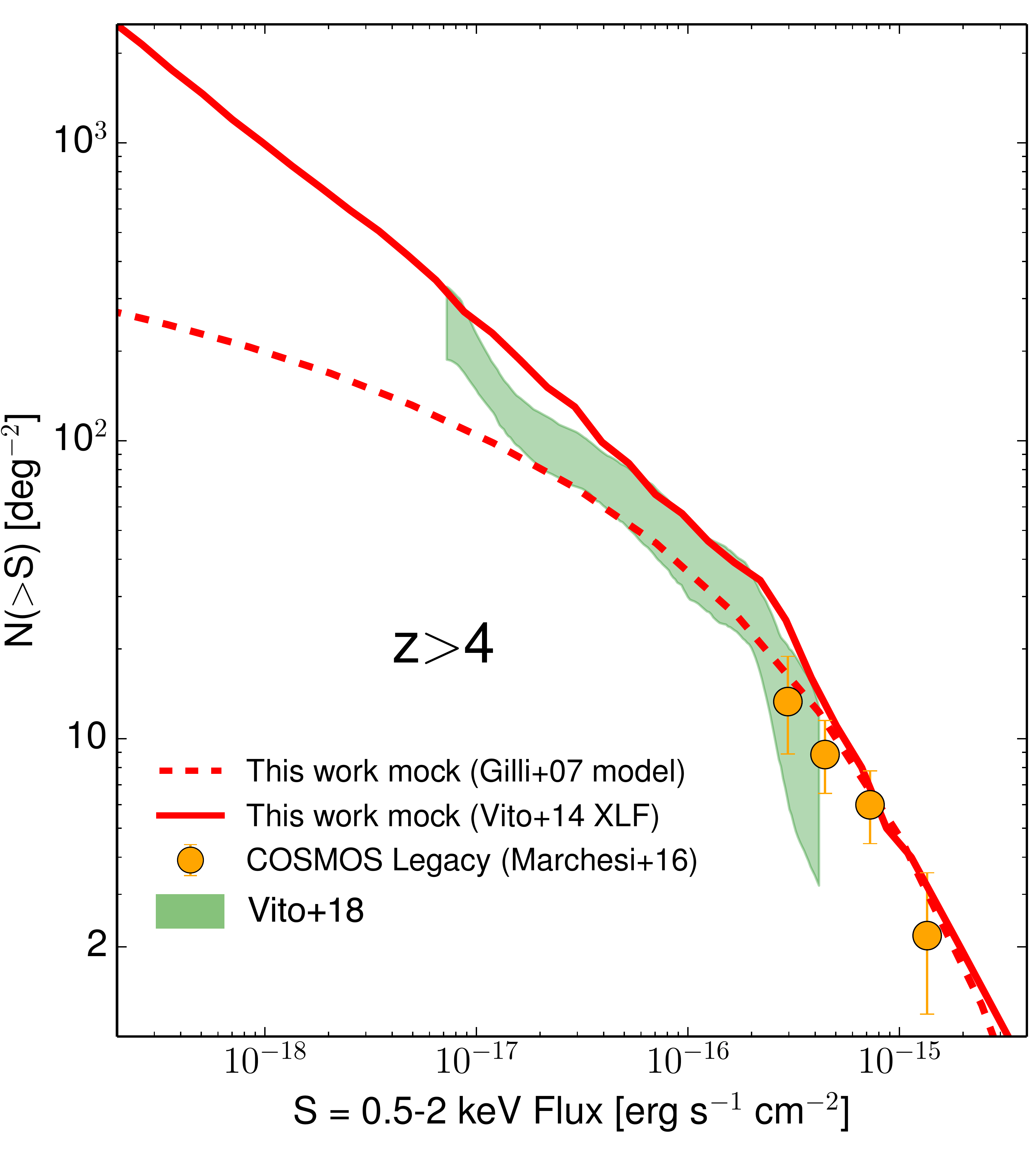}
  \end{minipage}
\caption{\normalsize Cumulative 0.5--2\,keV number counts at redshift $z>$3 (left) and $z>$4 (right). The number counts of the mock catalog we derived from the \citet{vito14} $z>$3 XLF are plotted as a solid red line, while those of the mock derived using the \citet{gilli07} CXB model are shown using a dashed red line. The $z$=[3--5] number counts derived with the \citet[][dashed black line]{ueda14} model are also shown for comparison. Finally, the observational results from \cha\ COSMOS Legacy \citep[orange circles,][]{marchesi16b} and from the CDF-S 7\,Ms and CDF-N 2\,Ms \citep[green shaded area highligting the  68\,\% confidence region;][]{vito18} are also plotted for comparision.
}\label{fig:logn-logs_zgt3}
\end{figure*}

In particular, as shown in Figure \ref{fig:logn-logs_zgt3}, at fluxes $<$10$^{-16}$\,erg\,s$^{-1}$\,cm$^{-2}$ the predictions of the \citet{gilli07} model lie below the number counts observed in the deepest X-ray surveys currently available \citep[the CDF-S 7\,Ms and the CDF-N 2\,Ms, see][]{vito18}. The \citet{vito18} data is instead in close agreement with the predictions of the X-ray luminosity function (XLF) by \citet[][]{vito14}, which was computed  using a sample of 141 X-ray selected high-$z$ AGNs. 

For this reason, we generated a second AGN mock catalog, containing only $z>$3 sources, based on the \citet[][]{vito14} XLF.  
More in detail, the new $z>$3 mock catalog is obtained by simulating AGN in the redshift range $z$=[3-20] and down to log(L$_{\rm 0.5-2keV}$)=40 extrapolating the XLFs of \citet{vito14}, which were originally derived in the redshift range $z$=[3-5] and for log(L$_{\rm 0.5-2keV}$)$>$43. With these redshift and luminosity limits, AGNs in the mock catalog reach fluxes below 2$\times$10$^{-20}$\cgs.

To generate the \citet[][]{vito14} mock catalog, AGNs with different intrinsic 2-10 keV luminosities, redshifts and column densities have been extracted by resampling the pure density evolution (PDE) model XLF corrected for redshift incompleteness \citep[see Figure 7 and Table 5 of][]{vito14} We assumed: $i$) a constant obscured AGN fraction with luminosity; $ii$) number ratios between unobscured, obscured Compton-thin, and obscured Compton-thick AGNs of 1:4:4; $iii$) the same column density distribution of \citet[][see also Figure \ref{fig:AGN_input_info}, right panel]{gilli07}.  All these assumptions are in agreement with the observational results reported in \citet{vito18}. Finally, to compute the number of sources per unit of solid angle, the source populations have been weighted for the volume element dV/d$z$/d$\Omega$. We use the same spectral templates used for the \citet{gilli07} mock and presented in Section \ref{sec:AGN_mock}, do no to include any dispersion in photon indices, and assume disk reflection in each template.

In Figure \ref{fig:logn-logs_zgt3}, left panel, we also report the $z>$3 number counts derived using the \citet{ueda14} XLF,  which are in excellent agreement with the predictions of our \citet{vito14} mock catalog. The discrepancy at 0.5--2 keV fluxes $<$10$^{-17}$\,\flu\ is mostly due to the \citet{ueda14} model not including AGNs at $z>$5. As a reference, the \citet{vito14} mock contains $\sim$3300 $z>$5 AGNs at fluxes $f_{\rm 0.5-2}<$10$^{-17}$\,\flu: adding this number of sources to the \citet{ueda14} number counts, the discrepancy between the two is $<$30\,\%, and is dominated by the uncertainties of the XLF models reported by \citet{ueda14} and \citet{vito14}.

Consequently, we expect that simulations with a mock generated using the \citet{ueda14} XLF would produce high-$z$ predictions in close agreement with those which we will present in Section \ref{sec:high-z} and we obtained using the \citet{vito14} mock.

\subsection{Galaxy mock catalog}\label{sec:galaxy_mock}
Galaxies (here and in the rest of the paper this is how we define point-like sources for which the X-ray emission is not caused by an accreting supermassive black hole) have been extracted by interpolating between the peakM and peakG model logN-logS by \citet{ranalli05}. Galaxy spectra have been assumed to be simple power-laws with photon index $\Gamma$=2, i.e. their band ratios do not depend on redshift \citep[see, e.g.,][]{sazonov17,barger19}. The Galactic absorption is the same used for AGNs, i.e., N$_{\rm H}$=1.8$\times$10$^{20}$\,cm$^{-2}$.

Galaxies have been simulated down to a 0.5--2\,keV flux limit $f_{0.5-2}$=10$^{-20}$\,erg\,s$^{-1}$\,cm$^{-2}$: the mock catalog we use in our analysis contains 294,000 sources per unit of deg$^2$.

We show in Figure \ref{fig:number_counts_AGN_clusters}, central panel,  the 0.5--2\,keV flux number counts of the simulated non-active galaxies, which are in good agreement with the most recent observational results in the CDF-S 7\,Ms \citep[][]{luo17}. Since the main focus of this work are active galactic nuclei, we do not associate a redshift (and therefore a luminosity) to the galaxies in our mock catalog. 

\subsection{Galaxy clusters mock catalog}\label{sec:clusters_mock}
To build a catalog of X-ray sources associated to galaxy clusters, we start with the predictions of dark matter halos extracted from a numerical mass function for a given set of cosmological parameters: we use the mass functions of both \cite{tinker08} and \cite{despali16}.
Specifically, we adopt $H_0=70$ km/s/Mpc, $\Omega_{\rm m}=1-\Omega_{\Lambda}=0.3$ and $\sigma_8 = 0.79$ \citep[see present cosmological constraints from galaxy clusters in][]{pratt19}.
The number density is estimated in the mass range $10^{12} - 4 \times 10^{15} M_{\odot}$
in 200 redshift bins between $z=0.03$ and $z=6$ using the python code {\tt Colossus}\footnote{http://www.benediktdiemer.com/code/colossus/} \citep{diemer18}.
Then, it is integrated over the cosmological volume to recover the number of haloes expected at given mass and redshift per square degree.
We also associate to each considered mass value, $M_{500}$, its corresponding radius $R_{500} = M_{500}^{1/3} / (4/3 \, \pi \, 500 \, \rho_{cz})^{1/3}$ in arcminutes\footnote{$\rho_{cz} = 3 H_0^2 E_z^2 / (8 \pi G)$ is the critical density of the Universe at given redshift, being $E_z = \left[ \Omega_{\Lambda} - \Omega_{\rm m} (1+z)^3 \right]^{0.5}$ and $G$ the universal gravitational constant.}, and an X-ray luminosity and temperature $kT$ as estimated from available scaling relations \citep[e.g.,][]{reichert11}. The clusters radial profile is assumed to be a simple $\beta$-profile.

We then use \texttt{XSPEC} to convert the X-ray luminosities into fluxes in the observed 0.5–2\,keV band and in the corresponding surface brightness, including the effect of Galactic absorption, in units of erg\,s$^{-1}$\,cm$^{-2}$\,arcmin$^{-2}$. The cluster spectral model we use is (in \texttt{XSPEC} nomenclature) $phabs*apec$, where $phabs$ is the same Galactic absorption used for AGNs and non-active galaxies, while $apec$ models emission from a collisionally-ionized diffuse gas with temperature $kT$ and metallicity $Z$=0.3\,$Z_\odot$, i.e.,  the average metallicity value for clusters as reported in the literature \citep[see, e.g.,][]{balestra07,maughan08,baldi12}. Finally, the cluster number counts are obtained by summing the estimated counts in each mass and redshift bin with an associated X-ray surface brightness above a given threshold and multiplying it by the explored area. We simulated clusters down to a 0.5--2\,keV surface brightness limit SB=10$^{-18}$\,erg\,s$^{-1}$\,cm$^{-2}$\,arcmin$^{-2}$.

As a consistency test, in Figure \ref{fig:number_counts_AGN_clusters}, right panel,  we report the cumulative number counts derived using our clusters mock catalog, and we compare them with results obtained from both real data and simulations. As it can be seen, the number counts derived using our mocks are consistent with the measurements from the ECDF-S \citep[][]{finoguenov15} and with those derived combining ROSAT, ASCA and \textit{Beppo}-SAX data \citep{rosati02}. We also plot the simulated number counts obtained using the \textit{Planck} cosmology \citep{planck14} and the \citet{leauthaud10} scaling relations: our data always lie slightly below the ones from the \textit{Planck} simulation, thus ensuring that we are not overpredicting the number of clusters in our mocks, at any flux.

Since the focus of this work is the detection of AGNs, in the rest of the paper we will not discuss the detection of galaxy clusters, which for the purposes of our analysis are therefore treated as fore/background emitters.

\subsection{Consistency with the cosmic X-ray background}\label{sec:cxb}
\begin{figure}[htbp]
\centering
\includegraphics[width=1.\linewidth]{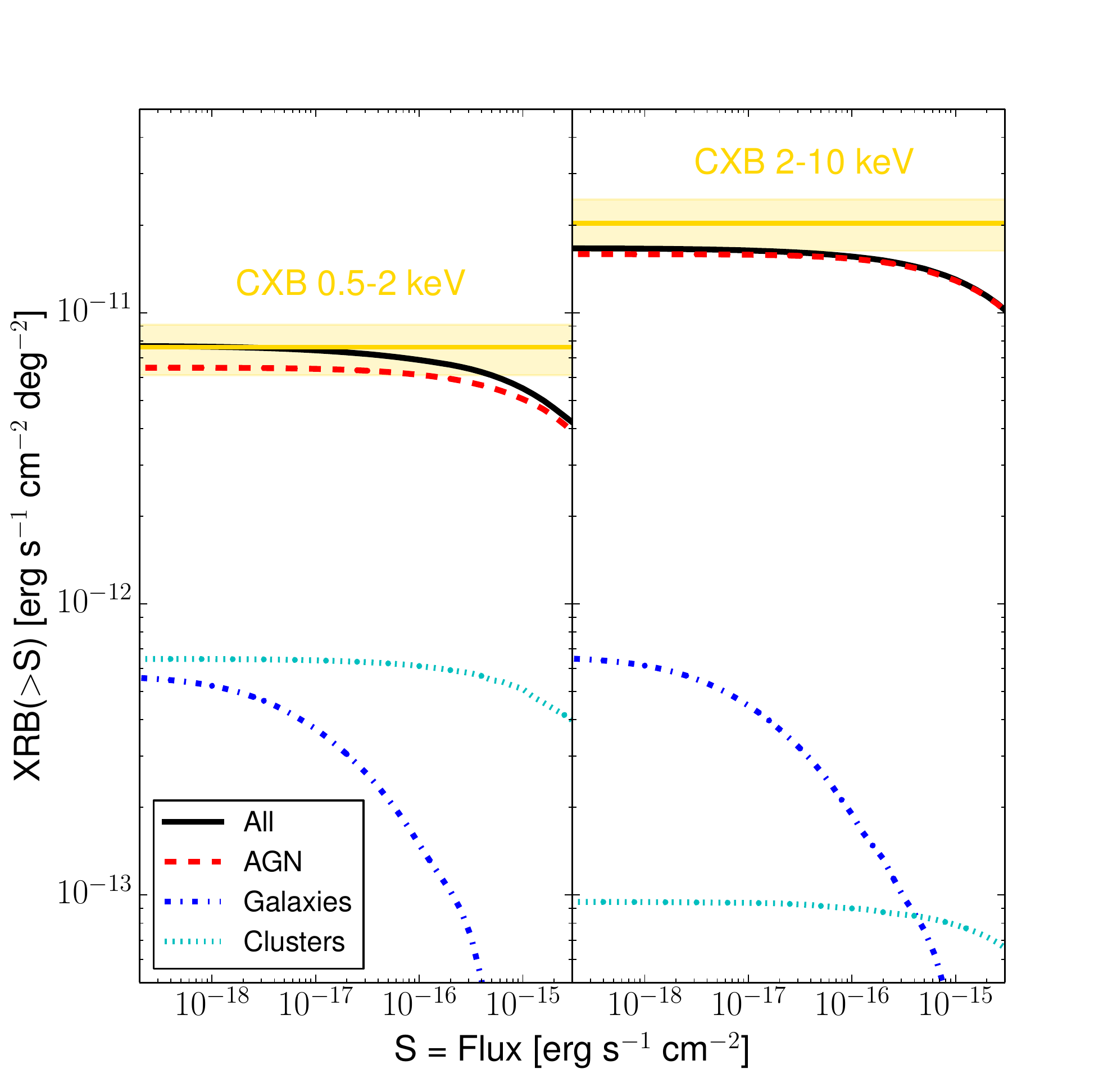}
\caption{\normalsize Cumulative 0.5--2\,keV (left) and 2--10\,keV (right) fluxes per unit of square degree of the AGNs (dashed red line), non-active galaxies (dash-dotted blue line) and galaxy clusters (dotted cyan line) simulated for the mocks used in this work. The overall distribution is plotted as a solid black line. The flux of the cosmic X-ray background measured by \citet{cappelluti17} is plotted in yellow, with 20\,\% errors to account for the uncertainties in the CXB absolute flux level measured by different instruments.}\label{fig:cxb}
\end{figure}

In Figure \ref{fig:cxb} we compare the cumulative 0.5--2\,keV and 2--10\,keV fluxes per unit of square degree of our mocks, divided by class of sources (i.e., AGNs, non-active galaxies and galaxy clusters), with the overall surface brightness of the extragalactic cosmic X-ray background (CXB) in the same bands \citep{cappelluti17}. Our mocks do not over-estimate the CXB in both the soft and the hard band, an evidence that further validates the reliability of the catalogs we generated.

\begin{figure}[htbp]
\centering
\includegraphics[width=1.\linewidth]{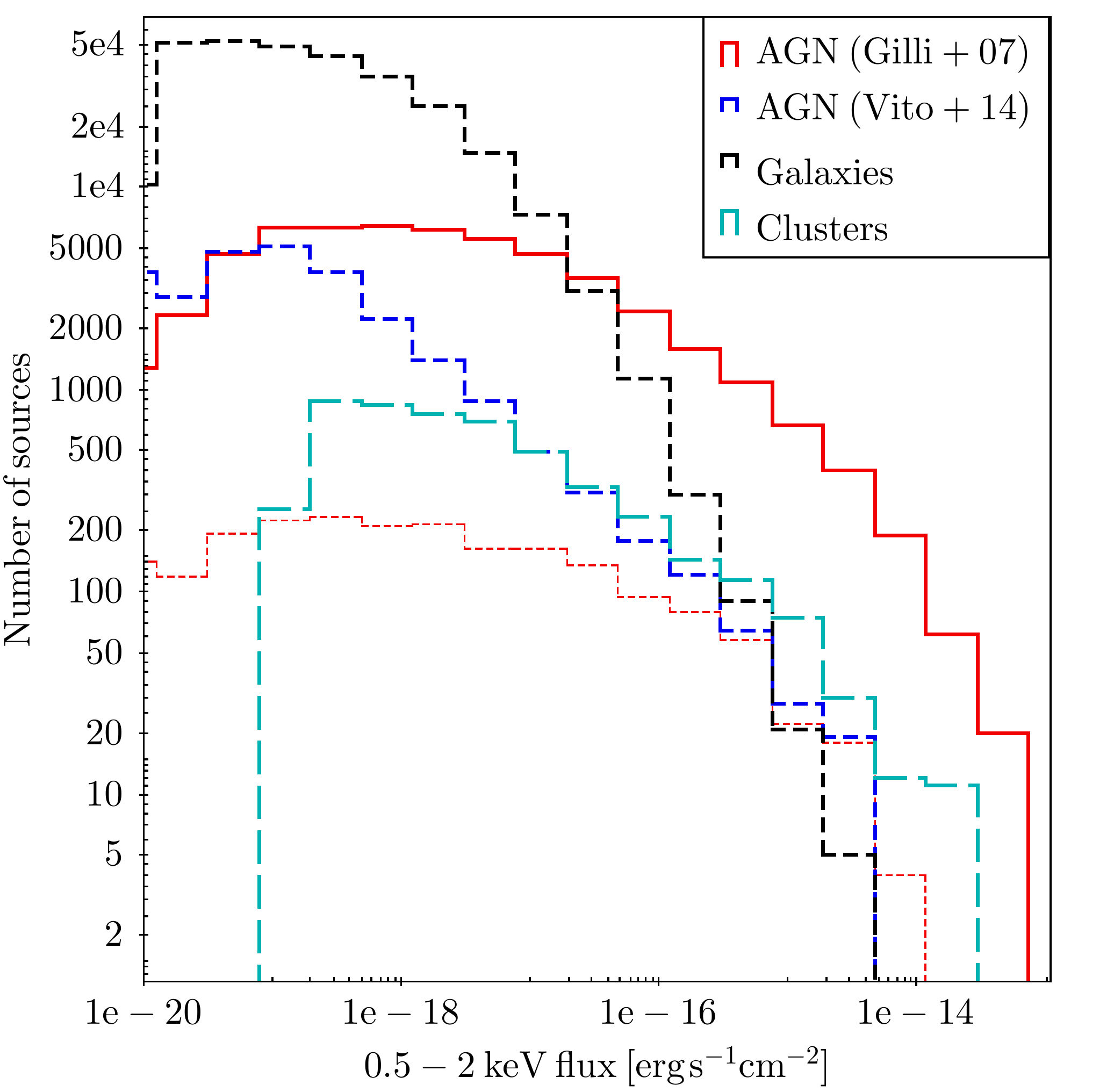}
\caption{\normalsize Distribution of the 0.5--2\,keV flux for the four mocks presented in this work: AGN from the \citet{gilli07} AGN population synthesis model (red solid line for the whole population; red dotted line for the $z>$3 subsample); $z>3$ AGN from the \citet{vito14} X-ray luminosity function (blue long-dashed line); non-active galaxies (black short-dashed line); and clusters of galaxies (cyan long dashed line). All histograms refer to a 1\,deg$^2$ field.}\label{fig:flux_052}
\end{figure}

To summarize the properties of the catalogs presented in this section, in Figure \ref{fig:flux_052} we plot the distribution of the 0.5--2\,keV observed flux for our four mocks. 
As can be seen, the faint end of the flux distribution (i.e., $f_{\rm 0.5-2}<$10$^{-17}$\,\flu, below the flux limit of currently available X-ray surveys) is dominated by non-active galaxies. However, we expect that a fraction of sources that would be classified as non-active galaxies are actually low-luminosity AGNs, i.e., sources where the luminosity of the accreting supermassive black hole is log(L$_{0.5-2}$)$<$42, and possibly even fainter than the log(L$_{0.5-2}$)=40 threshold we adopted in this work. In such sources, the AGN contribution to the overall X-ray luminosity might not be the dominant one, since other processes (e.g., star-formation, diffuse gas emission, emission from ultra-luminous X-ray sources...) can produce larger X-ray luminosities, up  to log(L$_{0.5-2}$)$\sim$42 \citep[see, e.g.,][]{ranalli05,lehmer16}.
We therefore expect a significant fraction of log(L$_{0.5-2}$)$<$40 AGNs to be included in our mock of non-active galaxies, as their X-ray emission would be dominated by the contribution of  the galactic sources mentioned above. The number of AGN detections reported in the next sections, as well as those of future simulations with our mocks, can then be treated as a lower limit.

It is also worth noting that, at low fluxes, the AGN population is dominated by low-luminosity AGNs (i.e., objects having 0.5--2\,keV luminosity below the 10$^{42}$\,\lu\ threshold), which account for $\sim$92\,\% of the $f_{\rm 0.5-2}<$10$^{-17}$\,\flu\ AGN population. A large part of  low-flux AGNs are also heavily obscured, CT-AGNs: more in detail, $\sim$52\,\% of the $f_{\rm 0.5-2}<$10$^{-17}$\,\flu\ AGN sample is made of CT sources, and 45\,\% of the same sample is made of low-luminosity CT-AGNs. These numbers underline how deep surveys with new X-ray facilities will allow us to detect a whole new AGN population of low-luminosity, heavily obscured sources. We will further discuss this topic in the next sections.

Notably, moving towards faint fluxes the high-$z$ AGNs in the \citet{vito14} mock significantly outnumber those in the \citet{gilli07} one, as shown also in Figure \ref{fig:logn-logs_zgt3}. Finally, clusters of galaxies contribute more significantly to the brightest end of the distribution ($f_{\rm 0.5-2}>$10$^{-15}$\,\cgs), where they even outnumber non-active galaxies.

\section{The AXIS probe}\label{sec:axis}
As an example of a practical application of our mock catalogs, we simulate a series of potential surveys with the Advanced X-ray Imaging Satellite \citep[AXIS;][]{mushotzky19}\footnote{\url{http://axis.astro.umd.edu}} mission concept. In this section, we briefly describe AXIS technical layout. 

AXIS is a probe-class mission proposed to the Decadal Survey on Astronomy and Astrophysics 2020\footnote{\url{https://www.nationalacademies.org/our-work/decadal-survey-on-astronomy-and-astrophysics-2020-astro2020}} with 0.3-10\,keV observing band pass.
The probe is designed to have a sub-arcsecond resolution over an area of over 500 square arcminutes, i.e., over a 24$^\prime\times$24$^\prime$ field of view\footnote{The final technical configuration of the AXIS CCDs has not yet been finalized. In this work, we simulate a single square CCD with a 24$^\prime$ side.}: this would be an improvement of a factor $\sim$100 with respect to \cha\ ACIS-I, which only has sub-arcsecond resolution at off-axis angles OAA$<$2$^\prime$. AXIS is planned to have a remarkably low detector background, thus increasing sensitivity to extended sources, and high observing efficiency. 

In Figure \ref{fig:AXIS_technical}, left panel, we plot the AXIS effective area as a function of energy. We also compare the AXIS effective area with those of \cha\ and \xmm\ pn, as well as with the predicted one for \textit{Athena}. AXIS is expected represent a major improvement with respect to current facilities in terms of collecting area. For example, at 1\,keV the planned AXIS effective is a factor $\sim$25 larger than the one of \cha\ as of 2020, and a factor $\sim$10 larger than the one of \cha\ at launch. AXIS is expected to produce a significant improvement, in terms of grasp, i.e., of effective area multiplied by field of view, even with respect to \xmm\, which is currently the best X-ray imaging telescope in that regard. At 1\,keV, AXIS would collect $\sim$5 times more counts than \xmm\ in the same amount of time.

A significant improvement in effective area would also take place at harder energies: at 6\,keV (i.e., at the energy of the Iron K$\alpha$ line), the AXIS effective area is expected to be a factor of $\sim$5 larger than the \cha\ one, and a factor of $\sim$2 larger than the \xmm\ one.

As previously stated, the most remarkable feature of AXIS with respect to \cha\ is its stable point spread function (PSF) as a function of the off-axis angle: as shown in Figure \ref{fig:AXIS_technical}, right panel, the AXIS PSF (here plotted as the half-energy width) is designed to be almost constant over the whole CCD, being $\lesssim$1$^{\prime\prime}$ even at the edge of the field of view. This would represent a large improvement with respect to \cha, whose HEW is $<$1$^{\prime\prime}$ on-axis, but becomes $>$5$^{\prime\prime}$ for off-axis angles (OAAs) $>$5$^{\prime}$. Such an angular resolution would allow AXIS surveys to, for example, locate AGNs within host galaxies even at high redshifts, where mergers are expected to be common, as well as to detect binary AGNs and runaway black holes. This type of science can be performed by \cha\ only with pointed, more time-consuming observations.

\subsection{The complementary strengths of AXIS and \athena}
If approved for funding, AXIS should be launched in the 2030s. This would allow it to work in synergy with the Advanced Telescope for High ENergy Astrophysics (\athena) mission.

\athena\ is the next ESA X-ray observatory mission, selected in the Cosmic Vision program to address the Hot and Energetic Universe scientific theme. It is the second L(arge)-class mission within that program and is due for launch in the early 2030s. 

\begin{figure*}
\begin{minipage}[b]{.52\textwidth}
  \centering
  \includegraphics[width=1\textwidth]
  {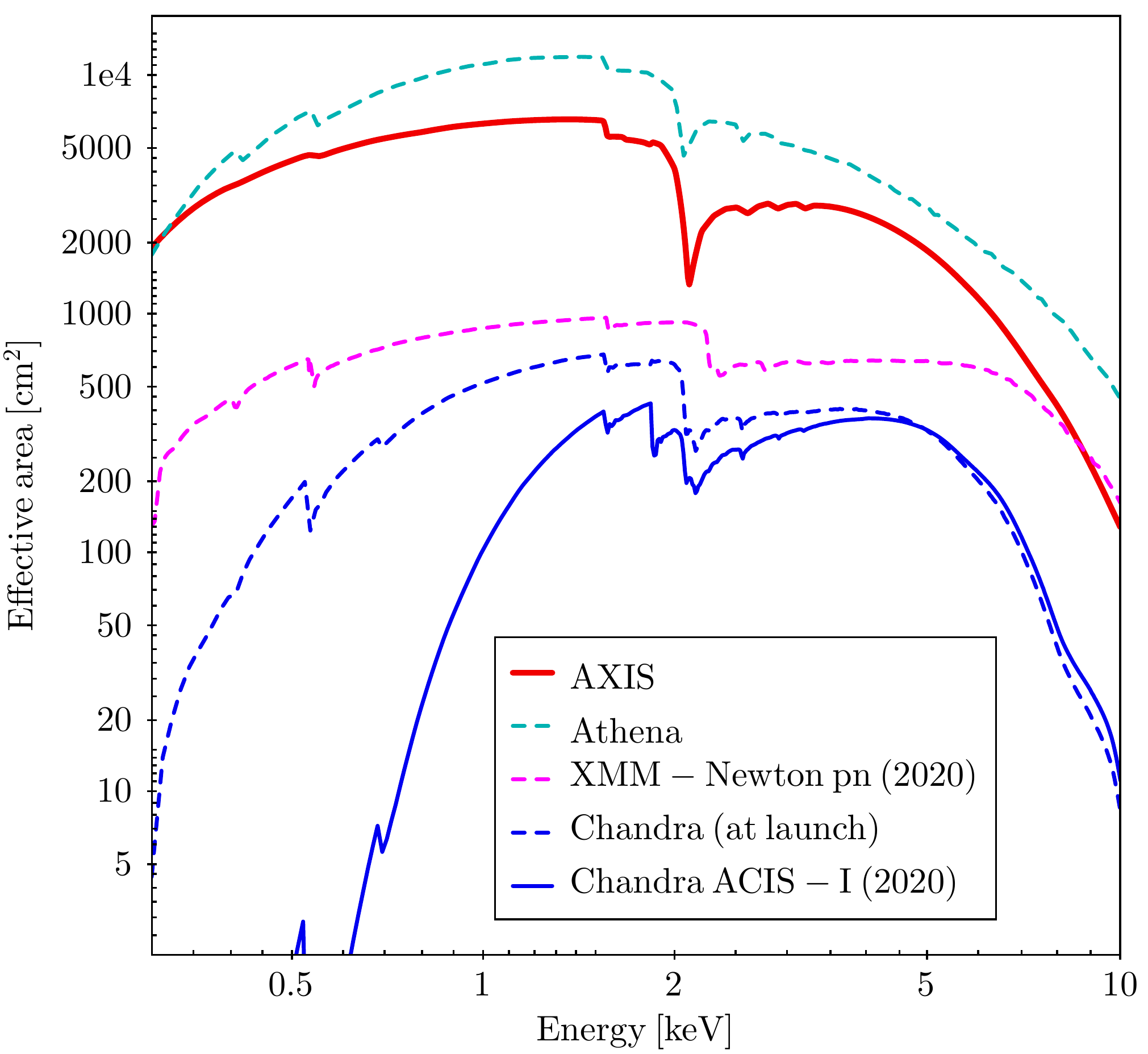}
  \end{minipage}
\begin{minipage}[b]{.47\textwidth}
  \centering
  \includegraphics[width=1\textwidth]{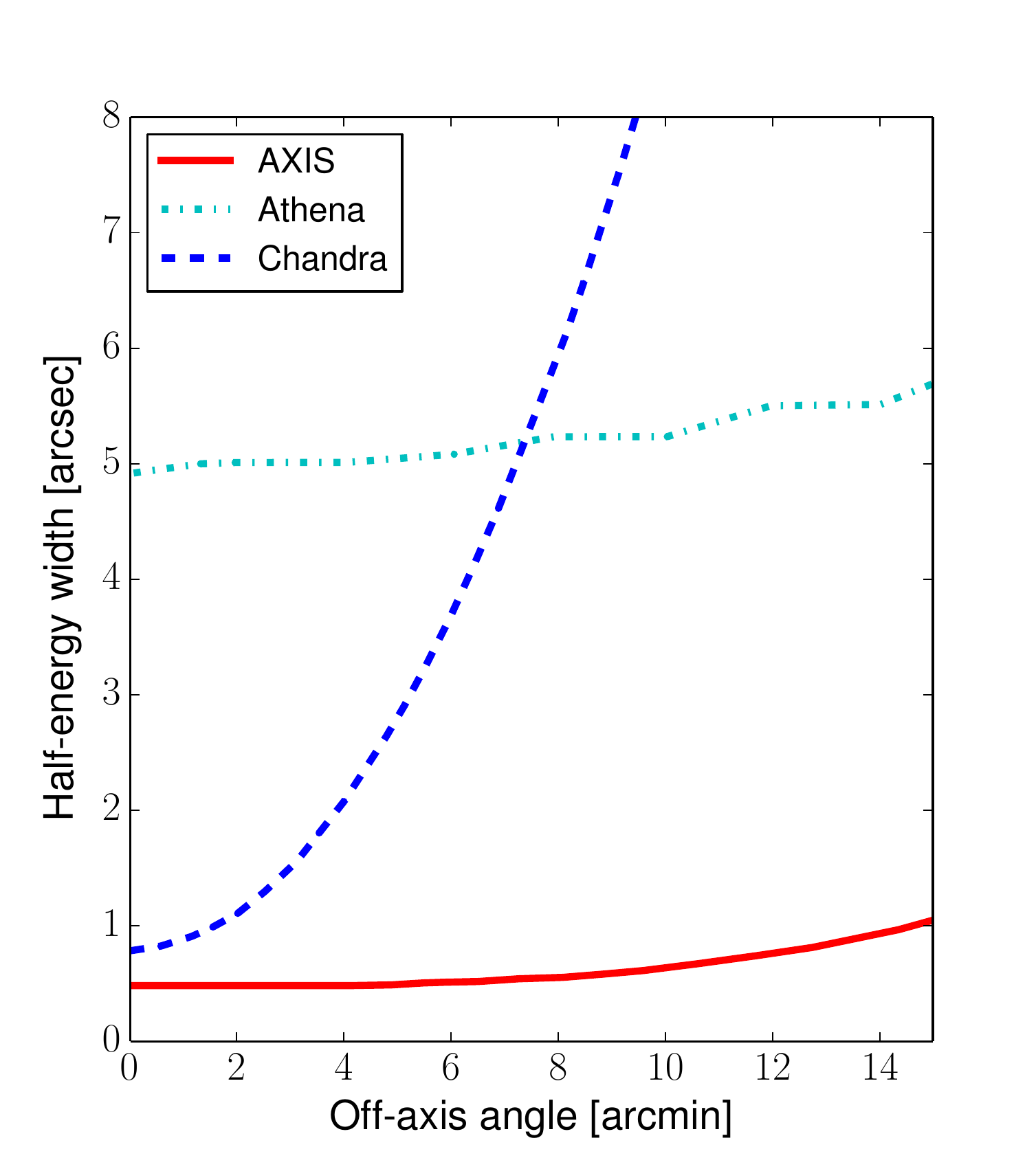}
  \end{minipage}
  \caption{\normalsize Left: AXIS planned effective area  as a function of energy (red solid red line), compared with those of the \xmm\ \texttt{pn} camera (dashed magenta line) and of the \cha\ ACIS-I camera (dashed blue line) as of early 2020, and with the ACIS-I one at the time \cha\ was launched (solid blue line). The \textit{Athena} effective area for a 5$^\prime$ field of view (dashed cyan line) is also plotted for comparison. Right: AXIS (solid red line) angular resolution as a function of the off-axis angle. The same quantities are also plotted for \athena--WFI \citep[dash-dotted cyan  line;][]{nandra13} and \cha\ ACIS-I (dashed blue line).}\label{fig:AXIS_technical}
\end{figure*}

\athena\ will mount two instruments. One is the X-ray Integral Field Unit, a cryogenic X-ray spectrometer with planned 2.5\,eV spectral resolution, 5$^{\prime\prime}$ pixels and a field of 5$^{\prime}\times$5$^{\prime}$ \citep{barret16}.
The other instrument is the Wide Field Imager (WFI), which will represent the ideal successor of \xmm\ and can be directly compared with AXIS. 
We therefore plot in both panels of Figure \ref{fig:AXIS_technical} the planned effective area as a function of energy and HEW as a function of OAA for \athena-WFI. 

The \athena\ WFI \citep{meidinger17}, with its 40$^\prime$--diameter DEPFET chips, will be the instrument with the largest effective area in the 0.3--10\,keV band: with respect to AXIS, the \athena\ collecting area will be a factor $\sim$2--3 times larger, thus making it the ideal instrument for X-ray spectroscopy\footnote{Although, thanks to its low background, AXIS would also be effective in performing X-ray spectroscopy, even for high-redshift, heavily obscured sources. We discuss this in detail in section \ref{sec:results-CT}.}. 

\athena, similarly to AXIS, is expected to have a remarkably stable PSF over the whole field of view, with a half-energy width HEW$\sim$5--6$^{\prime\prime}$. This is a factor of $\sim$10 larger than the AXIS one, which implies that AXIS, being significantly less affected by confusion issues than \athena, would be able to perform deeper surveys, as we will show in Section \ref{sec:results}. The largest field of view and effective area of \athena\ will instead make it ideal for large area ($>$10\,deg$^2$) surveys: we will further discuss the complementarity between the two missions in Section \ref{sec:high-z}.

\section{Results of the simulations of AGN surveys with AXIS}\label{sec:results}
Using as an input the mock catalogs presented in Section \ref{sec:software_and_catalogs}, we simulate three different type of AXIS surveys, following the approach presented in the AXIS white paper \citep{mushotzky19}. Overall, each survey requires 5\,Ms of AXIS observing time: notably, these combined 15\,Ms of AXIS time would amount to only 15\,\% of the planned 5 years of observations with AXIS. 

The proposed surveys are the following.

\begin{enumerate}
    \item A deep, pencil-beam survey, i.e., a 5\,Ms observation of a single AXIS pointing ($\sim$0.16\,deg$^2$, i.e., $\sim$24$\times$24 square arcmin).
    \item A moderate--area, moderate--depth survey, i.e., 2.5 deg$^2$ uniformly covered with a tiling of 300\,ks AXIS observations.
    \item A large-area, shallow-depth survey, i.e., 50 deg$^2$ covered with a tiling of 15\,ks AXIS observations.
\end{enumerate}

In Table \ref{tab:survey_layout} we report a summary of the properties of the three surveys, as well as the expected number of detections in each of it, based on the results reported in this work and extensively discussed in the following sections.

\subsection{Simulations and source detection}\label{sec:detection}
Since performing \texttt{SIXTE} simulations is a time- and machine-consuming task, we chose to simulate a 0.95\,deg$^2$ field of view\footnote{Due to the technical setup we adopted in \texttt{SIXTE}, the simulated area is $\sim$5\,\% smaller than the mock one.} with the same exposure of the proposed intermediate survey (i.e., 300\,ks) and a 9.5\,deg$^2$ field of view with the same exposure of the proposed wide survey (i.e., 15\,ks). We then extrapolate the results obtained on this smaller field to the proposed areas. The deep-area survey is instead simulated in its entirety.

The AGN, galaxy and cluster mock catalogs we presented in the previous sections all have flux limits well below the threshold of the deepest surveys planned with current or future instrumentation, and therefore cause the majority of the unresolved cosmic background in our analysis. The only additional component we include in our simulations is the Galactic diffuse foreground, which we model with a thermal component having surface brightness 1.8\,$\times$10$^{-15}$\,erg\,s$^{-1}$\,cm$^{-2}$\,arcmin$^{-2}$, following the same approach described in \citet{cucchetti18}.

The particle background at 1\,keV is assumed to be 7$\times$10$^{-5}$\,cts\,keV$^{-1}$\,s$^{-1}$\,arcmin$^{-2}$ for AXIS \citep{mushotzky19}, and 6$\times$10$^{-4}$\,cts\,keV$^{-1}$\,s$^{-1}$\,arcmin$^{-2}$ for \athena\ \citep{nandra13}.

The source detection was performed using the CIAO tool \texttt{wavdetect} on the 0.5--7\,keV (Full), 0.5--2\,keV (Soft) and 2--7\,keV (Hard) images. These energy ranges are fairly conservative, and allow us to make a direct comparison with the results obtained by the \cha\ surveys. It is worth noting, however, that AXIS would collect a significant amount of photons in the 0.3--0.5\,keV and in the 7--10\,keV energy ranges. The false-probability detection rate \texttt{SIGTHRESH} was set to a fairly conservative value, $sig$=5$\times$10$^{-8}$, i.e., $\sim$1/n$_{pix}$=1/4096$^2$, where n$_{pix}$ is the number of pixels in our CCD. We then used a  $\sqrt{2}$ wavelet sequence (i.e., 1, 1.4, 2, 2.83, 4). 

We note that \texttt{wavdetect} is commonly used to perform a preliminary source detection and generate a first catalog of candidate objects which is then given as an input to other, more refined detection tools such as ACIS Extract \citep[][for practical applications of this tool see, e.g., \citealt{luo17,nanni20}]{broos12}. In this work we do not perform a second level source detection, but we keep only those sources with significance $\sigma>$3\footnote{$\sigma$ is the \texttt{SRC\_SIGNIFICANCE} parameter computed by \texttt{wavdetect} and can be treated as the source signal-to-noise ratio.}. Thanks to the low background and the excellent PSF over the whole field of view expected for AXIS, the choice of a $\sigma$=3 threshold allows us to keep the number of spurious sources well below 0.5\,\% of the total number of detections.

In the following sections, we report the results of our detection in the three simulated fields. The matches between the output catalogs obtained using \texttt{wavdetect} and the AGNs and galaxies input ones were performed assuming a maximum positional offset $d$=1$^{\prime\prime}$. We first matched the output catalog with the AGN one, and then matched the sources with no AGN counterpart with the non-active galaxies catalog. The sources with no match in either of the two input catalogs are classified as spurious and are caused by random background fluctuations.

Thanks to the PSF quality, even at larger off-axis-angles (see Figure \ref{fig:AXIS_technical}, right panel), the chosen maximum distance, while small, is actually a fairly conservative one: for example, in our simulated wide area survey the average distance between the output position and the input one is $d_{AGN}$=0.12$^{\prime\prime}$ (with standard deviation $\sigma_{d,AGN}$=0.11$^{\prime\prime}$) for AGNs and  $d_{gal}$=0.15$^{\prime\prime}$ (with standard deviation $\sigma_{d,gal}$=0.10$^{\prime\prime}$) for non-active galaxies.

\begingroup
\renewcommand*{\arraystretch}{1.5}
\begin{table*}
\centering
\scalebox{1}{
\vspace{.1cm}
  \begin{tabular}{ccccccc}
       \hline
       \hline      
Survey & Area & Tile exposure & Total exposure & Flux limit (0.5--2\,keV) & \multicolumn{2}{c}{Number of detections}\\
       & deg$^2$ & ks & Ms & erg s$^{-1}$ cm$^{-2}$ & AGN & Galaxies\\ 
       \hline
Deep &  0.16 &  5000 & 5 & 5$\times$10$^{-19}$ & 3496 & 5387 \\ 
\hline
Intermediate     &  2.5 & 300 & 5 & 3$\times$10$^{-18}$ & 32655 & 22071 \\ 
\hline
Wide &  50 &  15 & 5 & 4$\times$10$^{-17}$ & 190149 & 21840 \\ 
    \hline
    \hline
	\vspace{0.02cm}
\end{tabular}}
	\caption{\normalsize Properties of three reference AXIS surveys simulated in this work. The flux limit corresponds to 1\,\% of the covered field (see Figure \ref{fig:ratio_vs_flux}), and is the flux at which 1\,\% of the input sources is detected. The number of detections is computed in the 0.5--7\,keV band.
	}
\label{tab:survey_layout}
\end{table*}
\endgroup

\subsection{AXIS Deep Field results}\label{sec:results_deep}
\begin{figure*}
\begin{minipage}[b]{0.75\textwidth}
  \centering
  \includegraphics[width=1\textwidth]{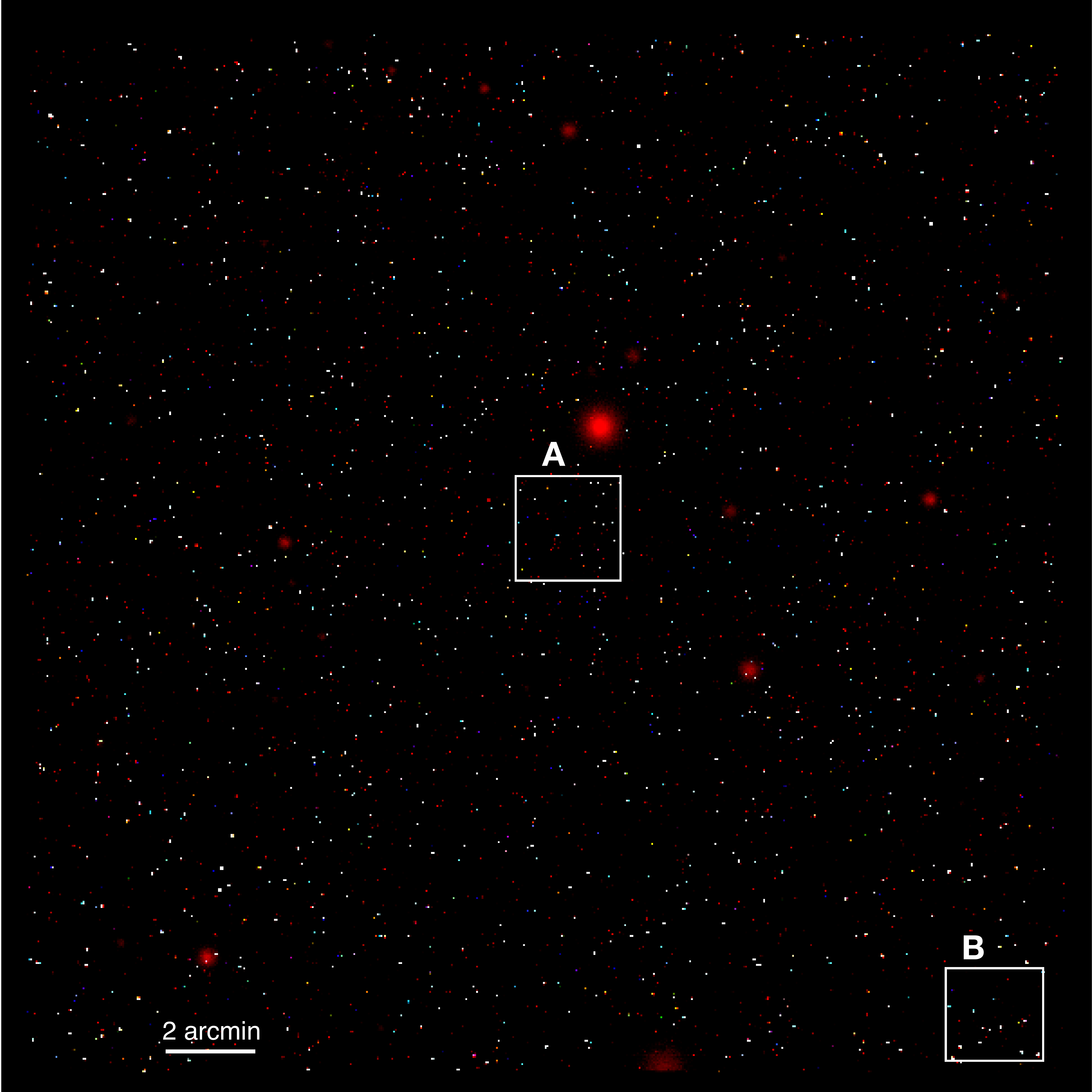}
\end{minipage}
\begin{minipage}[b]{.25\textwidth}
  \centering
  \includegraphics[width=1\textwidth]{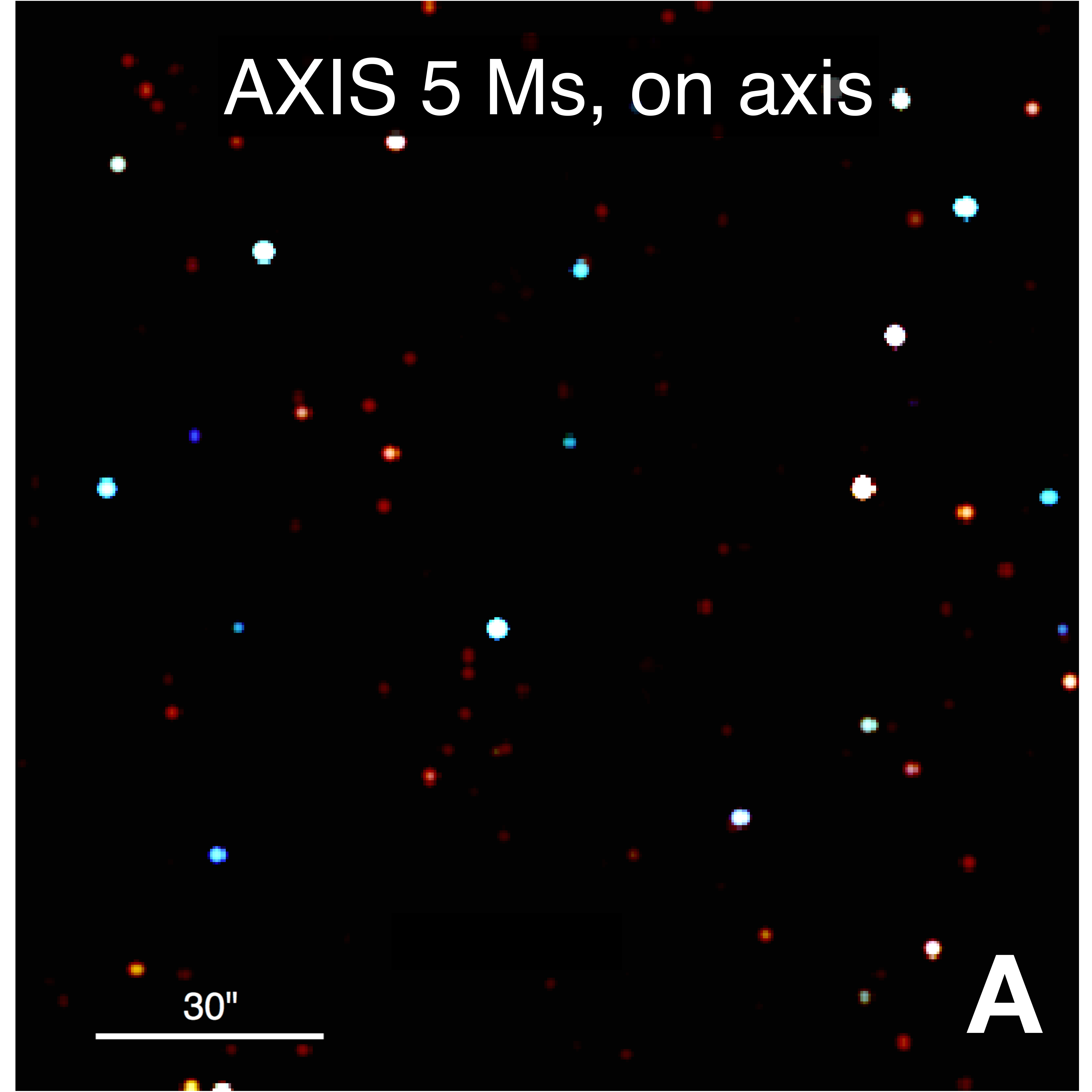}
\vfill
\centering
  \includegraphics[width=1\textwidth]{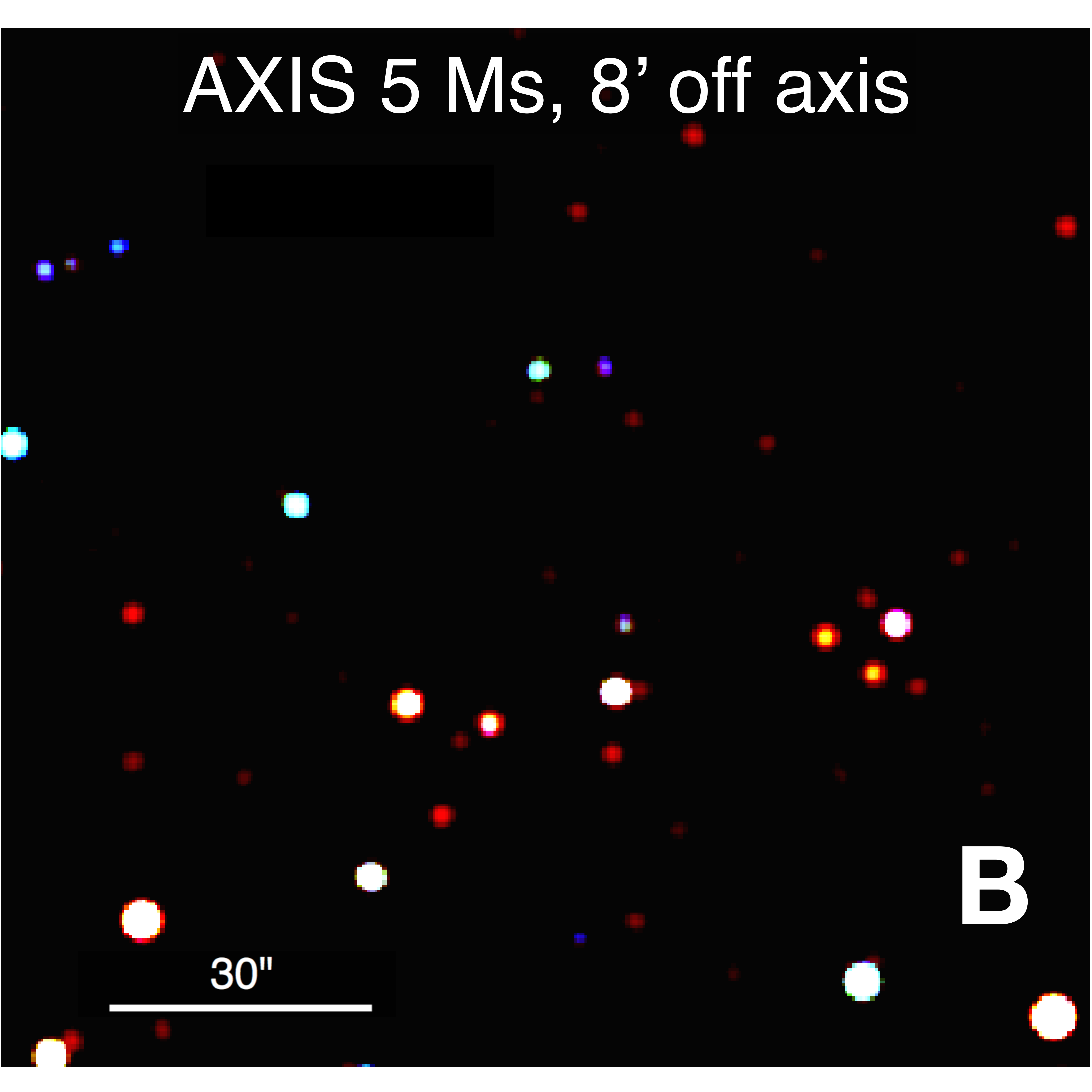}
\vfill
\centering
  \includegraphics[width=1\textwidth]{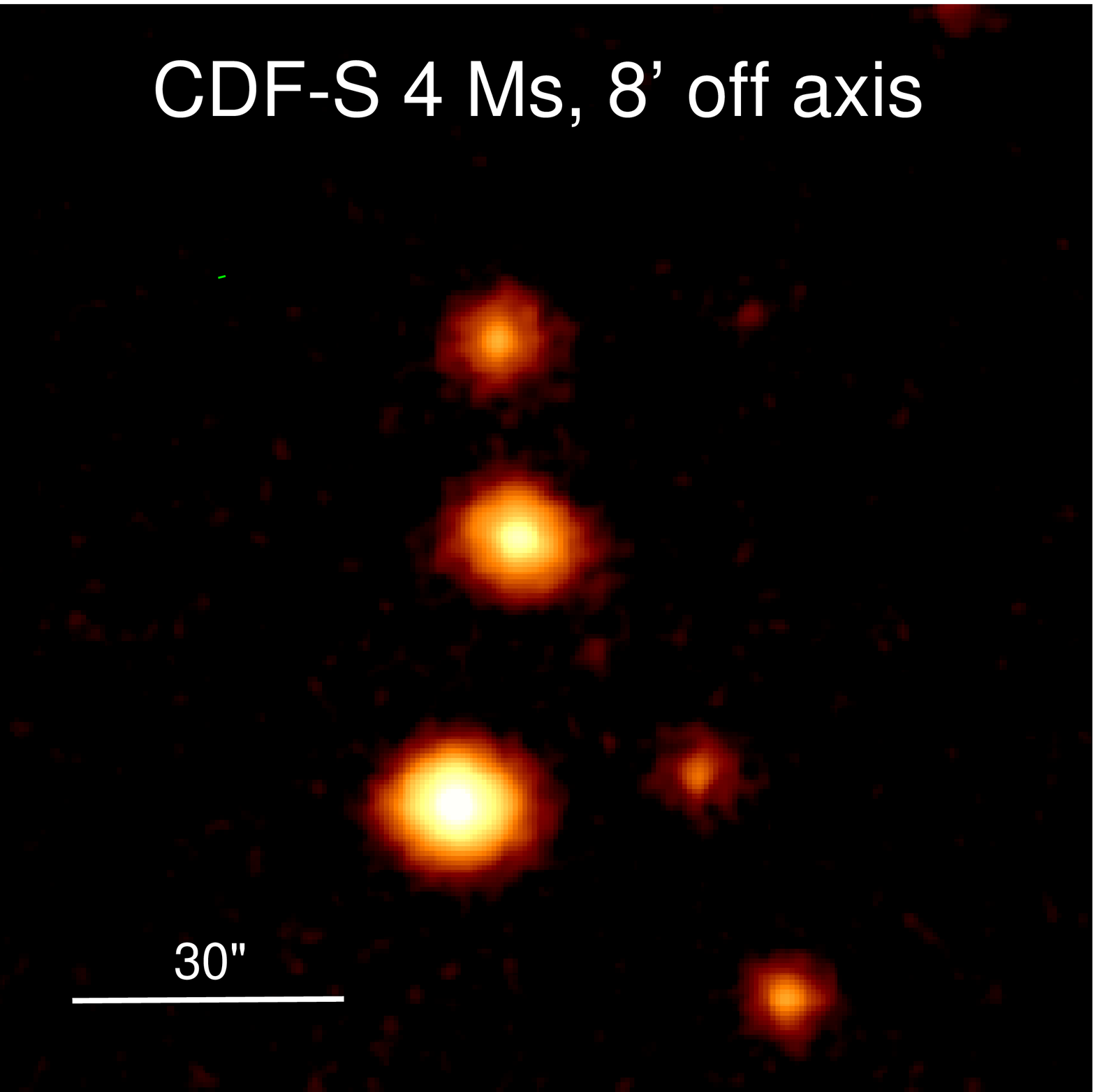}
\end{minipage}
  \caption{\normalsize \textit{Left}; Smoothed three-color image (red: 0.5-2\,keV; green: 2-4.5\,keV blue: 7\,keV) of a simulated 5-Ms deep field (24$^\prime$x24$^\prime$) with AXIS. The extended red structures are galaxy clusters. The white boxes highlight two regions, one on-axis and the one at the boundary of the field of view: we show a zoom-in of these regions in the top and central right panels. The AXIS PSF is expected to have remarkably small degradation as a function of the off-axis angle. In the bottom right panel, we show as a reference a $\sim$2$^\prime$x2$^\prime$, 8$^\prime$ off-axis 0.5--7\,keV image of the 4\,Ms CDF-S.}\label{fig:deep5ms_RGB}
\end{figure*}

We report in Table \ref{tab:results_all} the results of the simulation of a 5\,Ms AXIS observation over a single pointing, i.e., an area of $\sim$0.16\,deg$^2$, while in Figure \ref{fig:deep5ms_RGB} we show the three-color image of the simulated AXIS Deep field. In the same figure we also show a zoom-in of two different regions of the field: one on-axis (left panel), and one 8$^\prime$ off-axis: a telescope designed to have sub-arcsec PSF at large off-axis angles would be an ideal survey instrument, since it would ensure high sensitivity over the whole field of view, and would make the counterpart identification process much easier.  
As a comparison, in the bottom right panel of the same figure we also show a 8$^\prime$ off-axis image of the 4\,Ms CDF-S \citep{xue11}, whose exposure is comparable to the simulated 5\,Ms AXIS one. At these off-axis radii, the \cha\ PSF is $>$5$^{\prime\prime}$ (see also Figure \ref{fig:AXIS_technical}, right panel), thus making the optical counterpart identification process more complicated.

Based on our simulations, a deep survey with AXIS would detect $\sim$9000 sources and reach a flux limit $f_{lim}\sim$5$\times$10$^{-19}$ erg\,s$^{-1}$\,cm$^{-2}$ in the 0.5--2\,keV band. Here and in the following sections, the ``flux limit'' is the flux at which 1\,\% of the survey area is covered (see Figure \ref{fig:ratio_vs_flux}). We also remind that, as mentioned in Section \ref{sec:detection}, in our analysis we chose a fairly conservative source detection threshold. It is therefore possible that the actual flux limits would be slightly fainter than the one we obtained. As a comparison, the CDF-S survey \citep[][]{luo17}, which covered a field of 0.135\,deg$^2$ with a combined 7\,Ms \cha\ observation, contains 1008 sources and reaches a flux limit in the 0.5--2\,keV band $f_{lim}\sim$6.5$\times$10$^{-18}$ erg\,s$^{-1}$\,cm$^{-2}$.

Out of these $\sim$9000 sources, $\sim$3500 (i.e., $\sim$39\,\% of the sample) are AGNs. In Figure \ref{fig:ratio_vs_flux}, left panel, we report the completeness of the survey as a function of the 0.5--10\,keV, 0.5--2\,keV and  2--10\,keV flux, i.e., the  ratio between the number of detections and the number of input sources. These curves are fully equivalent to the so-called survey sensitivity curves, and can also be used to estimate the fraction of survey area covered at a given flux. In the 0.5--2\,keV band  50\,\% of the field is covered down to a flux $\sim$3$\times$10$^{-18}$\,erg\,s$^{-1}$\,cm$^{-2}$, while in the 0.5--10 and 2--10\,keV bands 50\,\% of the field is covered down to a flux limit $\sim$2$\times$10$^{-17}$\,erg\,s$^{-1}$\,cm$^{-2}$.

As already mentioned in Section \ref{sec:AGN_mock}, our mock does not include AGNs with luminosities L$_{\rm 0.5-2}<$10$^{40}$\,\lu, corresponding to the flux limit of the AXIS deep survey at redshift $z\sim$1.1. Consequently, it is reasonable to assume that the number of AGNs detected in the AXIS deep field would be even larger than the one we obtained from our simulations at $z<$1.1, while we do not expect the adopted luminosity threshold to impact the number of AGNs detected at higher redshift. More in detail, working under the simple assumption that the number and redshift distribution of AGNs in a dex of luminosity in the range log(L$_{\rm 0.5-2}$)=[39-40] are equal to those in the log(L$_{\rm 0.5-2}$)=[40-41] bin, we would expect to detect in the AXIS deep field $\sim$300-400 additional AGNs at L$_{\rm 0.5-2}<$10$^{40}$\,\lu, although their identification as AGN would possibly be challenging, since non-AGN processes can produce similar, if not higher, luminosities. As a reminder, based on our simulations the AXIS Deep survey would contain $\sim$3500 AGNs with luminosities down to log(L$_{\rm 0.5-2}$)=40.

As shown in deep \cha\ surveys like the CDF-S 7\,Ms one \citep{luo17}, the deeper an X-ray survey is, the larger is the fraction of non-active galaxies it detects. As a consequence, in the simulated AXIS deep field the fraction of non-AGN sources detected in the 0.5--7\,keV and 0.5--2\,keV is $\sim$60\,\%, i.e., as opposed to shallower X-ray surveys, the majority of detected objects are not be AGNs, allowing a direct measurement of the star formation rate over a wide redshift range. In the hard X-ray band, instead, the majority of the detected sources are expected to be AGNs, but it would still be possible to detect a significant number of galaxies ($>$2000).

\begingroup
\renewcommand*{\arraystretch}{1.7}
\begin{table*}
\centering
\scalebox{1}{
\vspace{.1cm}
  \begin{tabular}{ccccccc}
       \hline
       \hline       
       Survey & Band & Total & AGN & Galaxies & Flux 20\,\% & Flux 80\,\% \\ 
        & & & & & \multicolumn{2}{c}{erg s$^{-1}$ cm$^{-2}$}\\
       \hline
\multirowcell{3}{Deep}    
&    0.5-7\,keV   &   9016 &  3496 (38.8\,\%)   &    5387 (59.7\,\%)     & 8.5$\times$10$^{-18}$ & 2.9$\times$10$^{-17}$\\ 
    &    0.5-2\,keV   &   8205 &  3051  (37.2\,\%)   &    5065  (61.7\,\%)     & 1.4$\times$10$^{-18}$ & 4.4$\times$10$^{-18}$ \\ 
    &    2-7\,keV	    &	5706 &  3521 (60.7\,\%)   &	2172  (37.5\,\%)     & 9.1$\times$10$^{-18}$ & 2.4$\times$10$^{-17}$\\ 
        \hline
        \hline
 \multirowcell{3}{Intermediate}     
 & 0.5-7\,keV	    &   54774  & 32655 (59.6\,\%)  &	22071 (40.3\,\%)    & 3.9$\times$10$^{-17}$ & 1.5$\times$10$^{-16}$\\ 
                                                            & 0.5-2\,keV	    & 47689 & 27555 (57.8\,\%)   &   20124 (42.2\,\%)	 & 8.8$\times$10$^{-18}$ & 2.8$\times$10$^{-17}$\\ 
                                                            & 2-7\,keV       &	26192 & 23553	(90.0\,\%)&   2605 (9.9\,\%)    & 7.1$\times$10$^{-17}$ & 2.0$\times$10$^{-16}$\\ 
        \hline
	    \hline
\multirowcell{3}{Wide} 
&   0.5-7\,keV	&   212058  &   190149 (89.7\,\%)  &   21840 (10.3\,\%)    & 3.6$\times$10$^{-16}$ & 1.5$\times$10$^{-15}$\\ 
                                                &   0.5-2\,keV	&	172690	&   153669 (89.0\,\%)	&   18974 (11.0\,\%)	 & 1.0$\times$10$^{-16}$ & 2.7$\times$10$^{-16}$\\ 
                                                &   2-7\,keV	    &	81443	&   80831 (99.2\,\%)	&    581 (0.7\,\%)	 & 1.2$\times$10$^{-15}$ & 3.1$\times$10$^{-15}$ \\ 
    \hline
    \hline
	\vspace{0.02cm}
\end{tabular}}
	\caption{\normalsize Number of sources, overall and divided by class, detected in the 0.5-7\,keV, 0.5-2\,keV and 2-7\,keV bands in each of the three reference AXIS surveys. 
	In parentheses, we report the fraction of sources belonging to each class with respect to the total number of sources detected in a given energy range. Spurious detections, not reported in this table, are always $<$0.5\,\% of the overall detections. 
	``Flux 20\,\%'' and ``Flux 80\,\%'' are the flux limits reached over 20\,\% and 80\,\% of the proposed survey area, respectively: the Full and Hard band fluxes are computed in the 0.5--10\,keV and 2--10\,keV bands for an easier comparison with previous works, but the detections are obtained in the 0.5--7\,keV and 2--7\,keV bands.}
\label{tab:results_all}
\end{table*}
\endgroup

\subsection{AXIS Intermediate Field results}\label{sec:results_intermediate}
We report in Table \ref{tab:results_all} the results of the simulation of a 300\,ks AXIS observation: the simulations were performed over a 0.95\,deg$^2$ mock field of view, and the numbers in the table have been obtained by rescaling those obtained from the simulations by 2.5/0.95, to estimate the actual number of objects which would be found by the proposed 2.5\,deg$^2$ survey.

\begin{figure*}
\begin{minipage}[b]{.33\textwidth}
  \centering
  \includegraphics[width=1\textwidth]{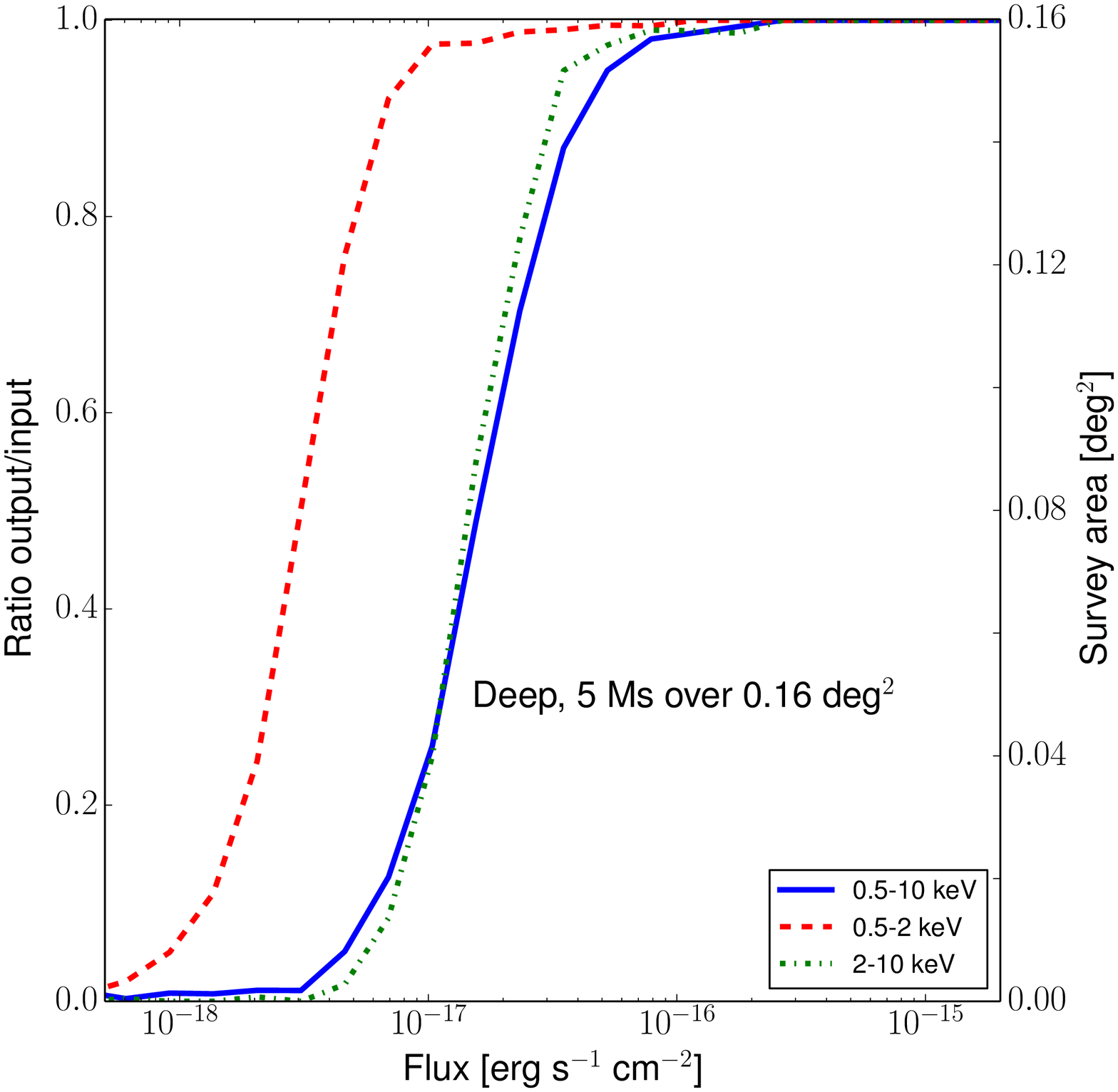}
  \end{minipage}
\begin{minipage}[b]{.33\textwidth}
  \centering
  \includegraphics[width=1\textwidth]{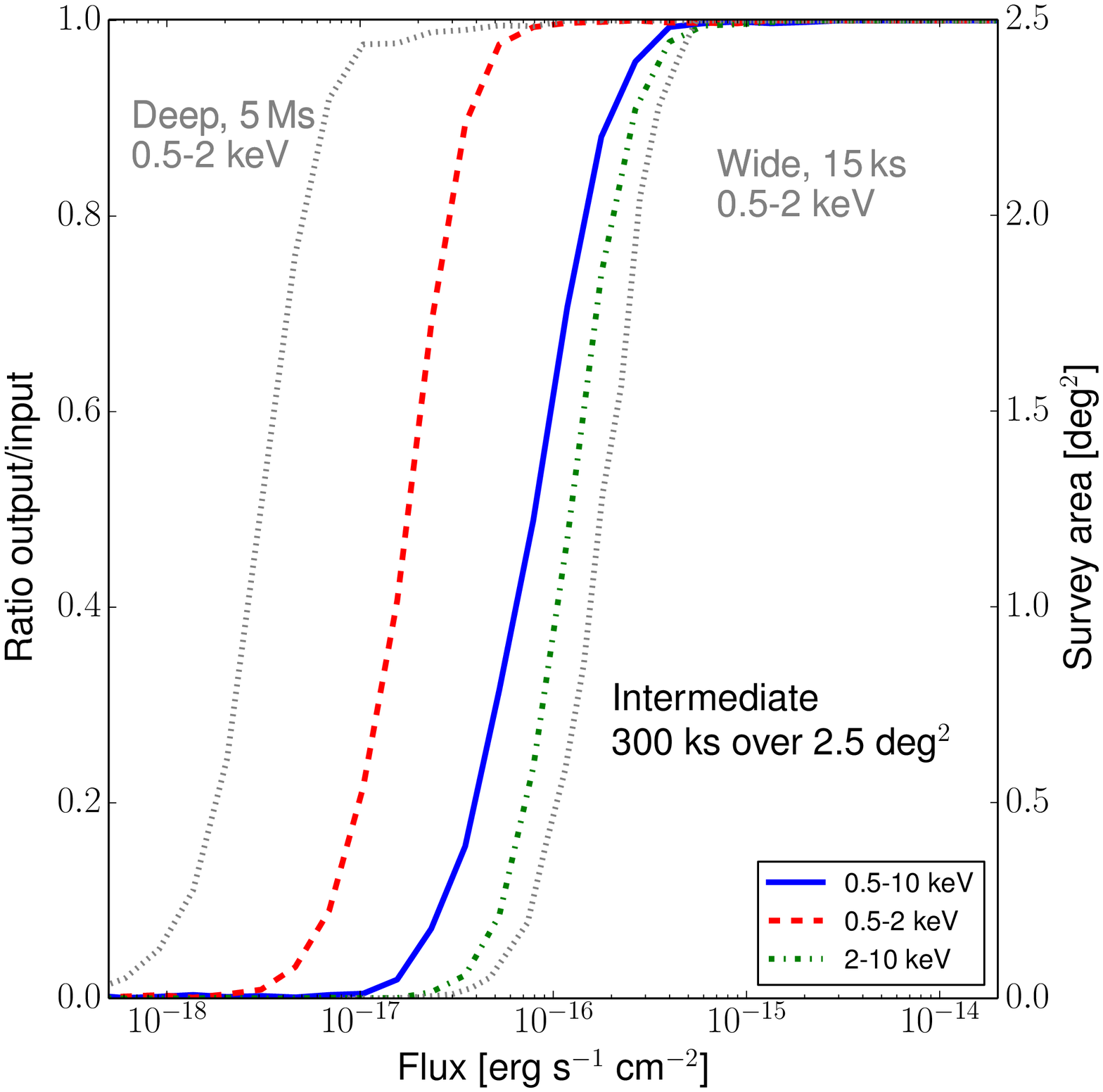}
  \end{minipage}
\begin{minipage}[b]{.32\textwidth}
  \centering
  \includegraphics[width=1\textwidth]{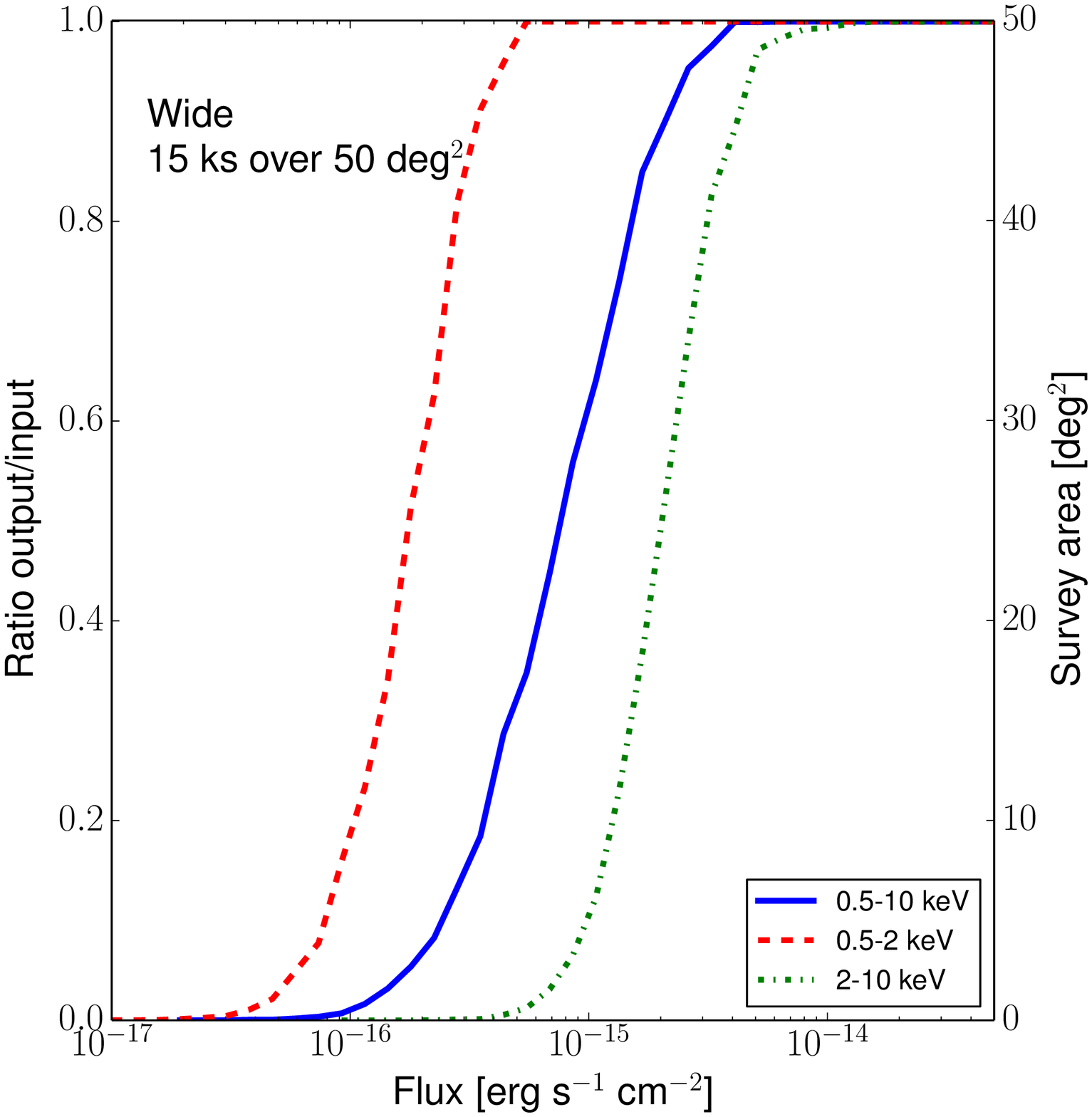}
  \end{minipage}
  \caption{\normalsize Survey completeness, i.e, ratio between number of detected sources and number of simulated sources,  and corresponding survey area for AGNs detected in the 0.5--7\,keV (solid blue line), 0.5--2\,keV (dashed red line) and 2--7\,keV (dash-dotted green line). In the left panel we report the curves obtained in the AXIS deep survey, in the central panel those obtained in the intermediate-area survey, and in the right panel those obtained in the wide-area survey. In the central panel, the 0.5--2\,keV completeness curves for the deep and wide surveys (dotted grey lines) are also plotted for comparison.}\label{fig:ratio_vs_flux}
\end{figure*}

Overall, we expect to detect $\sim$55,000 ($\sim$48,000) sources in the 0.5--7\,keV (0.5--2\,keV), reaching a flux limit in that same band $f_{lim}\sim$10$^{-17}$ erg\,s$^{-1}$\,cm$^{-2}$ ($\sim$3$\times$10$^{-18}$ erg\,s$^{-1}$\,cm$^{-2}$) (see Figure \ref{fig:ratio_vs_flux}, central panel). 
As a comparison, the \cha\ COSMOS Legacy survey, which covered 2.2\,deg$^2$ with 4.6\,Ms of \cha\ time, contained 4016 sources and reached a flux limit in the 0.5--2\,keV band $f_{lim}\sim$2$\times$10$^{-16}$ erg\,s$^{-1}$\,cm$^{-2}$.

In this intermediate survey, the majority of detected sources would be AGNs: more in detail, in the Full band we should detect more than 30,000 AGNs, which account for $\sim$60\,\% of the overall detections in the same band. In Figure \ref{fig:ratio_vs_flux}, central panel, we report the completeness of the survey as a function of the 0.5--10\,keV, 0.5--2\,keV and  2--10\,keV flux: in the 0.5--2\,keV band  50\,\% of the field is covered down to a flux limit $\sim$2$\times$10$^{-17}$\,erg\,s$^{-1}$\,cm$^{-2}$, while in the 0.5--10 and 2--10\,keV bands 50\,\% of the field is covered down to a flux limit $\sim$8$\times$10$^{-17}$\,erg\,s$^{-1}$\,cm$^{-2}$ and $\sim$10$^{-16}$\,erg\,s$^{-1}$\,cm$^{-2}$, respectively.

\subsection{AXIS Wide Field results}\label{sec:results_wide}
We report in Table \ref{tab:results_all} the results of the simulation of a 15\,ks AXIS pointing: the simulations were performed over a 9.5\,deg$^2$ mock field of view. The numbers in the table have then been obtained by rescaling the simulations one by 50/9.5, to estimate the actual number of objects which would be found by the proposed 50\,deg$^2$ survey.

Overall, we expect to detect more than 210,000 (170,000) sources in the 0.5-7\,keV (0.5-2\,keV) band, reaching a flux limit in the same band $f_{lim}\sim$10$^{-16}$ erg\,s$^{-1}$\,cm$^{-2}$ ($\sim$4$\times$10$^{-17}$ erg\,s$^{-1}$\,cm$^{-2}$). As a comparison, the XMM-XXL Survey \citep{pierre16,chiappetti18} covered 50 deg$^2$ with a total of over 6\,Ms of \xmm\ time to detect 26,056 sources down to a flux limit $f_{lim}\sim$10$^{-15}$ erg\,s$^{-1}$\,cm$^{-2}$ in the 0.5-2\,keV band. The all-sky X-ray instrument eROSITA \citep{merloni12}, which was launched in July 2019 and covers the 0.5--10\,keV energy range, is instead expected to detect 2.7 millions of AGNs band over four years of observations: the planned flux limit at 1\,\% of the area covered (i.e., $\sim$40\,deg$^2$) is $\sim$2$\times$10$^{-15}$\,\flu\ in the 0.5--2\,keV band \citep{comparat19}.

While the AXIS survey plan does not include, as of today, multiple pointings of the same field, it is worth noting that the AXIS wide field would be an ideal region for follow-up observations aimed at finding AGN variability. For example, 1\,Ms of AXIS time could be spent covering several square degrees of the AXIS wide field with the same depth of the original survey, one or more times. This would allow us to study AGN variability in samples of thousands of AGNs, at fluxes $>$1\,dex fainter than those sampled by the all-sky telescope eROSITA.

\section{The high-redshift Universe as seen by \athena\ and AXIS}\label{sec:high-z}
\subsection{Extragalactic surveys with \athena--WFI}\label{sec:athena_surveys}
As already mentioned in Section \ref{sec:axis}, the technical specifics of \athena-WFI, in particular its large field of view and effective area, and its stable PSF, make it an ideal survey instrument. A total of 23.6Ms (nearly 25\,\% of the expected four years of mission lifetime) has been devoted in the Mock Observing Plan for two extragalactic surveys. 

These surveys have been designed since the mission proposal \citep{aird13,nandra13} to address several open topics on early SMBH-galaxy evolution. For example, \athena\ will ($i$) study the mass accretion of the earliest growing SMBHs at $z>$6; ($ii$) find distant evolved groups of galaxies with formed hot gaseous atmosphere at $z>$6; ($iii$) determine the accretion energy density in the Universe, by measuring the X-ray luminosity function and obscuration properties of the AGN population up to $z$=4; ($iv$) determine the incidence of strong and ionized outflowing absorbers (including ultra-fast outflows) among the luminous AGN population from $z$=1 to $z$=4.

For each of these scientific goals a specific requirement has been derived: for example, to address scientific goal ($i$) one needs to  detect at least 10 AGNs with 43$<$log L$_{0.5-2keV}<$ 43.5  at $z$=6--7 and at least 10 AGNs with 44$<$log L$_{0.5-2keV}<$44.5  at $z$=7--8. Such a requirement can then be translated into a survey sensitivity goal (in terms of point source sensitivity vs area) for the given mission specifications. 

All together these requirements set the exposure request for the two planned surveys: i.e., 4$\times$1.5\,Ms + 3$\times$1.05\,Ms + 5$\times$950\,ks pointings over the 5.28\,deg$^2$ of the deep survey (mainly driven by the high-$z$ AGN search and Compton-thick AGN characterization), and 108$\times$90\,ks pointings over the 47.52\,deg$^2$ of the wide one (mainly driven by high-z AGN and first groups search).

To make a comparison with the AXIS results, we have performed \sixte\ simulations with the same mock catalogs described in Section \ref{sec:software_and_catalogs}, coupled with the most up-to-date calibration files and matrices available for \textit{Athena}. The cosmic background and galactic foreground are the same used for AXIS, while the particle background is modeled using a flat power law with normalization 6$\times10^{-4}$\,cts\,keV$^{-1}$\,s$^{-1}$\,arcmin$^{-2}$, as defined in the instrument scientific goals \citep{nandra13}.

We performed a full \sixte\ simulation of the deep tier of the survey (12 pointings in total), while for the wide tier we simulated only 10 fields, and rescaled the result to the planned 108 pointings. 
The detection process is the same described in Section \ref{sec:results}, except for the false-probability detection 
rate \texttt{SIGTHRESH} in \texttt{wavdetect}, which was set to 10$^{-6}$ , i.e., $\sim$1/n$_{\rm pix}$=1/1024$^2$. 
The detection has been performed in the 0.5--7, 0.5--2, and 2--7\,keV bands. 

\begin{figure*}[htbp]
\begin{minipage}[b]{.55\textwidth}
  \centering
\includegraphics[width=1.\linewidth]{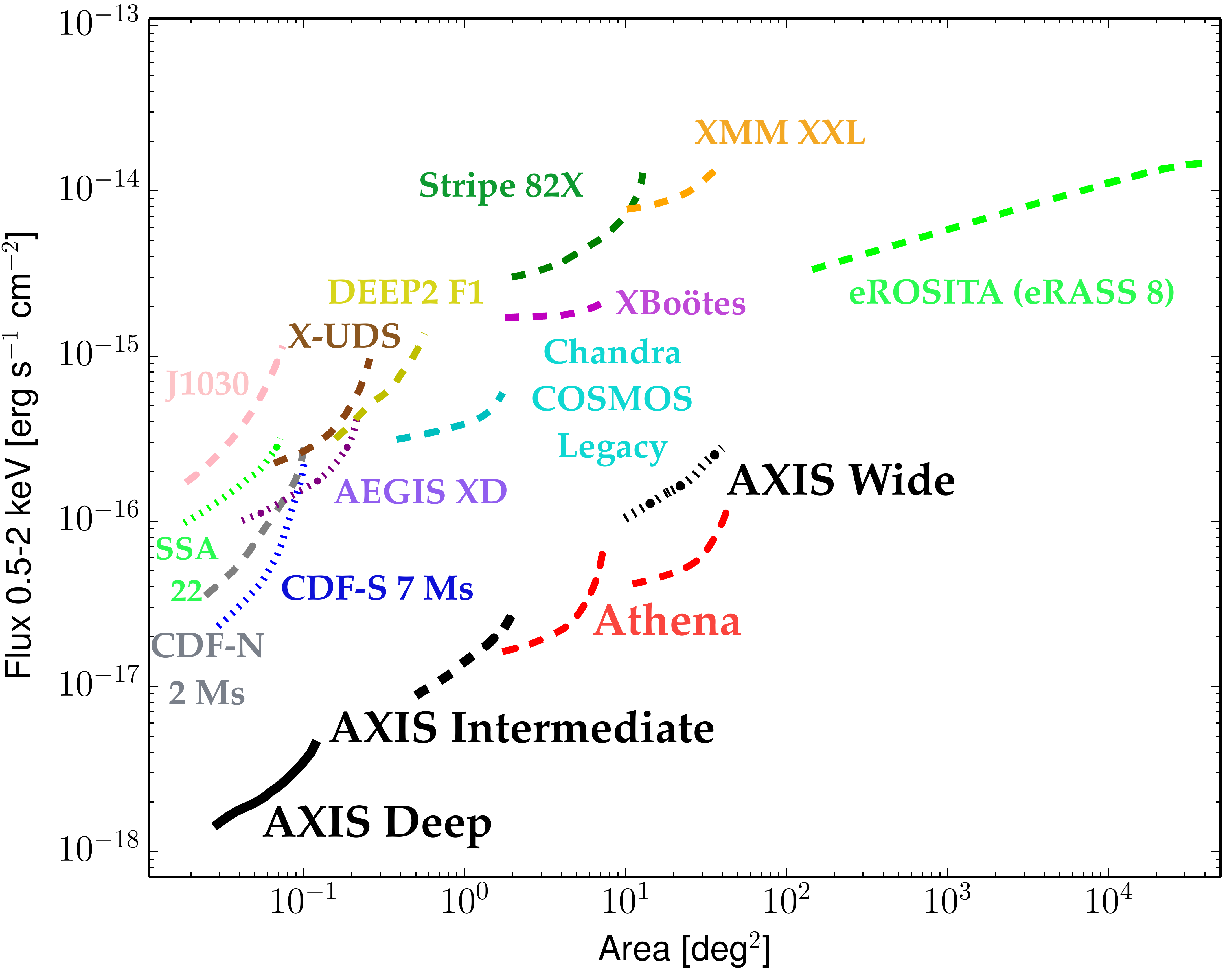}
\end{minipage}
\begin{minipage}[b]{.44\textwidth}
  \centering
\includegraphics[width=1.\linewidth]{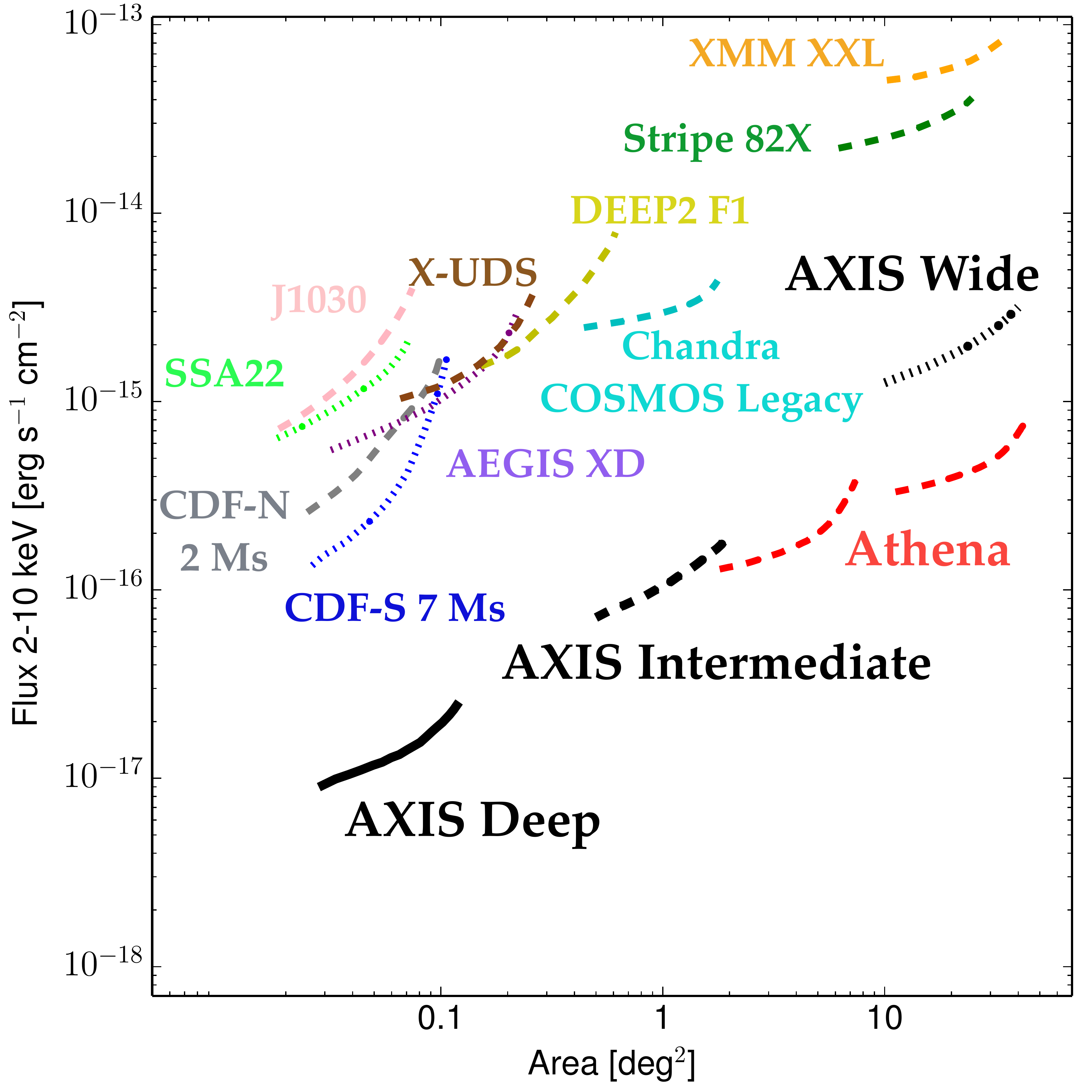}
\end{minipage}
\caption{\normalsize 0.5--2\,keV (left) and 2--10\,keV (right) area-flux curves for the AXIS deep, 5\,Ms (solid black line), intermediate, 300\,ks (dashed black line) and wide, 15\,ks (dotted black line) reference surveys. For comparison, we show the area-flux curves of several existing X-ray surveys: CDF-S 7\,Ms \citep[][dotted blue line]{luo17}; CDF-N 2\,Ms \citep[][dashed grey line]{xue16}; AEGIS XD \citep[][dotted purple line]{nandra15}; SSA22 \citep[dotted light green line]{lehmer09}; XDEEP-2 F1 \citep[][dashed yellow line]{goulding12}; J1030 \citep[dashed pink line;][]{nanni20}; X-UDS \citep[][dashed brown line]{kocevski18}; \cha\ COSMOS Legacy \citep[][dashed cyan line]{civano16}; X-Bootes \citep[][dashed magenta line]{murray05}; Stripe 82X \citep[][dashed green line]{lamassa13a,lamassa13b,lamassa16}; and XMM-XXL \citep[dashed orange line]{pierre16}. We also show the predictions made for the \textit{Athena} deep-- and wide--area survey (dashed red lines; see the text for more details), and those for four years of eROSITA observations \citep[i.e., eRASS:8;][dashed light green line]{merloni12,comparat19}.
The plotted lines have been derived from the 0.5-2\,keV survey sensitivity
curves in an area range that starts at 20\% and stops at 80\% of the area covered by the survey.}\label{fig:area_flux}
\end{figure*}

In Figure \ref{fig:area_flux} we report the area versus 0.5--2\,keV (left panel) and 2--10\,keV (right panel) flux curve for the simulated AXIS (black lines) and \athena\ (red lines) surveys, as well as for several existing \cha\ and \xmm\ surveys, namely CDF-S 7\,Ms \citep{luo17}, CDF-N 2\,Ms \citep[][]{xue16}, AEGIS XD \citep{nandra15}, SSA22 \citep{lehmer09}, J1030 \citep{nanni20}, XDEEP-2 F1 \citep{goulding12}, \cha\ COSMOS Legacy \citep{civano16}, X-Bootes \citep{murray05}, Stripe 82X \citep{lamassa13a,lamassa13b,lamassa16} and XMM-XXL \citep{pierre16}. In the left panel, we also report the 0.5--2\,keV predictions for four years of eROSITA observations \citep[i.e., eRASS:8;][dashed light green line]{merloni12,comparat19}.

As it can be seen, a survey program such as those planned for \athena\ and AXIS would represent a major improvement with respect to currently available surveys: particularly, AXIS surveys would be able to sample fluxes $\sim$25 and $\sim$50 times deeper than those reached by currently available surveys at 1 and 50 deg$^2$, respectively. AXIS and \athena\ would complement each other: the larger collecting and effective area of \athena\ would allow one to reach unprecedentedly deep fluxes over 10s of square degrees, while the AXIS PSF quality ($<$1$^{\prime\prime}$ over the whole field of view) would represent a major step forward for deep-- and intermediate--area surveys.

\subsection{A new X-ray view of the high-redshift Universe}\label{sec:results_high-z}

In Table \ref{tab:results_highz} we report the expected number of high-$z$ sources expected in each AXIS survey in the two sets of simulations we performed (i.e., one using the \citealt{gilli07} mock, the other using the \citealt{vito14} mock).

Regardless of which of the two mock catalogs will turn out to be the most reliable one, all surveys would provide a significant improvement to the $z>$3 statistic of X-ray selected AGNs: we in fact may expect to detect $\sim$130--350 $z>$3 AGNs in the deep survey, $\sim$1000--2000 $z>$3 AGNs in the intermediate survey and $\sim$6000--6700 $z>$3 AGNs in the wide-area survey. 
This would be a major leap forward with respect to currently available X-ray datasets. 
For example, the deepest \cha\ survey,  the CDF-S 7\,Ms one, contains $\sim$70 $z>$3 AGNs \citep{vito18} over an area of $\sim$330\,arcmin$^2$; the 2.2\,deg$^2$ \cha\ COSMOS Legacy survey, which required an overall 4.6\,Ms \cha\ exposure, contains 174 $z>$3 AGNs \citep{marchesi16b}; and the 31\,deg$^2$ Stripe 82X survey, which combined $\sim$500\,ks of\cha\ time and $\sim$1\,Ms of \xmm\ time for an overall 1.5\,Ms X-ray exposure, contains 45 $z>3$ \citet{ananna17}.

Even more importantly, the AXIS surveys would allow us to detect for the first time a population of X-ray-selected $z>$6 AGNs: we expect to detect a total of $\sim$30--100 of these primordial accreting supermassive black holes from the three surveys. If the predictions of the \citet{vito14} XLF are confirmed, we also expect to detect sources up to $z\sim$8, possibly enabling the direct detection of late-stage accreting  SMBHs seeds.

\begingroup
\renewcommand*{\arraystretch}{1.5}
\begin{table*}
\centering
\scalebox{0.95}{
\vspace{.1cm}
  \begin{tabular}{c cccccccccc c ccccc}
       \hline
       \hline
       & \multicolumn{10}{c}{\citet{gilli07} mock} & & \multicolumn{5}{c}{\citet{vito14} mock}\\
       \hline       
       AXIS Survey & \multicolumn{2}{c}{$z>$3} & \multicolumn{2}{c}{$z>$4} & \multicolumn{2}{c}{$z>$5} & \multicolumn{2}{c}{$z>$6} & \multicolumn{2}{c}{$z>$7} & & $z>$3 & $z>$4 & $z>$5 & $z>$6 & $z>$7 \\
       & $n_{\rm src}$ & $L_{\rm lim}$ & $n_{\rm src}$ & $L_{\rm lim}$ & $n_{\rm src}$ & $L_{\rm lim}$ & $n_{\rm src}$ & $L_{\rm lim}$ & $n_{\rm src}$ & $L_{\rm lim}$ & & $n_{\rm src}$ & $n_{\rm src}$ & $n_{\rm src}$ & $n_{\rm src}$ & $n_{\rm src}$ \\ 
       \hline
        Deep            & 127   & 41.1 & 27 & 41.3 & 7 & 41.5 & 2 & 41.7  & 1 & 41.9 & & 343   &	63  &	15  &	8   &	3 \\
        Intermediate    & 1066  & 41.8 & 200 & 42.1 & 61 &	42.3 & 11 & 42.5 & 2 & 42.6 & & 1948	&   389 &   103 &   29  &   8 \\
        Wide	        & 5997  & 42.9 & 966 & 43.2  & 127 & 43.4 &	16 & 43.6 & 11 & 43.7 & & 6690	&   1291    &   321 &   63  &   23 \\
       \hline
	    \hline
\end{tabular}}
	\caption{\normalsize Number of high-redshift sources detected in each AXIS simulation, and logarithm of the 0.5--2\,keV luminosity corresponding to the flux at which 20\,\% of the survey area is covered, see Table \ref{tab:results_all}, at different high-$z$ thresholds. 
	The number of detections are computed rescaling the numbers obtained in the simulations to the survey proposed area (2.5\,deg$^2$ for the 300\,ks, intermediate-area survey and 50\,deg$^2$ for the 15\,ks, wide-area survey). On the left, we report the results obtained using the \citet{gilli07} mock, while on the right we report the results obtained using the \citet{vito14} mock (more details in the text).
	}
\label{tab:results_highz}
\end{table*}
\endgroup

In Table \ref{tab:athena_high-z} we report the total number of high-$z$ sources detected in the \athena\ simulations. Overall, \athena\ will detect 9000--11,500 $z>$3 AGNs, i.e., about 20\,\% more than what is expected from AXIS surveys (whose reference survey plan is currently 50\,\% shorter).

\begingroup
\renewcommand*{\arraystretch}{1.5}
\begin{table*}
\begin{center}
\scalebox{0.94}{
\begin{tabular}{c c c c c c c c c c c c c c c c c}
\hline \hline
         &  \multicolumn{10}{c}{\citet{gilli07} mock}    & & \multicolumn{5}{c}{\citet{vito14} mock}  \\
\hline          
\athena\ survey   & \multicolumn{2}{c}{$z>$3} & \multicolumn{2}{c}{$z>$4} & \multicolumn{2}{c}{$z>$5} & \multicolumn{2}{c}{$z>$6} & \multicolumn{2}{c}{$z>$7} & & $z>$3 & $z>$4 & $z>$5 & $z>$6 & $z>$7 \\ 
& $n_{\rm src}$ & $L_{\rm lim}$ & $n_{\rm src}$ & $L_{\rm lim}$ & $n_{\rm src}$ & $L_{\rm lim}$ & $n_{\rm src}$ & $L_{\rm lim}$ & $n_{\rm src}$ & $L_{\rm lim}$ & & $n_{\rm src}$ & $n_{\rm src}$ & $n_{\rm src}$ & $n_{\rm src}$ & $n_{\rm src}$ \\

\hline 
Deep     &   1744    & 42.1 &   326    &  42.4 &   68 & 42.6   &  15 & 42.8    &    6  & 42.9  &  & 2667       &  534     &  147     &   38    &  15    \\
Wide     &   7344  & 42.5  &   1350 & 42.8    &    259 & 43.0   &  54 & 43.2     &   22  & 43.3  &  & 8878       &  1831     &  416     &  85     & 23    \\
\hline
\hline
\end{tabular}}
\caption{\normalsize Number of expected high-redshift detections in the deep and wide tier of the \athena--WFI survey simulation, and logarithm of the 0.5--2\,keV luminosity limit (computed using the flux limit at 20\,\% of the survey area).}\label{tab:athena_high-z}
\end{center}
\end{table*}
\endgroup

\subsubsection{Constraints on theoretical models of early black hole accretion}
\begin{figure*}[htbp]
  \centering
\includegraphics[width=0.8\linewidth]{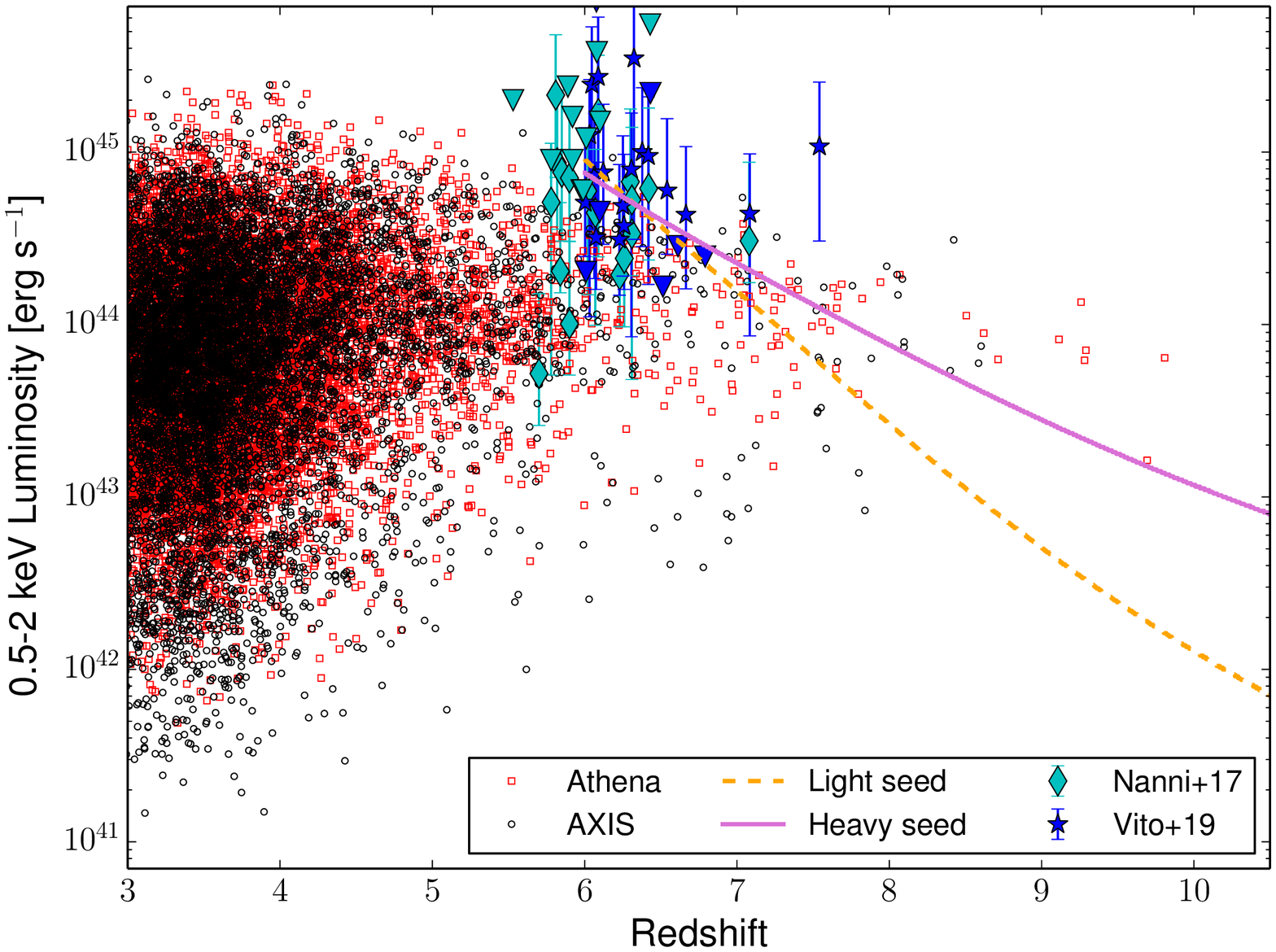}
\caption{\normalsize 
0.5--2\,keV luminosity as a function of redshift for the $z>$3 sources detected in the simulated AXIS (black circles) and \athena\ (red squares) surveys. 1\,$\sigma$ uncertainties are computed using the \citet{gehrels86} equations. The $z>$5.5 AGNs detected in the X-rays by currently available facilities are also plot for comparison \citep[][blue stars; \citealt{nanni17}, cyan diamonds]{vito19}.
Two different models of black hole seed accretion are also shown (light SMBH seed:  dashed yellow line; heavy SMBH seed: solid magenta line; see the text for more details).
}\label{fig:z_gt3_vs_Lx}
\end{figure*}

In Figure \ref{fig:z_gt3_vs_Lx} we plot the AGN 0.5--2\,keV luminosity as a function of redshift for the AXIS and \athena\ $z>$3 samples. While \athena\ will collect more sources, AXIS is expected to sample luminosities $\sim$1\,dex fainter up to redshift $\sim$8, once again highlighting the complementarity between the two instruments.

In Figure \ref{fig:z_gt3_vs_Lx} we also report the evolution with redshift of the X-ray luminosity of two idealized BHs growing to Log(M$_{\rm BH}$/M$_{\odot}$)=9.1 at z=6 through continuous accretion. Such large SMBHs are commonly found to power luminous ($L_X\sim$ 10$^{45}$\,erg\,s$^{-1}$) quasars at $z$=6 in current wide-area optical surveys, such as SDSS \citep{fan06} and PanSTARRS \citep{banados16}.

In the general case of an accreting BH radiating at a given fraction $\lambda\equiv L_{\rm bol}/L_E$ of its Eddington luminosity $L_E$ and with constant efficiency $\epsilon$, the BH mass grows as $M(t)=M_{\rm seed} e^{t/t_{\rm Sal}}$ , where $M_{\rm seed}$ is the BH seed mass and $t_{\rm Sal}$ is the Salpeter e-folding time: $t_{\rm Sal} = 50 \; {\rm Myr} \; \left(\frac{9\epsilon}{1-\epsilon}\right)\lambda^{-1}$. 
Under these assumptions, the QSO bolometric luminosity also grows exponentially, i.e., $L_{\rm bol}(t) =   L_{\rm bol,0} e^{t/t_{\rm Sal}}$ , where $L_{\rm bol,0}$ is the bolometric luminosity at the beginning of the accretion. We assume $\epsilon=0.1$ and converted from bolometric to 0.5-2\,keV luminosities using the recent AGN bolometric corrections of \citet{duras20}. We rescale the 2-10\,keV band luminosities considered by \citet{duras20} to the 0.5-2\,keV band using a photon index of $\Gamma=1.9$.

We consider two different seed masses that bracket the range proposed by theory: $i$) a light seed with mass M$_{\rm seed}$=10$^2$\,M$_{\odot}$, similar to those expected from the remnants of the first, PopIII stars, and $ii$) a heavy seed with M$_{\rm seed}$=10$^5$\,M$_{\odot}$. Under favorable environmental conditions, such massive seeds may form through the direct collapse of large, pristine gas clouds \citep[see][for a recent review on the formation of early BHs]{inayoshi20}.
In order to produce a $10^9\;M_{\odot}$ BH by z=6, light seeds must be continuously accreting at their Eddington limit ($\lambda=1.0$) for $\sim 820$ Myr (i.e. since $z_{start}=30$; yellow dashed curve). Trivially, heavy seeds are able to produce the same $10^9\;M_{\odot}$ masses at z=6 by accreting over a shorter period of time ($\sim 670$ Myr; $z_{start}=16$) and with lower Eddington ratios ($\lambda=0.7$; magenta solid curve). Clearly, these growth models are oversimplified, but, as shown in Figure \ref{fig:z_gt3_vs_Lx}, in principle both \athena\ and AXIS surveys would have the sensitivity to detect the progenitors of SDSS QSOs.  
Based on our simulations, these surveys are expected to detect $\sim$80 AGNs at $z\geq$7. Among this high-$z$ AGN population, the two instruments might be able to track the progenitors of SDSS QSOs up to $z\sim 8$ ($z\sim 9$) if they grow from light (heavy) seeds. 

\begin{figure*}[htbp]
\begin{minipage}[b]{.5\textwidth}
  \centering
\includegraphics[width=1.\linewidth]{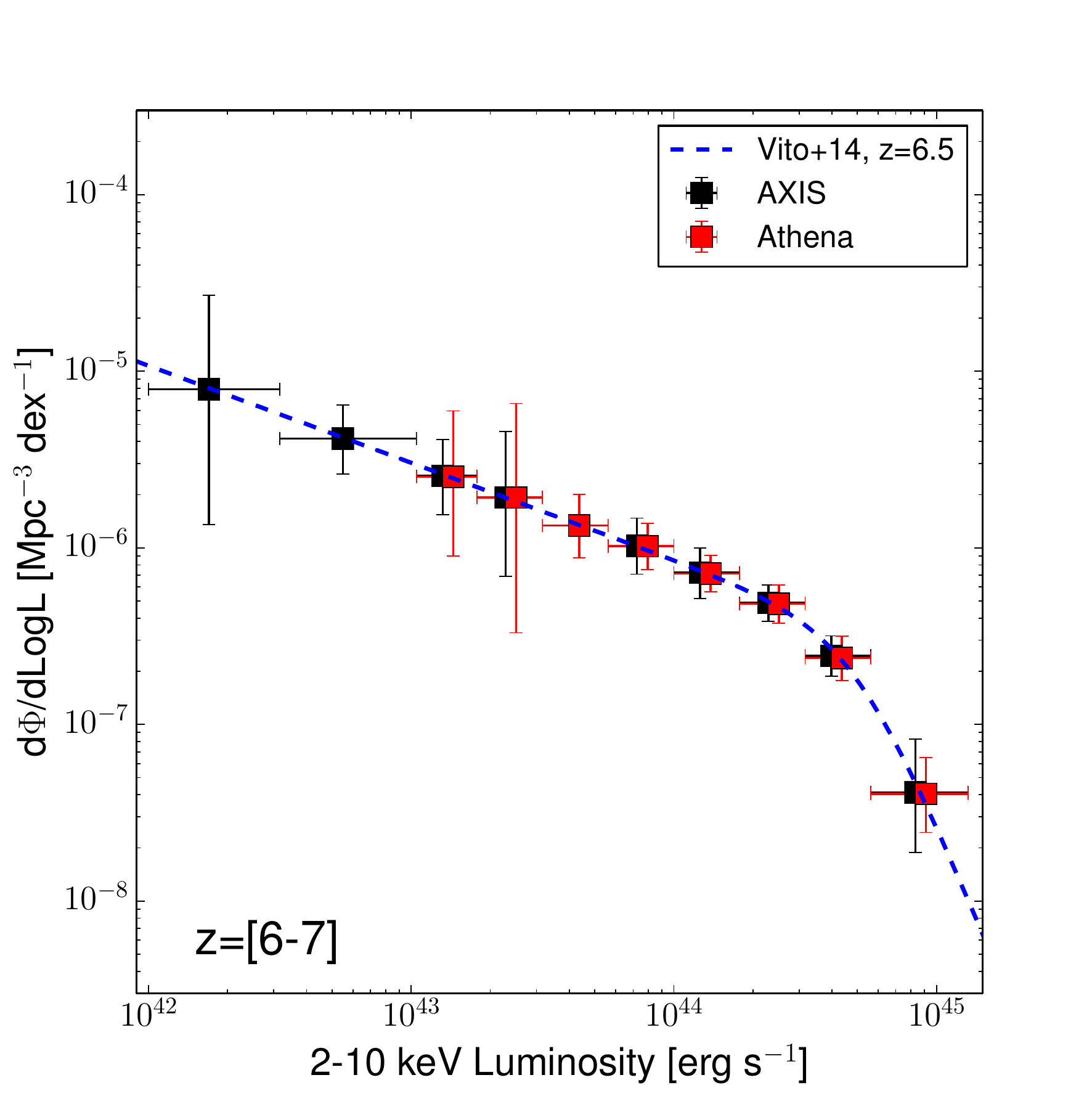}
\end{minipage}
\begin{minipage}[b]{.49\textwidth}
  \centering
\includegraphics[width=1.\linewidth]{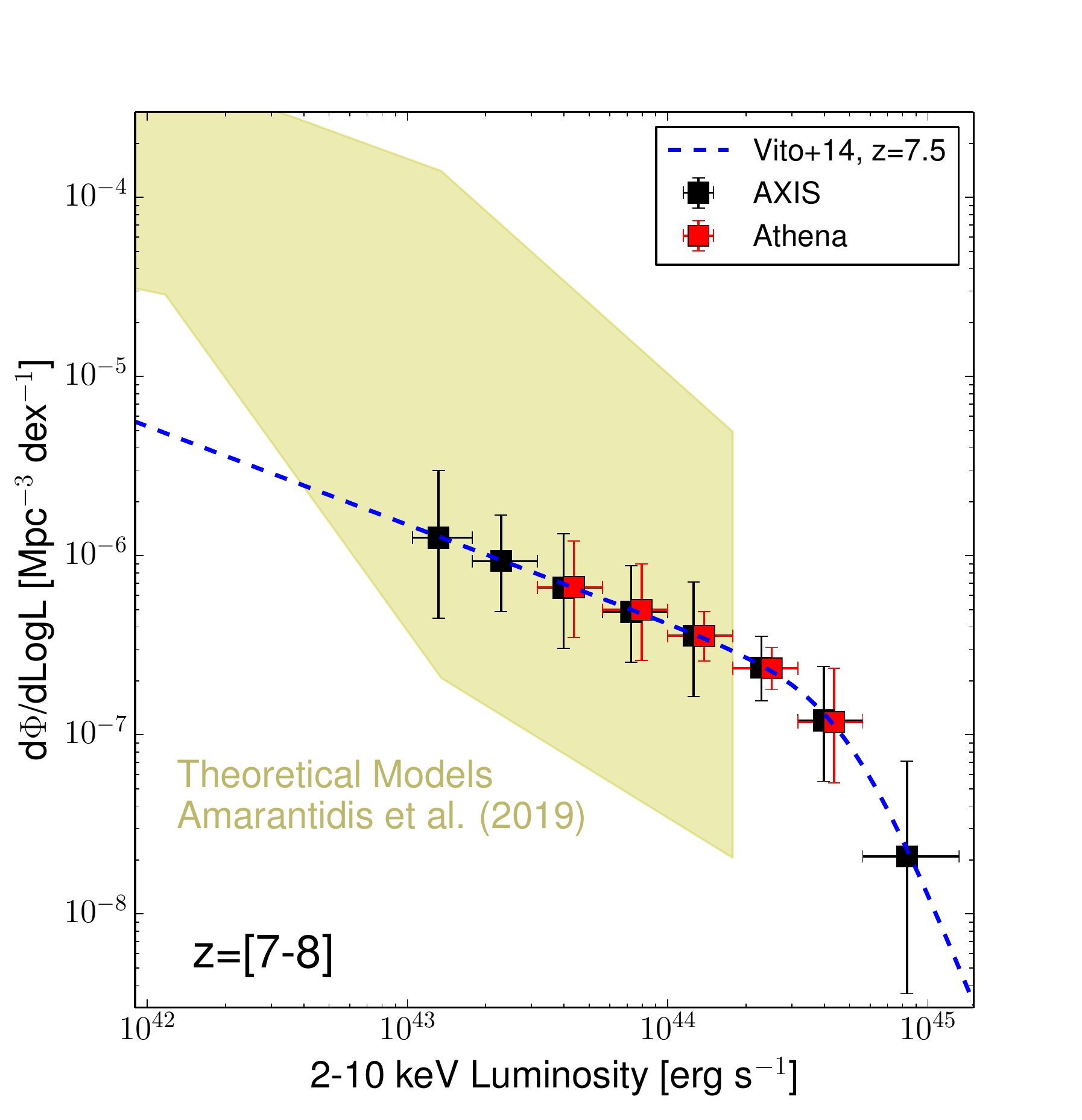}
\end{minipage}
\caption{\normalsize 
Expected X-ray luminosity function at $z\sim$6.5 (left) and $z\sim$7.5 (right), of AXIS (black) and \athena\ (red), respectively. The \citet{vito14} XLFs at the same redshifts are plotted as a blue dashed line and a solid cyan line, respectively. In the right panel, the range of predictions from the hydro-dynamical simulations and semi-analytical models discussed in \citet[][khaki area]{amarantidis19} is also shown for comparison.
}\label{fig:XLF_z_gt6}
\end{figure*}

In Figure \ref{fig:XLF_z_gt6}, we show the AXIS and \athena\ X-ray luminosity functions at $z\sim$6.5 and $z\sim$7.5. As it can be seen, the two surveys would nicely complement each other, with AXIS reaching $\sim$1\,dex lower luminosities (down to log(L$_{\rm 2-10keV}\sim$42 at $z$=6.5) and \athena\ achieving a better source statistic at 2--10\,keV luminosities $\gtrsim$5\,$\times$10$^{43}$\,erg\,s$^{-1}$.

In Figure \ref{fig:XLF_z_gt6}, we also plot the XLF predictions from the \citet{vito14} model, that we used as a reference for our mock: consequently, the simulated XLFs closely match the \citet{vito14} one. In the same figure, we show the range of predictions of several different models of SMBH early accretion, both hydro-dynamical \citep[namely, EAGLE,][Horizon-AGN, \citealt{dubois14}, Illustris, \citealt{vogelsberger14}, and MassiveBlackII, \citealt{khandai15} ]{crain15,schaye15,mcalpine16} and semi-analytical \citep[GALFORM,][L-Galaxies \citealt{guo11,henriques15}, MERAXES \citealt{mutch16,qin17}, and SHARK \citealt{lagos18}]{cole00,lacey16}. The range of predictions has been taken from \citet{amarantidis19}, to which we refer for a complete description of the models \citep[see also][for another theoretical prediction]{ni20}.

As it can be seen, the basically non-existent observational evidence at $z>$7 is reflected in the large discrepancy ($\sim$3\,dex in the luminosity range 10$^{43}$--10$^{44}$\,erg\,s$^{-1}$) between the predictions of the theoretical models. This issue would be significantly addressed by the launch of AXIS and \athena: in fact, while these two instruments are not expected to directly detect the first BH seeds, they would allow us to constrain the AGN $z>$7 XLF with uncertainties $<$50\,\% over a wide range of luminosities.  This would enable an unprecedented tuning of the theoretical SMBH accretion models, allowing us to rule out many combinations of parameters.

\subsection{Heavily obscured black hole accretion in the early Universe}\label{sec:results-CT}
Based on all models of AGN population synthesis \citep[see, e.g.,][]{gilli07,ballantyne11,ananna19} and on the available observational evidence, it is well established that most of the mass accretion in SMBHs takes place in an obscured phase, where the column density of the material surrounding the SMBH on pc scales is N$_{\rm H}>$10$^{22}$\,cm$^{-2}$. Indeed, a large part of AGNs are expected to be Compton thick (CT-), i.e., with column density N$_{\rm H}>$10$^{24}$\,cm$^{-2}$: as shown in Figure~\ref{fig:AGN_input_info}, right panel, in the mock we used in this work $\sim$45\,\% of the AGNs are Compton thick \citep[see also, e.g.,][]{georgakakis13}.

While, based on this information, we expect to detect a large number of CT-AGNs within the AXIS surveys, for a proper characterization of these objects the $\sim$5--10 0.5--7\,keV net counts needed for a detection are not enough. Instead, one needs to detect $>$100 net counts to perform a X-ray spectral analysis and constrain with reliable uncertainties the obscuring material column density.

In Table \ref{tab:CT-detections} we report the expected number of CT-AGNs detected in each of the three AXIS surveys, both in the overall sample and at $z>$3: the results are derived using the \citet{vito14} high-$z$ mock catalog. While we expect to detect most CT-AGNs in the wide area survey, only the intermediate and the deep X-ray surveys would allow us to properly characterize a significant number of CT sources through X-ray spectroscopy. Overall, the AXIS surveys are expected to contain $\sim$850 CT-AGNs with $>$100 net counts in the 0.5--7\,keV band, $\sim$40 of which at $z>$3: in currently available \cha\ and \xmm\ surveys, fewer than five $z>$3 CT-AGN are detected with $>$100 net counts in the 0.5--7\,keV band \citep[see, e.g.,][]{brightman14,lanzuisi18,corral19}. Similar surveys would then allow a full X-ray characterization of heavily obscured sources at the peak of the black hole accretion history.

In Table \ref{tab:CT-detections} we also report the number of Compton thick AGNs detected in the two proposed \athena\ surveys. The large \athena\ grasp, combined with its planned deep exposure, will be particularly handy to find and characterize large numbers of these otherwise elusive objects. In particular, we expect to detect $\sim$130 $z>$3 CT-AGN with more than 100 0.5--7\,keV net counts in the two \athena\ surveys. While AXIS is expected to detect fewer high-$z$ CT-AGN with $>$100 0.5--7\,keV net counts, it would also generally sample a population of obscured sources intrinsically less luminous than those detected by \athena.

As we mentioned earlier in the text, and as shown in Figure \ref{fig:AGN_spectra}, in heavily obscured sources the $>$10\,keV flux predicted by the physically motivated \borus\ model is 30--60\,\% fainter than the one predicted by the \texttt{pexmon} one. Consequently, we expect that the number of CT sources reported in Table \ref{tab:CT-detections} might be somewhat over-estimated. 

More quantitatively, and focusing at first on the high-$z$ subsample, the observed 0.5--2\,keV (2--7\,keV) band corresponds, at $z$=3, to the 2--8\,keV (8--28\,keV) rest-frame one. In these energy ranges, the ratio between the \borus\ and  \texttt{pexmon} fluxes are $r_{\rm 0.5-2}$=1.07 and $r_{\rm 2-10}$=0.47.
As a consequence, we expect that the choice of \texttt{pexmon} instead of \borus\ does not affect the number of CT-AGN detections in the 0.5--2\,keV band in our simulation. To instead check how significant is the effect in the 2--10\,keV band, we rescale by $r_{\rm 2-10}$ the 2--10\,keV fluxes of the $z>$3 CT sources detected in our simulated surveys. We find that in the AXIS intermediate and wide surveys $\sim$10--15\,\% of the CT-AGNs originally detected would have a flux fainter than the one corresponding to 20\,\% of the area covered by the survey (see Table \ref{tab:results_all}) and would likely be missed. In the AXIS deep field, the fraction of missed objects is even smaller, being $\sim$5\,\%. In \athena\, we expect to miss $\sim$10\,\% of the $z>$3 CT-AGNs in the deep survey, and $\sim$15\,\% in the wide one.

Finally, we note that, at $z\sim$1 (i.e., the average redshift of the CT-AGNs detected in AXIS and \athena, and the redshift where the bulk of CT-AGNs would be detected) the ratio between the rest-frame \borus\ and \texttt{pexmon} 0.5--7\,keV flux is $\sim$1 and we therefore do not expect the overall number of CT detections to change significantly. 

\begin{figure}[htbp]
\centering
\includegraphics[width=1.\linewidth]{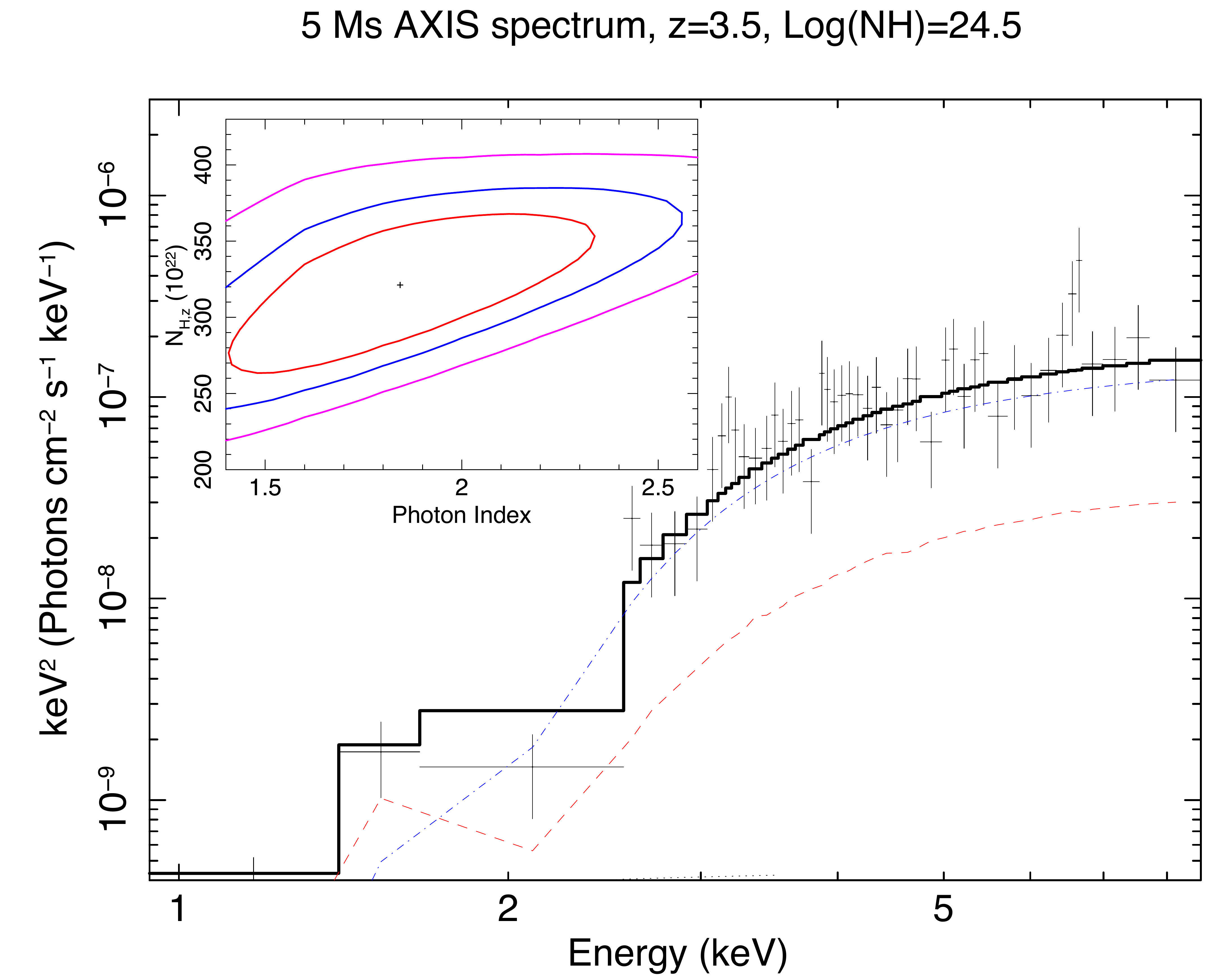}
\caption{\normalsize Spectrum of a $z$=3.5 CT-AGN with Log($N_{\rm H}$)=24.5 and 0.5--2\,keV intrinsic luminosity Log(L$_{\rm 0.5-2}$)=44.5, as it would be detected in the AXIS deep, 5\,Ms survey. The spectrum is rebinned with a minimum of 5 counts per bin. 
In the insets, we show the 68, 90 and 99\,\% confidence level contours for the power law photon index $\Gamma$ and the line-of-sight column density $N_{\rm H}$.
}\label{fig:CT_spectra}
\end{figure}

In Figure \ref{fig:CT_spectra} we show the AXIS spectrum of a Compton thick AGNs at $z$=3.5 simulated using \borus, as it would be observed in the AXIS deep, 5\,Ms survey. It is worth pointing out, however, that \borus\ does not take into account off-nuclear absorption, caused by the interstellar medium in the host galaxy, which can be significant in high-$z$ AGNs \citep[see, e.g.,][]{circosta19,damato20,ni20}.

The source has column density Log($N_{\rm H}$=24.5 and 0.5--2\,keV rest-frame, absorption corrected luminosity Log(L$_{0.5-2}$=44.5). Such a luminosity corresponds to an observed flux in the same band $f_{\rm 0.5-2}\sim$2\,$\times$\,10$^{-16}$, and would lead to the detection of $\sim$250 net counts in the 0.5--7\,keV band in a 5\,Ms AXIS observations. With such a count statistic, it would be possible to measure the line-of-sight column density with 90\,\% confidence uncertainties $\leq$30\,\%, as shown in the inset of Figure \ref{fig:CT_spectra}. Based on our simulations, the deep and intermediate AXIS surveys should detect $\sim$20 CT-AGN at $z>$3 at least as bright as the one shown in Figure \ref{fig:CT_spectra}.

\begingroup
\renewcommand*{\arraystretch}{1.5}
\begin{table*}
\centering
\scalebox{0.95}{
\vspace{.1cm}
  \begin{tabular}{c ccccccc c ccccccc}
       \hline
       \hline
                &  \multicolumn{7}{c}{AXIS} & & \multicolumn{7}{c}{\athena}  \\
       & \multicolumn{4}{c}{Whole sample} & & \multicolumn{2}{c}{$>$100 net counts} &  & \multicolumn{4}{c}{Whole sample} & & \multicolumn{2}{c}{$>$100 net counts}\\
       \hline       
       Survey & \multicolumn{2}{c}{All} & \multicolumn{2}{c}{$z>$3} & & All & $z>$3 & & \multicolumn{2}{c}{All} & \multicolumn{2}{c}{$z>3$} & & All  & $z>3$ \\
       & $n_{\rm src}$ & $L_{\rm lim}$ & $n_{\rm src}$ & $L_{\rm lim}$ & & $n_{\rm src}$ & $n_{\rm src}$ & & $n_{\rm src}$ & $L_{\rm lim}$ & $n_{\rm src}$ & $L_{\rm lim}$ & & $n_{\rm src}$ & $n_{\rm src}$\\
       \hline
        Deep            &   850  & 37.0 & 62 & 41.6 & & 452 & 31 & &  4476     & 38.1 & 218 & 43.8 & & 2821      & 131    \\
        Intermediate    &   4396 & 37.8 & 268 & 42.3 & & 335  & 10 & & -- & -- & -- & -- & & -- & --\\
        Wide	        &   6113 &  38.9 & 172 & 44.5 & & 69 & 0 &  & 7236  & 38.5  &   236  & 44.2 &  &  594      &  0    \\
       \hline
	    \hline
\end{tabular}}
	\caption{\normalsize Expected number of Compton thick AGN (i.e., sources having column density N$_{\rm H}>$10$^{24}$\,cm$^{-2}$) in each of the planned AXIS and \athena\ surveys. 
	We report both the overall results and the number of sources for which more than 100 0.5--7\,keV net counts are detected, thus making possible a reliable X-ray spectral analysis. The logarithm of the 0.5--2\,keV luminosity is computed using the spectral model described in Section \ref{sec:results-CT} and a 0.5--2\,keV flux corresponding to the one at which 20\,\% of the survey area is covered (see Table \ref{tab:results_all}).
	}
\label{tab:CT-detections}
\end{table*}
\endgroup

\section{Summary and conclusions}\label{sec:summary}
In this paper, we presented new mock catalogs of active galactic nuclei, non-active galaxies and clusters of galaxies for the simulations of X-ray surveys. The mocks are made available online also in the \sixte\ format, which makes them ready-to-use for any type of simulation.

All the mocks are derived from the most up-to-date observational evidence and its extrapolations to fluxes not yet sampled by current X-ray facilities. In particular, two different AGN mocks have been derived using the \citet{gilli07} AGN population synthesis model and, at redshift $z>$3, the \citet{vito14} X-ray luminosity function. These mocks reach 0.5--2\,keV luminosities $L_{\rm 0.5-2}$=10$^{40}$\,erg\,s$^{-1}$ and fluxes in the same band $f_{\rm 0.5-2}$=10$^{-20}$\,\cgs, i.e., way below the flux and luminosity limits of current X-ray facilities. Therefore, our mocks can be easily used both to simulate X-ray surveys with current facilities, such as \cha, \xmm, or the eROSITA all-sky mission, and to make predictions for future instruments, such as the forthcoming \athena\ mission, or the AXIS and \textit{Lynx} mission concepts.

In particular, in this work we used our mocks to simulate a set of surveys with \athena\ and the proposed AXIS probe. We find that these future, next generation surveys may transform our knowledge of the deep X-ray Universe. Some examples are as follows.

\begin{enumerate}
    \item As shown in Table \ref{tab:results_all}, AXIS would lead to the detection of over 275,000 X-ray sources at a $>$3\,$\sigma$ significance level. About 80\,\% of these objects are expected to be AGNs, while the remaining 20\,\% (i.e., $\sim$50,000 sources), are expected to be non-active galaxies. Based on a comparison with currently available X-ray surveys, $\sim$90\,\% of these objects would be detected in the X-rays for the first time.
    \item The combination of \athena\ and AXIS would be strategical to improve our knowledge of the high-$z$ redshift Universe, which has not been significantly explored by current X-ray facilities. Based on our simulations, \athena\ and AXIS are expected to detect $\sim$20,000 $z>$3 AGNs, i.e., a factor $\sim$60 more than those detected with current X-ray facilities. This would allow us to investigate for the first time with unprecedented statistics the history of SMBH accretion before the peak of AGN activity. The two instruments would also detect $\gtrsim$70 $z>$7 AGNs, and would make it possible to constrain the AGN X-ray luminosity function at $z\sim$7 down to luminosities L$_{\rm 2-10keV}\sim$10$^{43}$\,erg\,s$^{-1}$.
    This would enable an unprecedented tuning of theoretical SMBH accretion models.
    \item \athena\ and AXIS are expected to detect thousands of heavily obscured AGNs at high redshift: these are the objects which are thought to accrete more efficiently, but their large column density has made them invisible to current X-ray facilities. For a subsample of over 150 of these $z>$3 CT-AGNs, the two instruments would detect more than 100 net counts, thus allowing us to reliably constrain the obscuring material column density.  This will enable an accurate census of the AGN population before the peak of AGN activity at $z\sim$2--3.
\end{enumerate}

\section*{Acknowledgements}
We thank the anonymous referee for the useful suggestions, that helped in significantly improving the paper, and 
Stephanie LaMassa for her helpful explanation of the Stripe 82X survey technical layout.
We acknowledge financial contribution from the agreement ASI-INAF n. 2017-14-H.O. S.E. acknowledges financial contribution from the contract ASI-INAF Athena 2015-046-R.0 and from INAF ``Call per interventi aggiuntivi a sostegno della ricerca di main stream di INAF''.

\bibliographystyle{aa}
\bibliography{axis_simulations}

\begin{thebibliography}{107}
\expandafter\ifx\csname natexlab\endcsname\relax\def\natexlab#1{#1}\fi

\bibitem[{{Aird} {et~al.}(2015){Aird}, {Coil}, {Georgakakis}, {Nandra},
  {Barro}, \& {P{\'e}rez-Gonz{\'a}lez}}]{aird15}
{Aird}, J., {Coil}, A.~L., {Georgakakis}, A., {et~al.} 2015, \mnras, 451, 1892

\bibitem[{{Aird} {et~al.}(2013){Aird}, {Comastri}, {Brusa}, {Cappelluti},
  {Moretti}, {Vanzella}, {Volonteri}, {Alexander}, {Afonso}, {Fiore},
  {Georgantopoulos}, {Iwasawa}, {Merloni}, {Nandra}, {Salvaterra}, {Salvato},
  {Severgnini}, {Schawinski}, {Shankar}, {Vignali}, \& {Vito}}]{aird13}
{Aird}, J., {Comastri}, A., {Brusa}, M., {et~al.} 2013, arXiv e-prints,
  arXiv:1306.2325

\bibitem[{{Amarantidis} {et~al.}(2019){Amarantidis}, {Afonso}, {Messias},
  {Henriques}, {Griffin}, {Lacey}, {Lagos}, {Gonzalez-Perez}, {Dubois},
  {Volonteri}, {Matute}, {Pappalardo}, {Qin}, {Chary}, \&
  {Norris}}]{amarantidis19}
{Amarantidis}, S., {Afonso}, J., {Messias}, H., {et~al.} 2019, \mnras, 485,
  2694

\bibitem[{{Ananna} {et~al.}(2017){Ananna}, {Salvato}, {LaMassa}, {Urry},
  {Cappelluti}, {Cardamone}, {Civano}, {Farrah}, {Gilfanov}, {Glikman},
  {Hamilton}, {Kirkpatrick}, {Lanzuisi}, {Marchesi}, {Merloni}, {Nandra},
  {Natarajan}, {Richards}, \& {Timlin}}]{ananna17}
{Ananna}, T.~T., {Salvato}, M., {LaMassa}, S., {et~al.} 2017, \apj, 850, 66

\bibitem[{{Ananna} {et~al.}(2019){Ananna}, {Treister}, {Urry}, {Ricci},
  {Kirkpatrick}, {LaMassa}, {Buchner}, {Civano}, {Tremmel}, \&
  {Marchesi}}]{ananna19}
{Ananna}, T.~T., {Treister}, E., {Urry}, C.~M., {et~al.} 2019, \apj, 871, 240

\bibitem[{{Arnaud}(1996)}]{arnaud96}
{Arnaud}, K.~A. 1996, in Astronomical Society of the Pacific Conference Series,
  Vol. 101, Astronomical Data Analysis Software and Systems V, ed. G.~H.
  {Jacoby} \& J.~{Barnes} (Astronomical Society of the Pacific), 17

\bibitem[{{Ba{\~n}ados} {et~al.}(2016){Ba{\~n}ados}, {Venemans}, {Decarli},
  {Farina}, {Mazzucchelli}, {Walter}, {Fan}, {Stern}, {Schlafly}, {Chambers},
  {Rix}, {Jiang}, {McGreer}, {Simcoe}, {Wang}, {Yang}, {Morganson}, {De Rosa},
  {Greiner}, {Balokovi{\'c}}, {Burgett}, {Cooper}, {Draper}, {Flewelling},
  {Hodapp}, {Jun}, {Kaiser}, {Kudritzki}, {Magnier}, {Metcalfe}, {Miller},
  {Schindler}, {Tonry}, {Wainscoat}, {Waters}, \& {Yang}}]{banados16}
{Ba{\~n}ados}, E., {Venemans}, B.~P., {Decarli}, R., {et~al.} 2016, \apjs, 227,
  11

\bibitem[{{Baldi} {et~al.}(2012){Baldi}, {Ettori}, {Molendi}, {Balestra},
  {Gastaldello}, \& {Tozzi}}]{baldi12}
{Baldi}, A., {Ettori}, S., {Molendi}, S., {et~al.} 2012, \aap, 537, A142

\bibitem[{{Balestra} {et~al.}(2007){Balestra}, {Tozzi}, {Ettori}, {Rosati},
  {Borgani}, {Mainieri}, {Norman}, \& {Viola}}]{balestra07}
{Balestra}, I., {Tozzi}, P., {Ettori}, S., {et~al.} 2007, \aap, 462, 429

\bibitem[{{Ballantyne} {et~al.}(2011){Ballantyne}, {Draper}, {Madsen}, {Rigby},
  \& {Treister}}]{ballantyne11}
{Ballantyne}, D.~R., {Draper}, A.~R., {Madsen}, K.~K., {Rigby}, J.~R., \&
  {Treister}, E. 2011, \apj, 736, 56

\bibitem[{{Balokovi{\'c}} {et~al.}(2018){Balokovi{\'c}}, {Brightman},
  {Harrison}, {Comastri}, {Ricci}, {Buchner}, {Gandhi}, {Farrah}, \&
  {Stern}}]{balokovic18}
{Balokovi{\'c}}, M., {Brightman}, M., {Harrison}, F.~A., {et~al.} 2018, \apj,
  854, 42

\bibitem[{{Barger} {et~al.}(2019){Barger}, {Cowie}, {Bauer}, \&
  {Gonz{\'a}lez-L{\'o}pez}}]{barger19}
{Barger}, A.~J., {Cowie}, L.~L., {Bauer}, F.~E., \& {Gonz{\'a}lez-L{\'o}pez},
  J. 2019, \apj, 887, 23

\bibitem[{{Barret} {et~al.}(2016){Barret}, {Lam Trong}, {den Herder}, {Piro},
  {Barcons}, {Huovelin}, {Kelley}, {Mas-Hesse}, {Mitsuda}, {Paltani}, {Rauw},
  {Ro{\.Z}anska}, {Wilms}, {Barbera}, {Bozzo}, {Ceballos}, {Charles},
  {Decourchelle}, {den Hartog}, {Duval}, {Fiore}, {Gatti}, {Goldwurm},
  {Jackson}, {Jonker}, {Kilbourne}, {Macculi}, {Mendez}, {Molendi},
  {Orleanski}, {Pajot}, {Pointecouteau}, {Porter}, {Pratt}, {Pr{\^e}le},
  {Ravera}, {Renotte}, {Schaye}, {Shinozaki}, {Valenziano}, {Vink}, {Webb},
  {Yamasaki}, {Delcelier-Douchin}, {Le Du}, {Mesnager}, {Pradines}, {Brand
  uardi-Raymont}, {Dadina}, {Finoguenov}, {Fukazawa}, {Janiuk}, {Miller},
  {Naz{\'e}}, {Nicastro}, {Sciortino}, {Torrejon}, {Geoffray}, {Hernandez},
  {Luno}, {Peille}, {Andr{\'e}}, {Daniel}, {Etcheverry}, {Gloaguen}, {Hassin},
  {Hervet}, {Maussang}, {Moueza}, {Paillet}, {Vella}, {Campos Garrido},
  {Damery}, {Panem}, {Panh}, {Band ler}, {Biffi}, {Boyce}, {Cl{\'e}net},
  {DiPirro}, {Jamotton}, {Lotti}, {Schwander}, {Smith}, {van Leeuwen}, {van
  Weers}, {Brand}, {Cobo}, {Dauser}, {de Plaa}, \& {Cucchetti}}]{barret16}
{Barret}, D., {Lam Trong}, T., {den Herder}, J.-W., {et~al.} 2016, in Society
  of Photo-Optical Instrumentation Engineers (SPIE) Conference Series, Vol.
  9905, \procspie, 99052F

\bibitem[{{Bi} {et~al.}(2020){Bi}, {Feng}, \& {Ho}}]{bi20}
{Bi}, S., {Feng}, H., \& {Ho}, L.~C. 2020, arXiv e-prints, arXiv:2007.06026

\bibitem[{{Brightman} {et~al.}(2014){Brightman}, {Nandra}, {Salvato}, {Hsu},
  {Aird}, \& {Rangel}}]{brightman14}
{Brightman}, M., {Nandra}, K., {Salvato}, M., {et~al.} 2014, \mnras, 443, 1999

\bibitem[{{Broos} {et~al.}(2012){Broos}, {Townsley}, {Getman}, \&
  {Bauer}}]{broos12}
{Broos}, P., {Townsley}, L., {Getman}, K., \& {Bauer}, F. 2012, {AE: ACIS
  Extract}

\bibitem[{{Buchner} {et~al.}(2015){Buchner}, {Georgakakis}, {Nandra},
  {Brightman}, {Menzel}, {Liu}, {Hsu}, {Salvato}, {Rangel}, {Aird}, {Merloni},
  \& {Ross}}]{buchner15}
{Buchner}, J., {Georgakakis}, A., {Nandra}, K., {et~al.} 2015, \apj, 802, 89

\bibitem[{{Capak} {et~al.}(2011){Capak}, {Riechers}, {Scoville}, {Carilli},
  {Cox}, {Neri}, {Robertson}, {Salvato}, {Schinnerer}, {Yan}, {Wilson}, {Yun},
  {Civano}, {Elvis}, {Karim}, {Mobasher}, \& {Staguhn}}]{capak11}
{Capak}, P.~L., {Riechers}, D., {Scoville}, N.~Z., {et~al.} 2011, \nat, 470,
  233

\bibitem[{{Cappelluti} {et~al.}(2017){Cappelluti}, {Li}, {Ricarte}, {Agarwal},
  {Allevato}, {Tasnim Ananna}, {Ajello}, {Civano}, {Comastri}, {Elvis},
  {Finoguenov}, {Gilli}, {Hasinger}, {Marchesi}, {Natarajan}, {Pacucci},
  {Treister}, \& {Urry}}]{cappelluti17}
{Cappelluti}, N., {Li}, Y., {Ricarte}, A., {et~al.} 2017, \apj, 837, 19

\bibitem[{{Chiappetti} {et~al.}(2018){Chiappetti}, {Fotopoulou}, {Lidman},
  {Faccioli}, {Pacaud}, {Elyiv}, {Paltani}, {Pierre}, {Plionis}, {Adami},
  {Alis}, {Altieri}, {Baldry}, {Bolzonella}, {Bongiorno}, {Brown}, {Driver},
  {Elmer}, {Franzetti}, {Grootes}, {Guglielmo}, {Iovino}, {Koulouridis},
  {Lef{\`e}vre}, {Liske}, {Maurogordato}, {Melnyk}, {Owers}, {Poggianti},
  {Polletta}, {Pompei}, {Ponman}, {Robotham}, {Sadibekova}, {Tuffs},
  {Valtchanov}, {Vignali}, \& {Wagner}}]{chiappetti18}
{Chiappetti}, L., {Fotopoulou}, S., {Lidman}, C., {et~al.} 2018, \aap, 620, A12

\bibitem[{{Circosta} {et~al.}(2019){Circosta}, {Vignali}, {Gilli}, {Feltre},
  {Vito}, {Calura}, {Mainieri}, {Massardi}, \& {Norman}}]{circosta19}
{Circosta}, C., {Vignali}, C., {Gilli}, R., {et~al.} 2019, \aap, 623, A172

\bibitem[{{Civano} {et~al.}(2016){Civano}, {Marchesi}, {Comastri}, {Urry},
  {Elvis}, {Cappelluti}, {Puccetti}, {Brusa}, {Zamorani}, {Hasinger},
  {Aldcroft}, {Alexand er}, {Allevato}, {Brunner}, {Capak}, {Finoguenov},
  {Fiore}, {Fruscione}, {Gilli}, {Glotfelty}, {Griffiths}, {Hao}, {Harrison},
  {Jahnke}, {Kartaltepe}, {Karim}, {LaMassa}, {Lanzuisi}, {Miyaji}, {Ranalli},
  {Salvato}, {Sargent}, {Scoville}, {Schawinski}, {Schinnerer}, {Silverman},
  {Smolcic}, {Stern}, {Toft}, {Trakhtenbrot}, {Treister}, \&
  {Vignali}}]{civano16}
{Civano}, F., {Marchesi}, S., {Comastri}, A., {et~al.} 2016, \apj, 819, 62

\bibitem[{{Cole} {et~al.}(2000){Cole}, {Lacey}, {Baugh}, \& {Frenk}}]{cole00}
{Cole}, S., {Lacey}, C.~G., {Baugh}, C.~M., \& {Frenk}, C.~S. 2000, \mnras,
  319, 168

\bibitem[{{Comastri} {et~al.}(2011){Comastri}, {Ranalli}, {Iwasawa}, {Vignali},
  {Gilli}, {Georgantopoulos}, {Barcons}, {Brand t}, {Brunner}, {Brusa},
  {Cappelluti}, {Carrera}, {Civano}, {Fiore}, {Hasinger}, {Mainieri},
  {Merloni}, {Nicastro}, {Paolillo}, {Puccetti}, {Rosati}, {Silverman},
  {Tozzi}, {Zamorani}, {Balestra}, {Bauer}, {Luo}, \& {Xue}}]{comastri11}
{Comastri}, A., {Ranalli}, P., {Iwasawa}, K., {et~al.} 2011, \aap, 526, L9

\bibitem[{{Comparat} {et~al.}(2019){Comparat}, {Merloni}, {Salvato}, {Nandra},
  {Boller}, {Georgakakis}, {Finoguenov}, {Dwelly}, {Buchner}, {Del Moro},
  {Clerc}, {Wang}, {Zhao}, {Prada}, {Yepes}, {Brusa}, {Krumpe}, \&
  {Liu}}]{comparat19}
{Comparat}, J., {Merloni}, A., {Salvato}, M., {et~al.} 2019, \mnras, 487, 2005

\bibitem[{{Corral} {et~al.}(2019){Corral}, {Georgantopoulos}, {Akylas}, \&
  {Ranalli}}]{corral19}
{Corral}, A., {Georgantopoulos}, I., {Akylas}, A., \& {Ranalli}, P. 2019, \aap,
  629, A133

\bibitem[{{Crain} {et~al.}(2015){Crain}, {Schaye}, {Bower}, {Furlong},
  {Schaller}, {Theuns}, {Dalla Vecchia}, {Frenk}, {McCarthy}, {Helly},
  {Jenkins}, {Rosas-Guevara}, {White}, \& {Trayford}}]{crain15}
{Crain}, R.~A., {Schaye}, J., {Bower}, R.~G., {et~al.} 2015, \mnras, 450, 1937

\bibitem[{{Cucchetti} {et~al.}(2018){Cucchetti}, {Pointecouteau}, {Peille},
  {Clerc}, {Rasia}, {Biffi}, {Borgani}, {Tornatore}, {Dolag}, {Roncarelli},
  {Gaspari}, {Ettori}, {Bulbul}, {Dauser}, {Wilms}, {Pajot}, \&
  {Barret}}]{cucchetti18}
{Cucchetti}, E., {Pointecouteau}, E., {Peille}, P., {et~al.} 2018, \aap, 620,
  A173

\bibitem[{{D'Amato} {et~al.}(2020){D'Amato}, {Gilli}, {Vignali}, {Massardi},
  {Pozzi}, {Zamorani}, {Circosta}, {Vito}, {Fritz}, {Cresci}, {Casasola},
  {Calura}, {Feltre}, {Manieri}, {Rigopoulou}, {Tozzi}, \& {Norman}}]{damato20}
{D'Amato}, Q., {Gilli}, R., {Vignali}, C., {et~al.} 2020, \aap, 636, A37

\bibitem[{{Dauser} {et~al.}(2019){Dauser}, {Falkner}, {Lorenz}, {Kirsch},
  {Peille}, {Cucchetti}, {Schmid}, {Brand}, {Oertel}, {Smith}, \&
  {Wilms}}]{dauser19}
{Dauser}, T., {Falkner}, S., {Lorenz}, M., {et~al.} 2019, \aap, 630, A66

\bibitem[{{Despali} {et~al.}(2016){Despali}, {Giocoli}, {Angulo}, {Tormen},
  {Sheth}, {Baso}, \& {Moscardini}}]{despali16}
{Despali}, G., {Giocoli}, C., {Angulo}, R.~E., {et~al.} 2016, \mnras, 456, 2486

\bibitem[{{Diemer}(2018)}]{diemer18}
{Diemer}, B. 2018, \apjs, 239, 35

\bibitem[{{Donley} {et~al.}(2012){Donley}, {Koekemoer}, {Brusa}, {Capak},
  {Cardamone}, {Civano}, {Ilbert}, {Impey}, {Kartaltepe}, {Miyaji}, {Salvato},
  {Sanders}, {Trump}, \& {Zamorani}}]{donley12}
{Donley}, J.~L., {Koekemoer}, A.~M., {Brusa}, M., {et~al.} 2012, \apj, 748, 142

\bibitem[{{Donley} {et~al.}(2008){Donley}, {Rieke}, {P{\'e}rez-Gonz{\'a}lez},
  \& {Barro}}]{donley08}
{Donley}, J.~L., {Rieke}, G.~H., {P{\'e}rez-Gonz{\'a}lez}, P.~G., \& {Barro},
  G. 2008, \apj, 687, 111

\bibitem[{{Dubois} {et~al.}(2014){Dubois}, {Pichon}, {Welker}, {Le Borgne},
  {Devriendt}, {Laigle}, {Codis}, {Pogosyan}, {Arnouts}, {Benabed}, {Bertin},
  {Blaizot}, {Bouchet}, {Cardoso}, {Colombi}, {de Lapparent}, {Desjacques},
  {Gavazzi}, {Kassin}, {Kimm}, {McCracken}, {Milliard}, {Peirani}, {Prunet},
  {Rouberol}, {Silk}, {Slyz}, {Sousbie}, {Teyssier}, {Tresse}, {Treyer},
  {Vibert}, \& {Volonteri}}]{dubois14}
{Dubois}, Y., {Pichon}, C., {Welker}, C., {et~al.} 2014, \mnras, 444, 1453

\bibitem[{{Duras} {et~al.}(2020){Duras}, {Bongiorno}, {Ricci}, {Piconcelli},
  {Shankar}, {Lusso}, {Bianchi}, {Fiore}, {Maiolino}, {Marconi}, {Onori},
  {Sani}, {Schneider}, {Vignali}, \& {La Franca}}]{duras20}
{Duras}, F., {Bongiorno}, A., {Ricci}, F., {et~al.} 2020, \aap, 636, A73

\bibitem[{{Fan} {et~al.}(2006){Fan}, {Strauss}, {Becker}, {White}, {Gunn},
  {Knapp}, {Richards}, {Schneider}, {Brinkmann}, \& {Fukugita}}]{fan06}
{Fan}, X., {Strauss}, M.~A., {Becker}, R.~H., {et~al.} 2006, \aj, 132, 117

\bibitem[{{Finoguenov} {et~al.}(2015){Finoguenov}, {Tanaka}, {Cooper},
  {Allevato}, {Cappelluti}, {Choi}, {Heymans}, {Bauer}, {Ziparo}, {Ranalli},
  {Silverman}, {Brandt}, {Xue}, {Mulchaey}, {Howes}, {Schmid}, {Wilman},
  {Comastri}, {Hasinger}, {Mainieri}, {Luo}, {Tozzi}, {Rosati}, {Capak}, \&
  {Popesso}}]{finoguenov15}
{Finoguenov}, A., {Tanaka}, M., {Cooper}, M., {et~al.} 2015, \aap, 576, A130

\bibitem[{{Gehrels}(1986)}]{gehrels86}
{Gehrels}, N. 1986, \apj, 303, 336

\bibitem[{{Georgakakis} {et~al.}(2013){Georgakakis}, {Carrera}, {Lanzuisi},
  {Brightman}, {Buchner}, {Aird}, {Page}, {Cappi}, {Afonso}, {Alonso-Herrero},
  {Ballo}, {Barcons}, {Ceballos}, {Comastri}, {Georgantopoulos}, {Mateos},
  {Nandra}, {Rosario}, {Salvato}, {Schawinski}, {Severgnini}, \&
  {Vignali}}]{georgakakis13}
{Georgakakis}, A., {Carrera}, F., {Lanzuisi}, G., {et~al.} 2013, arXiv
  e-prints, arXiv:1306.2328

\bibitem[{{Georgantopoulos} {et~al.}(2013){Georgantopoulos}, {Comastri},
  {Vignali}, {Ranalli}, {Rovilos}, {Iwasawa}, {Gilli}, {Cappelluti}, {Carrera},
  {Fritz}, {Brusa}, {Elbaz}, {Mullaney}, {Castello-Mor}, {Barcons}, {Tozzi},
  {Balestra}, \& {Falocco}}]{georgantopoulos13}
{Georgantopoulos}, I., {Comastri}, A., {Vignali}, C., {et~al.} 2013, \aap, 555,
  A43

\bibitem[{{Gilli} {et~al.}(2007){Gilli}, {Comastri}, \& {Hasinger}}]{gilli07}
{Gilli}, R., {Comastri}, A., \& {Hasinger}, G. 2007, \aap, 463, 79

\bibitem[{{Gilli} {et~al.}(2014){Gilli}, {Norman}, {Vignali}, {Vanzella},
  {Calura}, {Pozzi}, {Massardi}, {Mignano}, {Casasola}, {Daddi}, {Elbaz},
  {Dickinson}, {Iwasawa}, {Maiolino}, {Brusa}, {Vito}, {Fritz}, {Feltre},
  {Cresci}, {Mignoli}, {Comastri}, \& {Zamorani}}]{gilli14}
{Gilli}, R., {Norman}, C., {Vignali}, C., {et~al.} 2014, \aap, 562, A67

\bibitem[{{Gilli} {et~al.}(2011){Gilli}, {Su}, {Norman}, {Vignali}, {Comastri},
  {Tozzi}, {Rosati}, {Stiavelli}, {Brandt}, {Xue}, {Luo}, {Castellano},
  {Fontana}, {Fiore}, {Mainieri}, \& {Ptak}}]{gilli11}
{Gilli}, R., {Su}, J., {Norman}, C., {et~al.} 2011, \apjl, 730, L28

\bibitem[{{Goulding} {et~al.}(2012){Goulding}, {Forman}, {Hickox}, {Jones},
  {Kraft}, {Murray}, {Vikhlinin}, {Coil}, {Cooper}, {Davis}, \&
  {Newman}}]{goulding12}
{Goulding}, A.~D., {Forman}, W.~R., {Hickox}, R.~C., {et~al.} 2012, \apjs, 202,
  6

\bibitem[{{Guo} {et~al.}(2011){Guo}, {White}, {Boylan-Kolchin}, {De Lucia},
  {Kauffmann}, {Lemson}, {Li}, {Springel}, \& {Weinmann}}]{guo11}
{Guo}, Q., {White}, S., {Boylan-Kolchin}, M., {et~al.} 2011, \mnras, 413, 101

\bibitem[{{Hasinger} {et~al.}(2005){Hasinger}, {Miyaji}, \&
  {Schmidt}}]{hasinger05}
{Hasinger}, G., {Miyaji}, T., \& {Schmidt}, M. 2005, \aap, 441, 417

\bibitem[{{Henriques} {et~al.}(2015){Henriques}, {White}, {Thomas}, {Angulo},
  {Guo}, {Lemson}, {Springel}, \& {Overzier}}]{henriques15}
{Henriques}, B. M.~B., {White}, S. D.~M., {Thomas}, P.~A., {et~al.} 2015,
  \mnras, 451, 2663

\bibitem[{{Hodges-Kluck} {et~al.}(2020){Hodges-Kluck}, {Gallo}, {Seth},
  {Greene}, \& {Baldassare}}]{hodges20}
{Hodges-Kluck}, E., {Gallo}, E., {Seth}, A., {Greene}, J., \& {Baldassare}, V.
  2020, arXiv e-prints, arXiv:2006.16342

\bibitem[{{Inayoshi} {et~al.}(2020){Inayoshi}, {Ichikawa}, \&
  {Ho}}]{inayoshi20}
{Inayoshi}, K., {Ichikawa}, K., \& {Ho}, L.~C. 2020, \apj, 894, 141

\bibitem[{{K{\"a}fer} {et~al.}(2020){K{\"a}fer}, {Finoguenov}, {Eckert},
  {Clerc}, {Ramos-Ceja}, {Sanders}, \& {Ghirardini}}]{kafer20}
{K{\"a}fer}, F., {Finoguenov}, A., {Eckert}, D., {et~al.} 2020, \aap, 634, A8

\bibitem[{{Khandai} {et~al.}(2015){Khandai}, {Di Matteo}, {Croft}, {Wilkins},
  {Feng}, {Tucker}, {DeGraf}, \& {Liu}}]{khandai15}
{Khandai}, N., {Di Matteo}, T., {Croft}, R., {et~al.} 2015, \mnras, 450, 1349

\bibitem[{{Kocevski} {et~al.}(2018){Kocevski}, {Hasinger}, {Brightman},
  {Nandra}, {Georgakakis}, {Cappelluti}, {Civano}, {Li}, {Li}, {Aird},
  {Alexander}, {Almaini}, {Brusa}, {Buchner}, {Comastri}, {Conselice},
  {Dickinson}, {Finoguenov}, {Gilli}, {Koekemoer}, {Miyaji}, {Mullaney},
  {Papovich}, {Rosario}, {Salvato}, {Silverman}, {Somerville}, \&
  {Ueda}}]{kocevski18}
{Kocevski}, D.~D., {Hasinger}, G., {Brightman}, M., {et~al.} 2018, \apjs, 236,
  48

\bibitem[{{Lacey} {et~al.}(2016){Lacey}, {Baugh}, {Frenk}, {Benson}, {Bower},
  {Cole}, {Gonzalez-Perez}, {Helly}, {Lagos}, \& {Mitchell}}]{lacey16}
{Lacey}, C.~G., {Baugh}, C.~M., {Frenk}, C.~S., {et~al.} 2016, \mnras, 462,
  3854

\bibitem[{{Lagos} {et~al.}(2018){Lagos}, {Tobar}, {Robotham}, {Obreschkow},
  {Mitchell}, {Power}, \& {Elahi}}]{lagos18}
{Lagos}, C. d.~P., {Tobar}, R.~J., {Robotham}, A. S.~G., {et~al.} 2018, \mnras,
  481, 3573

\bibitem[{{LaMassa} {et~al.}(2016){LaMassa}, {Urry}, {Cappelluti},
  {B{\"o}hringer}, {Comastri}, {Glikman}, {Richards}, {Ananna}, {Brusa},
  {Cardamone}, {Chon}, {Civano}, {Farrah}, {Gilfanov}, {Green}, {Komossa},
  {Lira}, {Makler}, {Marchesi}, {Pecoraro}, {Ranalli}, {Salvato}, {Schawinski},
  {Stern}, {Treister}, \& {Viero}}]{lamassa16}
{LaMassa}, S.~M., {Urry}, C.~M., {Cappelluti}, N., {et~al.} 2016, \apj, 817,
  172

\bibitem[{{LaMassa} {et~al.}(2013{\natexlab{a}}){LaMassa}, {Urry},
  {Cappelluti}, {Civano}, {Ranalli}, {Glikman}, {Treister}, {Richards},
  {Ballantyne}, {Stern}, {Comastri}, {Cardamone}, {Schawinski},
  {B{\"o}hringer}, {Chon}, {Murray}, {Green}, \& {Nandra}}]{lamassa13b}
{LaMassa}, S.~M., {Urry}, C.~M., {Cappelluti}, N., {et~al.} 2013{\natexlab{a}},
  \mnras, 436, 3581

\bibitem[{{LaMassa} {et~al.}(2013{\natexlab{b}}){LaMassa}, {Urry}, {Glikman},
  {Cappelluti}, {Civano}, {Comastri}, {Treister}, {B{\"o}hringer}, {Cardamone},
  {Chon}, {Kephart}, {Murray}, {Richards}, {Ross}, {Rozner}, \&
  {Schawinski}}]{lamassa13a}
{LaMassa}, S.~M., {Urry}, C.~M., {Glikman}, E., {et~al.} 2013{\natexlab{b}},
  \mnras, 432, 1351

\bibitem[{{Lanzuisi} {et~al.}(2018){Lanzuisi}, {Civano}, {Marchesi},
  {Comastri}, {Brusa}, {Gilli}, {Vignali}, {Zamorani}, {Brightman},
  {Griffiths}, \& {Koekemoer}}]{lanzuisi18}
{Lanzuisi}, G., {Civano}, F., {Marchesi}, S., {et~al.} 2018, \mnras, 480, 2578

\bibitem[{{Lanzuisi} {et~al.}(2015){Lanzuisi}, {Ranalli}, {Georgantopoulos},
  {Georgakakis}, {Delvecchio}, {Akylas}, {Berta}, {Bongiorno}, {Brusa},
  {Cappelluti}, {Civano}, {Comastri}, {Gilli}, {Gruppioni}, {Hasinger},
  {Iwasawa}, {Koekemoer}, {Lusso}, {Marchesi}, {Mainieri}, {Merloni},
  {Mignoli}, {Piconcelli}, {Pozzi}, {Rosario}, {Salvato}, {Silverman},
  {Trakhtenbrot}, {Vignali}, \& {Zamorani}}]{lanzuisi15}
{Lanzuisi}, G., {Ranalli}, P., {Georgantopoulos}, I., {et~al.} 2015, \aap, 573,
  A137

\bibitem[{{Leauthaud} {et~al.}(2010){Leauthaud}, {Finoguenov}, {Kneib},
  {Taylor}, {Massey}, {Rhodes}, {Ilbert}, {Bundy}, {Tinker}, {George}, {Capak},
  {Koekemoer}, {Johnston}, {Zhang}, {Cappelluti}, {Ellis}, {Elvis}, {Giodini},
  {Heymans}, {Le F{\`e}vre}, {Lilly}, {McCracken}, {Mellier},
  {R{\'e}fr{\'e}gier}, {Salvato}, {Scoville}, {Smoot}, {Tanaka}, {Van
  Waerbeke}, \& {Wolk}}]{leauthaud10}
{Leauthaud}, A., {Finoguenov}, A., {Kneib}, J.-P., {et~al.} 2010, \apj, 709, 97

\bibitem[{{Lehmer} {et~al.}(2009){Lehmer}, {Alexander}, {Chapman}, {Smail},
  {Bauer}, {Brandt}, {Geach}, {Matsuda}, {Mullaney}, \& {Swinbank}}]{lehmer09}
{Lehmer}, B.~D., {Alexander}, D.~M., {Chapman}, S.~C., {et~al.} 2009, \mnras,
  400, 299

\bibitem[{{Lehmer} {et~al.}(2016){Lehmer}, {Basu-Zych}, {Mineo}, {Brand t},
  {Eufrasio}, {Fragos}, {Hornschemeier}, {Luo}, {Xue}, {Bauer}, {Gilfanov},
  {Ranalli}, {Schneider}, {Shemmer}, {Tozzi}, {Trump}, {Vignali}, {Wang},
  {Yukita}, \& {Zezas}}]{lehmer16}
{Lehmer}, B.~D., {Basu-Zych}, A.~R., {Mineo}, S., {et~al.} 2016, \apj, 825, 7

\bibitem[{{Luo} {et~al.}(2017){Luo}, {Brandt}, {Xue}, {Lehmer}, {Alexander},
  {Bauer}, {Vito}, {Yang}, {Basu-Zych}, {Comastri}, {Gilli}, {Gu},
  {Hornschemeier}, {Koekemoer}, {Liu}, {Mainieri}, {Paolillo}, {Ranalli},
  {Rosati}, {Schneider}, {Shemmer}, {Smail}, {Sun}, {Tozzi}, {Vignali}, \&
  {Wang}}]{luo17}
{Luo}, B., {Brandt}, W.~N., {Xue}, Y.~Q., {et~al.} 2017, \apjs, 228, 2

\bibitem[{{Marchesi} {et~al.}(2018){Marchesi}, {Ajello}, {Marcotulli},
  {Comastri}, {Lanzuisi}, \& {Vignali}}]{marchesi18}
{Marchesi}, S., {Ajello}, M., {Marcotulli}, L., {et~al.} 2018, \apj, 854, 49

\bibitem[{{Marchesi} {et~al.}(2019){Marchesi}, {Ajello}, {Zhao}, {Marcotulli},
  {Balokovi{\'c}}, {Brightman}, {Comastri}, {Cusumano}, {Lanzuisi}, {La
  Parola}, {Segreto}, \& {Vignali}}]{marchesi19}
{Marchesi}, S., {Ajello}, M., {Zhao}, X., {et~al.} 2019, \apj, 872, 8

\bibitem[{{Marchesi} {et~al.}(2016{\natexlab{a}}){Marchesi}, {Civano}, {Elvis},
  {Salvato}, {Brusa}, {Comastri}, {Gilli}, {Hasinger}, {Lanzuisi}, {Miyaji},
  {Treister}, {Urry}, {Vignali}, {Zamorani}, {Allevato}, {Cappelluti},
  {Cardamone}, {Finoguenov}, {Griffiths}, {Karim}, {Laigle}, {LaMassa},
  {Jahnke}, {Ranalli}, {Schawinski}, {Schinnerer}, {Silverman}, {Smolcic},
  {Suh}, \& {Trakhtenbrot}}]{marchesi16a}
{Marchesi}, S., {Civano}, F., {Elvis}, M., {et~al.} 2016{\natexlab{a}}, \apj,
  817, 34

\bibitem[{{Marchesi} {et~al.}(2016{\natexlab{b}}){Marchesi}, {Civano},
  {Salvato}, {Shankar}, {Comastri}, {Elvis}, {Lanzuisi}, {Trakhtenbrot},
  {Vignali}, {Zamorani}, {Allevato}, {Brusa}, {Fiore}, {Gilli}, {Griffiths},
  {Hasinger}, {Miyaji}, {Schawinski}, {Treister}, \& {Urry}}]{marchesi16b}
{Marchesi}, S., {Civano}, F., {Salvato}, M., {et~al.} 2016{\natexlab{b}}, \apj,
  827, 150

\bibitem[{{Marchesi} {et~al.}(2016{\natexlab{c}}){Marchesi}, {Lanzuisi},
  {Civano}, {Iwasawa}, {Suh}, {Comastri}, {Zamorani}, {Allevato}, {Griffiths},
  {Miyaji}, {Ranalli}, {Salvato}, {Schawinski}, {Silverman}, {Treister},
  {Urry}, \& {Vignali}}]{marchesi16c}
{Marchesi}, S., {Lanzuisi}, G., {Civano}, F., {et~al.} 2016{\natexlab{c}},
  \apj, 830, 100

\bibitem[{{Maughan} {et~al.}(2008){Maughan}, {Jones}, {Forman}, \& {Van
  Speybroeck}}]{maughan08}
{Maughan}, B.~J., {Jones}, C., {Forman}, W., \& {Van Speybroeck}, L. 2008,
  \apjs, 174, 117

\bibitem[{{McAlpine} {et~al.}(2016){McAlpine}, {Helly}, {Schaller}, {Trayford},
  {Qu}, {Furlong}, {Bower}, {Crain}, {Schaye}, {Theuns}, {Dalla Vecchia},
  {Frenk}, {McCarthy}, {Jenkins}, {Rosas-Guevara}, {White}, {Baes}, {Camps}, \&
  {Lemson}}]{mcalpine16}
{McAlpine}, S., {Helly}, J.~C., {Schaller}, M., {et~al.} 2016, Astronomy and
  Computing, 15, 72

\bibitem[{{Meidinger} {et~al.}(2017){Meidinger}, {Barbera}, {Emberger},
  {F{\"u}rmetz}, {Manhart}, {M{\"u}ller-Seidlitz}, {Nandra}, {Plattner}, {Rau},
  \& {Treberspurg}}]{meidinger17}
{Meidinger}, N., {Barbera}, M., {Emberger}, V., {et~al.} 2017, in Society of
  Photo-Optical Instrumentation Engineers (SPIE) Conference Series, Vol. 10397,
  \procspie, 103970V

\bibitem[{{Merloni} {et~al.}(2012){Merloni}, {Predehl}, {Becker},
  {B{\"o}hringer}, {Boller}, {Brunner}, {Brusa}, {Dennerl}, {Freyberg},
  {Friedrich}, {Georgakakis}, {Haberl}, {Hasinger}, {Meidinger}, {Mohr},
  {Nandra}, {Rau}, {Reiprich}, {Robrade}, {Salvato}, {Santangelo}, {Sasaki},
  {Schwope}, {Wilms}, \& {German eROSITA Consortium}}]{merloni12}
{Merloni}, A., {Predehl}, P., {Becker}, W., {et~al.} 2012, arXiv e-prints,
  arXiv:1209.3114

\bibitem[{{Miyaji} {et~al.}(2015){Miyaji}, {Hasinger}, {Salvato}, {Brusa},
  {Cappelluti}, {Civano}, {Puccetti}, {Elvis}, {Brunner}, {Fotopoulou}, {Ueda},
  {Griffiths}, {Koekemoer}, {Akiyama}, {Comastri}, {Gilli}, {Lanzuisi},
  {Merloni}, \& {Vignali}}]{miyaji15}
{Miyaji}, T., {Hasinger}, G., {Salvato}, M., {et~al.} 2015, \apj, 804, 104

\bibitem[{{Murphy} \& {Yaqoob}(2009)}]{murphy09}
{Murphy}, K.~D. \& {Yaqoob}, T. 2009, \mnras, 397, 1549

\bibitem[{{Murray} {et~al.}(2005){Murray}, {Kenter}, {Forman}, {Jones},
  {Green}, {Kochanek}, {Vikhlinin}, {Fabricant}, {Fazio}, {Brand}, {Brown},
  {Dey}, {Jannuzi}, {Najita}, {McNamara}, {Shields}, \& {Rieke}}]{murray05}
{Murray}, S.~S., {Kenter}, A., {Forman}, W.~R., {et~al.} 2005, \apjs, 161, 1

\bibitem[{{Mushotzky} {et~al.}(2019){Mushotzky}, {Aird}, {Barger},
  {Cappelluti}, {Chartas}, {Corrales}, {Eufrasio}, {Fabian}, {Falcone},
  {Gallo}, {Gilli}, {Grant}, {Hardcastle}, {Hodges-Kluck}, {Kara}, {Koss},
  {Li}, {Lisse}, {Loewenstein}, {Markevitch}, {Meyer}, {Miller}, {Mulchaey},
  {Petre}, {Ptak}, {Reynolds}, {Russell}, {Safi-Harb}, {Smith}, {Snios},
  {Tombesi}, {Valencic}, {Walker}, {Williams}, {Winter}, {Yamaguchi}, {Zhang},
  {Arenberg}, {Brand t}, {Burrows}, {Georganopoulos}, {Miller}, {Norman}, \&
  {Rosati}}]{mushotzky19}
{Mushotzky}, R.~F., {Aird}, J., {Barger}, A.~J., {et~al.} 2019, arXiv e-prints,
  arXiv:1903.04083

\bibitem[{{Mutch} {et~al.}(2016){Mutch}, {Geil}, {Poole}, {Angel}, {Duffy},
  {Mesinger}, \& {Wyithe}}]{mutch16}
{Mutch}, S.~J., {Geil}, P.~M., {Poole}, G.~B., {et~al.} 2016, \mnras, 462, 250

\bibitem[{{Nandra} {et~al.}(2013){Nandra}, {Barret}, {Barcons}, {Fabian}, {den
  Herder}, {Piro}, {Watson}, {Adami}, {Aird}, {Afonso}, {Alexander},
  {Argiroffi}, {Amati}, {Arnaud}, {Atteia}, {Audard}, {Badenes}, {Ballet},
  {Ballo}, {Bamba}, {Bhardwaj}, {Stefano Battistelli}, {Becker}, {De Becker},
  {Behar}, {Bianchi}, {Biffi}, {B{\^\i}rzan}, {Bocchino}, {Bogdanov}, {Boirin},
  {Boller}, {Borgani}, {Borm}, {Bouch{\'e}}, {Bourdin}, {Bower}, {Braito},
  {Branchini}, {Branduardi-Raymont}, {Bregman}, {Brenneman}, {Brightman},
  {Br{\"u}ggen}, {Buchner}, {Bulbul}, {Brusa}, {Bursa}, {Caccianiga},
  {Cackett}, {Campana}, {Cappelluti}, {Cappi}, {Carrera}, {Ceballos},
  {Christensen}, {Chu}, {Churazov}, {Clerc}, {Corbel}, {Corral}, {Comastri},
  {Costantini}, {Croston}, {Dadina}, {D'Ai}, {Decourchelle}, {Della Ceca},
  {Dennerl}, {Dolag}, {Done}, {Dovciak}, {Drake}, {Eckert}, {Edge}, {Ettori},
  {Ezoe}, {Feigelson}, {Fender}, {Feruglio}, {Finoguenov}, {Fiore}, {Galeazzi},
  {Gallagher}, {Gandhi}, {Gaspari}, {Gastaldello}, {Georgakakis},
  {Georgantopoulos}, {Gilfanov}, {Gitti}, {Gladstone}, {Goosmann}, {Gosset},
  {Grosso}, {Guedel}, {Guerrero}, {Haberl}, {Hardcastle}, {Heinz}, {Alonso
  Herrero}, {Herv{\'e}}, {Holmstrom}, {Iwasawa}, {Jonker}, {Kaastra}, {Kara},
  {Karas}, {Kastner}, {King}, {Kosenko}, {Koutroumpa}, {Kraft}, {Kreykenbohm},
  {Lallement}, {Lanzuisi}, {Lee}, {Lemoine-Goumard}, {Lobban}, {Lodato},
  {Lovisari}, {Lotti}, {McCharthy}, {McNamara}, {Maggio}, {Maiolino}, {De
  Marco}, {de Martino}, {Mateos}, {Matt}, {Maughan}, {Mazzotta}, {Mendez},
  {Merloni}, {Micela}, {Miceli}, {Mignani}, {Miller}, {Miniutti}, {Molendi},
  {Montez}, {Moretti}, {Motch}, {Naz{\'e}}, {Nevalainen}, {Nicastro}, {Nulsen},
  {Ohashi}, {O'Brien}, {Osborne}, {Oskinova}, {Pacaud}, {Paerels}, {Page},
  {Papadakis}, {Pareschi}, {Petre}, {Petrucci}, {Piconcelli}, {Pillitteri},
  {Pinto}, {de Plaa}, {Pointecouteau}, {Ponman}, {Ponti}, {Porquet}, {Pounds},
  {Pratt}, {Predehl}, {Proga}, {Psaltis}, {Rafferty}, {Ramos-Ceja}, {Ranalli},
  {Rasia}, {Rau}, {Rauw}, {Rea}, {Read}, {Reeves}, {Reiprich}, {Renaud},
  {Reynolds}, {Risaliti}, {Rodriguez}, {Rodriguez Hidalgo}, {Roncarelli},
  {Rosario}, {Rossetti}, {Rozanska}, {Rovilos}, {Salvaterra}, {Salvato}, {Di
  Salvo}, {Sanders}, {Sanz-Forcada}, {Schawinski}, {Schaye}, {Schwope},
  {Sciortino}, {Severgnini}, {Shankar}, {Sijacki}, {Sim}, {Schmid}, {Smith},
  {Steiner}, {Stelzer}, {Stewart}, {Strohmayer}, {Str{\"u}der}, {Sun}, {Takei},
  {Tatischeff}, {Tiengo}, {Tombesi}, {Trinchieri}, {Tsuru}, {Ud-Doula},
  {Ursino}, {Valencic}, {Vanzella}, {Vaughan}, {Vignali}, {Vink}, {Vito},
  {Volonteri}, {Wang}, {Webb}, {Willingale}, {Wilms}, {Wise}, {Worrall},
  {Young}, {Zampieri}, {In't Zand}, {Zane}, {Zezas}, {Zhang}, \&
  {Zhuravleva}}]{nandra13}
{Nandra}, K., {Barret}, D., {Barcons}, X., {et~al.} 2013, arXiv e-prints,
  arXiv:1306.2307

\bibitem[{{Nandra} {et~al.}(2015){Nandra}, {Laird}, {Aird}, {Salvato},
  {Georgakakis}, {Barro}, {Perez-Gonzalez}, {Barmby}, {Chary}, {Coil},
  {Cooper}, {Davis}, {Dickinson}, {Faber}, {Fazio}, {Guhathakurta}, {Gwyn},
  {Hsu}, {Huang}, {Ivison}, {Koo}, {Newman}, {Rangel}, {Yamada}, \&
  {Willmer}}]{nandra15}
{Nandra}, K., {Laird}, E.~S., {Aird}, J.~A., {et~al.} 2015, \apjs, 220, 10

\bibitem[{{Nandra} {et~al.}(2007){Nandra}, {O'Neill}, {George}, \&
  {Reeves}}]{nandra07}
{Nandra}, K., {O'Neill}, P.~M., {George}, I.~M., \& {Reeves}, J.~N. 2007,
  \mnras, 382, 194

\bibitem[{{Nanni} {et~al.}(2020){Nanni}, {Gilli}, {Vignali}, {Mignoli}, {Peca},
  {Marchesi}, {Annunziatella}, {Brusa}, {Calura}, {Cappelluti}, {Chiaberge},
  {Comastri}, {Iwasawa}, {Lanzuisi}, {Liuzzo}, {Marchesini}, {Prandoni},
  {Tozzi}, {Vito}, {Zamorani}, \& {Norman}}]{nanni20}
{Nanni}, R., {Gilli}, R., {Vignali}, C., {et~al.} 2020, \aap, 637, A52

\bibitem[{{Nanni} {et~al.}(2017){Nanni}, {Vignali}, {Gilli}, {Moretti}, \&
  {Brand t}}]{nanni17}
{Nanni}, R., {Vignali}, C., {Gilli}, R., {Moretti}, A., \& {Brand t}, W.~N.
  2017, \aap, 603, A128

\bibitem[{{Ni} {et~al.}(2020){Ni}, {Matteo}, {Gilli}, {Croft}, {Feng}, \&
  {Norman}}]{ni20}
{Ni}, Y., {Matteo}, T.~D., {Gilli}, R., {et~al.} 2020, \mnras
  [\eprint[arXiv]{1912.03780}]

\bibitem[{{Pierre} {et~al.}(2016){Pierre}, {Pacaud}, {Adami}, {Alis},
  {Altieri}, {Baran}, {Benoist}, {Birkinshaw}, {Bongiorno}, {Bremer}, {Brusa},
  {Butler}, {Ciliegi}, {Chiappetti}, {Clerc}, {Corasaniti}, {Coupon}, {De
  Breuck}, {Democles}, {Desai}, {Delhaize}, {Devriendt}, {Dubois}, {Eckert},
  {Elyiv}, {Ettori}, {Evrard}, {Faccioli}, {Farahi}, {Ferrari}, {Finet},
  {Fotopoulou}, {Fourmanoit}, {Gandhi}, {Gastaldello}, {Gastaud},
  {Georgantopoulos}, {Giles}, {Guennou}, {Guglielmo}, {Horellou}, {Husband},
  {Huynh}, {Iovino}, {Kilbinger}, {Koulouridis}, {Lavoie}, {Le Brun}, {Le
  Fevre}, {Lidman}, {Lieu}, {Lin}, {Mantz}, {Maughan}, {Maurogordato},
  {McCarthy}, {McGee}, {Melin}, {Melnyk}, {Menanteau}, {Novak}, {Paltani},
  {Plionis}, {Poggianti}, {Pomarede}, {Pompei}, {Ponman}, {Ramos-Ceja},
  {Ranalli}, {Rapetti}, {Raychaudury}, {Reiprich}, {Rottgering}, {Rozo},
  {Rykoff}, {Sadibekova}, {Santos}, {Sauvageot}, {Schimd}, {Sereno}, {Smith},
  {Smol{\v{c}}i{\'c}}, {Snowden}, {Spergel}, {Stanford}, {Surdej}, {Valageas},
  {Valotti}, {Valtchanov}, {Vignali}, {Willis}, \& {Ziparo}}]{pierre16}
{Pierre}, M., {Pacaud}, F., {Adami}, C., {et~al.} 2016, \aap, 592, A1

\bibitem[{{Planck Collaboration} {et~al.}(2014){Planck Collaboration}, {Ade},
  {Aghanim}, {Armitage-Caplan}, {Arnaud}, {Ashdown}, {Atrio-Barand ela},
  {Aumont}, {Baccigalupi}, {Banday}, {Barreiro}, {Bartlett}, {Battaner},
  {Benabed}, {Beno{\^\i}t}, {Benoit-L{\'e}vy}, {Bernard}, {Bersanelli},
  {Bielewicz}, {Bobin}, {Bock}, {Bonaldi}, {Bond}, {Borrill}, {Bouchet},
  {Bridges}, {Bucher}, {Burigana}, {Butler}, {Calabrese}, {Cappellini},
  {Cardoso}, {Catalano}, {Challinor}, {Chamballu}, {Chary}, {Chen}, {Chiang},
  {Chiang}, {Christensen}, {Church}, {Clements}, {Colombi}, {Colombo},
  {Couchot}, {Coulais}, {Crill}, {Curto}, {Cuttaia}, {Danese}, {Davies},
  {Davis}, {de Bernardis}, {de Rosa}, {de Zotti}, {Delabrouille}, {Delouis},
  {D{\'e}sert}, {Dickinson}, {Diego}, {Dolag}, {Dole}, {Donzelli}, {Dor{\'e}},
  {Douspis}, {Dunkley}, {Dupac}, {Efstathiou}, {Elsner}, {En{\ss}lin},
  {Eriksen}, {Finelli}, {Forni}, {Frailis}, {Fraisse}, {Franceschi}, {Gaier},
  {Galeotta}, {Galli}, {Ganga}, {Giard}, {Giardino}, {Giraud-H{\'e}raud},
  {Gjerl{\o}w}, {Gonz{\'a}lez-Nuevo}, {G{\'o}rski}, {Gratton}, {Gregorio},
  {Gruppuso}, {Gudmundsson}, {Haissinski}, {Hamann}, {Hansen}, {Hanson},
  {Harrison}, {Henrot-Versill{\'e}}, {Hern{\'a}ndez-Monteagudo}, {Herranz},
  {Hildebrand t}, {Hivon}, {Hobson}, {Holmes}, {Hornstrup}, {Hou}, {Hovest},
  {Huffenberger}, {Jaffe}, {Jaffe}, {Jewell}, {Jones}, {Juvela},
  {Keih{\"a}nen}, {Keskitalo}, {Kisner}, {Kneissl}, {Knoche}, {Knox}, {Kunz},
  {Kurki-Suonio}, {Lagache}, {L{\"a}hteenm{\"a}ki}, {Lamarre}, {Lasenby},
  {Lattanzi}, {Laureijs}, {Lawrence}, {Leach}, {Leahy}, {Leonardi},
  {Le{\'o}n-Tavares}, {Lesgourgues}, {Lewis}, {Liguori}, {Lilje},
  {Linden-V{\o}rnle}, {L{\'o}pez-Caniego}, {Lubin}, {Mac{\'\i}as-P{\'e}rez},
  {Maffei}, {Maino}, {Mand olesi}, {Maris}, {Marshall}, {Martin},
  {Mart{\'\i}nez-Gonz{\'a}lez}, {Masi}, {Massardi}, {Matarrese}, {Matthai},
  {Mazzotta}, {Meinhold}, {Melchiorri}, {Melin}, {Mendes}, {Menegoni},
  {Mennella}, {Migliaccio}, {Millea}, {Mitra}, {Miville-Desch{\^e}nes},
  {Moneti}, {Montier}, {Morgante}, {Mortlock}, {Moss}, {Munshi}, {Murphy},
  {Naselsky}, {Nati}, {Natoli}, {Netterfield}, {N{\o}rgaard-Nielsen},
  {Noviello}, {Novikov}, {Novikov}, {O'Dwyer}, {Osborne}, {Oxborrow}, {Paci},
  {Pagano}, {Pajot}, {Paladini}, {Paoletti}, {Partridge}, {Pasian},
  {Patanchon}, {Pearson}, {Pearson}, {Peiris}, {Perdereau}, {Perotto},
  {Perrotta}, {Pettorino}, {Piacentini}, {Piat}, {Pierpaoli}, {Pietrobon},
  {Plaszczynski}, {Platania}, {Pointecouteau}, {Polenta}, {Ponthieu}, {Popa},
  {Poutanen}, {Pratt}, {Pr{\'e}zeau}, {Prunet}, {Puget}, {Rachen}, {Reach},
  {Rebolo}, {Reinecke}, {Remazeilles}, {Renault}, {Ricciardi}, {Riller},
  {Ristorcelli}, {Rocha}, {Rosset}, {Roudier}, {Rowan-Robinson},
  {Rubi{\~n}o-Mart{\'\i}n}, {Rusholme}, {Sandri}, {Santos}, {Savelainen},
  {Savini}, {Scott}, {Seiffert}, {Shellard}, {Spencer}, {Starck}, {Stolyarov},
  {Stompor}, {Sudiwala}, {Sunyaev}, {Sureau}, {Sutton}, {Suur-Uski}, {Sygnet},
  {Tauber}, {Tavagnacco}, {Terenzi}, {Toffolatti}, {Tomasi}, {Tristram},
  {Tucci}, {Tuovinen}, {T{\"u}rler}, {Umana}, {Valenziano}, {Valiviita}, {Van
  Tent}, {Vielva}, {Villa}, {Vittorio}, {Wade}, {Wandelt}, {Wehus}, {White},
  {White}, {Wilkinson}, {Yvon}, {Zacchei}, \& {Zonca}}]{planck14}
{Planck Collaboration}, {Ade}, P.~A.~R., {Aghanim}, N., {et~al.} 2014, \aap,
  571, A16

\bibitem[{{Pratt} {et~al.}(2019){Pratt}, {Arnaud}, {Biviano}, {Eckert},
  {Ettori}, {Nagai}, {Okabe}, \& {Reiprich}}]{pratt19}
{Pratt}, G.~W., {Arnaud}, M., {Biviano}, A., {et~al.} 2019, \ssr, 215, 25

\bibitem[{{Qin} {et~al.}(2017){Qin}, {Mutch}, {Poole}, {Liu}, {Angel}, {Duffy},
  {Geil}, {Mesinger}, \& {Wyithe}}]{qin17}
{Qin}, Y., {Mutch}, S.~J., {Poole}, G.~B., {et~al.} 2017, \mnras, 472, 2009

\bibitem[{{Ranalli} {et~al.}(2005){Ranalli}, {Comastri}, \&
  {Setti}}]{ranalli05}
{Ranalli}, P., {Comastri}, A., \& {Setti}, G. 2005, \aap, 440, 23

\bibitem[{{Reichert} {et~al.}(2011){Reichert}, {B{\"o}hringer}, {Fassbender},
  \& {M{\"u}hlegger}}]{reichert11}
{Reichert}, A., {B{\"o}hringer}, H., {Fassbender}, R., \& {M{\"u}hlegger}, M.
  2011, \aap, 535, A4

\bibitem[{{Ricci} {et~al.}(2015){Ricci}, {Ueda}, {Koss}, {Trakhtenbrot},
  {Bauer}, \& {Gandhi}}]{ricci15}
{Ricci}, C., {Ueda}, Y., {Koss}, M.~J., {et~al.} 2015, \apjl, 815, L13

\bibitem[{{Rosati} {et~al.}(2002){Rosati}, {Borgani}, \& {Norman}}]{rosati02}
{Rosati}, P., {Borgani}, S., \& {Norman}, C. 2002, \araa, 40, 539

\bibitem[{{Sazonov} \& {Khabibullin}(2017)}]{sazonov17}
{Sazonov}, S. \& {Khabibullin}, I. 2017, \mnras, 468, 2249

\bibitem[{{Schaye} {et~al.}(2015){Schaye}, {Crain}, {Bower}, {Furlong},
  {Schaller}, {Theuns}, {Dalla Vecchia}, {Frenk}, {McCarthy}, {Helly},
  {Jenkins}, {Rosas-Guevara}, {White}, {Baes}, {Booth}, {Camps}, {Navarro},
  {Qu}, {Rahmati}, {Sawala}, {Thomas}, \& {Trayford}}]{schaye15}
{Schaye}, J., {Crain}, R.~A., {Bower}, R.~G., {et~al.} 2015, \mnras, 446, 521

\bibitem[{{Schmidt} {et~al.}(1995){Schmidt}, {Schneider}, \&
  {Gunn}}]{schmidt95}
{Schmidt}, M., {Schneider}, D.~P., \& {Gunn}, J.~E. 1995, \aj, 110, 68

\bibitem[{{Stern} {et~al.}(2012){Stern}, {Assef}, {Benford}, {Blain}, {Cutri},
  {Dey}, {Eisenhardt}, {Griffith}, {Jarrett}, {Lake}, {Masci}, {Petty},
  {Stanford}, {Tsai}, {Wright}, {Yan}, {Harrison}, \& {Madsen}}]{stern12}
{Stern}, D., {Assef}, R.~J., {Benford}, D.~J., {et~al.} 2012, \apj, 753, 30

\bibitem[{{The Lynx Team}(2018)}]{lynx18}
{The Lynx Team}. 2018, arXiv e-prints, arXiv:1809.09642

\bibitem[{{Tinker} {et~al.}(2008){Tinker}, {Kravtsov}, {Klypin}, {Abazajian},
  {Warren}, {Yepes}, {Gottl{\"o}ber}, \& {Holz}}]{tinker08}
{Tinker}, J., {Kravtsov}, A.~V., {Klypin}, A., {et~al.} 2008, \apj, 688, 709

\bibitem[{{Treister} {et~al.}(2013){Treister}, {Schawinski}, {Volonteri}, \&
  {Natarajan}}]{treister13}
{Treister}, E., {Schawinski}, K., {Volonteri}, M., \& {Natarajan}, P. 2013,
  \apj, 778, 130

\bibitem[{{Ueda} {et~al.}(2014){Ueda}, {Akiyama}, {Hasinger}, {Miyaji}, \&
  {Watson}}]{ueda14}
{Ueda}, Y., {Akiyama}, M., {Hasinger}, G., {Miyaji}, T., \& {Watson}, M.~G.
  2014, \apj, 786, 104

\bibitem[{{Vito} {et~al.}(2019){Vito}, {Brandt}, {Bauer}, {Calura}, {Gilli},
  {Luo}, {Shemmer}, {Vignali}, {Zamorani}, {Brusa}, {Civano}, {Comastri}, \&
  {Nanni}}]{vito19}
{Vito}, F., {Brandt}, W.~N., {Bauer}, F.~E., {et~al.} 2019, \aap, 630, A118

\bibitem[{{Vito} {et~al.}(2018){Vito}, {Brandt}, {Yang}, {Gilli}, {Luo},
  {Vignali}, {Xue}, {Comastri}, {Koekemoer}, {Lehmer}, {Liu}, {Paolillo},
  {Ranalli}, {Schneider}, {Shemmer}, {Volonteri}, \& {Wang}}]{vito18}
{Vito}, F., {Brandt}, W.~N., {Yang}, G., {et~al.} 2018, \mnras, 473, 2378

\bibitem[{{Vito} {et~al.}(2014){Vito}, {Gilli}, {Vignali}, {Comastri}, {Brusa},
  {Cappelluti}, \& {Iwasawa}}]{vito14}
{Vito}, F., {Gilli}, R., {Vignali}, C., {et~al.} 2014, \mnras, 445, 3557

\bibitem[{{Vogelsberger} {et~al.}(2014){Vogelsberger}, {Genel}, {Springel},
  {Torrey}, {Sijacki}, {Xu}, {Snyder}, {Nelson}, \&
  {Hernquist}}]{vogelsberger14}
{Vogelsberger}, M., {Genel}, S., {Springel}, V., {et~al.} 2014, \mnras, 444,
  1518

\bibitem[{{Wang} {et~al.}(2017){Wang}, {Fan}, {Yang}, {Wu}, {Yang}, {Bian},
  {McGreer}, {Li}, {Li}, {Ding}, {Dey}, {Dye}, {Findlay}, {Green}, {James},
  {Jiang}, {Lang}, {Lawrence}, {Myers}, {Ross}, {Schlegel}, \&
  {Shanks}}]{wang17}
{Wang}, F., {Fan}, X., {Yang}, J., {et~al.} 2017, \apj, 839, 27

\bibitem[{{Xue} {et~al.}(2016){Xue}, {Luo}, {Brandt}, {Alexander}, {Bauer},
  {Lehmer}, \& {Yang}}]{xue16}
{Xue}, Y.~Q., {Luo}, B., {Brandt}, W.~N., {et~al.} 2016, \apjs, 224, 15

\bibitem[{{Xue} {et~al.}(2011){Xue}, {Luo}, {Brandt}, {Bauer}, {Lehmer},
  {Broos}, {Schneider}, {Alexand er}, {Brusa}, {Comastri}, {Fabian}, {Gilli},
  {Hasinger}, {Hornschemeier}, {Koekemoer}, {Liu}, {Mainieri}, {Paolillo},
  {Rafferty}, {Rosati}, {Shemmer}, {Silverman}, {Smail}, {Tozzi}, \&
  {Vignali}}]{xue11}
{Xue}, Y.~Q., {Luo}, B., {Brandt}, W.~N., {et~al.} 2011, \apjs, 195, 10

\end{thebibliography}

\end{document}